\documentclass[11pt]{article}
\newtheorem{definition}{Definition}[section]
\newtheorem{proposition}{Proposition}[section]

\usepackage[bookmarks=false]{hyperref}
\usepackage{amssymb}
\usepackage{amsmath}
\usepackage{color}
\usepackage{graphicx}
\usepackage{rotating} 
\usepackage{multirow} 
\usepackage{bm}
\usepackage{hyperref}
\usepackage{epsfig}
\usepackage{graphics}
\usepackage{subcaption}
\begin{document}
\begin{center}
{\LARGE \bf{alphastable}: An \verb|R| Package for Modelling Multivariate Stable and Mixture of Symmetric Stable Distributions}\\
\vspace{.5cm} 
 Mahdi Teimouri\\
  Department of Statistics, Faculty of Science, Gonbad Kavous University\\
  No.163, Basirat Blvd, Gonbad Kavous, Iran\\
   teimouri@aut.ac.ir\\[2ex]
Mahdi Torshizi\\
Department of Computer Engineering, Faculty of Science, Gonbad Kavous University\\
  No.163, Basirat Blvd, Gonbad Kavous, Iran\\
   mtorshizi@gonbad.ac.ir\\[2ex]
Adel Mohammadpour\\
  Department of Statistics, Faculty of Mathematics and Computer Science, Amirkabir University of Technology (Tehran Polytechnic)\\
  424 Hafez Ave., Tehran 15914, Iran\\
adel@aut.ac.ir\\[2ex]
Saralees Nadarajah\\
  School of Mathematics, University of Manchester\\
  Manchester M13 9PL, UK\\
  mbbsssn2@manchester.ac.uk
\end{center}
%\maketitle
\vspace{1.5cm}
{\bf Abstract:}~~
The family of stable distributions received extensive applications in many fields of studies since it incorporates both the skewness and heavy tails. In this paper, we introduce a package written in the \verb|R| language called {\bf{alphastable}}. The {\bf{alphastable}} performs a variety of tasks including: 1- generating random numbers from univariate, truncated, and multivariate stable distributions. 2- computing the probability density function of univariate and multivariate elliptically contoured stable distributions, 3- computing the distribution function of univariate stable distributions, 4- estimating the parameters of univariate symmetric stable, univariate Cauchy, mixture of Cauchy, mixture of univariate symmetric stable, multivariate elliptically contoured stable, and multivariate strictly stable distributions. This package, as it will be shown, is very useful for modelling data in univariate and multivariate cases that arise in the fields of finance and economics.

\vspace{.5cm}\noindent
\section{Introduction}
Different types of data found in applications arise from a distribution whose probability density function (pdf) is heavy-tailed, asymmetric, or both heavy-tailed and asymmetric. Despite lack of closed-form expression for the pdf, the class of stable distributions is the most appropriate candidate for modelling processes whose frequency curve is heavy-tailed and skewed for which, the normal distribution is quite inappropriate. The family of stable distributions is becoming increasingly popular in some fields such as biology, ecology, economics, finance, genetics, insurance, physics, physiology, and telecommunications. Details for the applications of stable distributions in aforementioned fields can be found in \cite{johnson2004information}, \cite{klebanov2006ill}, \cite{mandelbrot2007misbehavior}, \cite{mittnik2003prediction}, \cite{nikias1995signal}, \cite{nolan2003modeling}, \cite{ortobelli2010risk}, \cite{rachev2000stable}, \cite{rachev2003handbook}, \cite{rachev2005fat}, and \cite{samorodnitsky1994stable}. As the central member of this family, the Cauchy distribution itself received many applications in a variety of such fields, for example, as applied physics (\cite{stapf1996proton, min1996levy}), electrical engineering \cite{winterton1992source}, seismology \cite{kagan1992correlations}, and theoretical physics \cite{lorentz1906absorption}. The class of stable distributions is introduced in terms of its characteristic function (chf). The chf of a univariate stable distribution takes different parameterizations. Among them, we shall refer to two important forms which, are known in the literature as $S_{0}$ and $S_{1}$, see (\cite{nolan1998parameterizations, zolotarev1986one}). In what follows, we review briefly $S_{0}$ and $S_{1}$ parameterizations. If random variable $Y$ follows a stable distribution, then the chf of $Y$, i.e. $\varphi_{Y}{(t)}=E \exp (jtY)$, in $S_{0}$ parameterization is given by
\begin{eqnarray} 
\label{chf0} 
\varphi_{Y}{(t)}=\left\{\begin{array}{*{20}c} 
\exp\biggl\{-\left| \sigma t \right|^\alpha\Big[1-j\beta~\mathrm{sgn}(t)\tan\left(\frac{\pi \alpha}{2}\right)\big[(\sigma |t|)^{1-\alpha}-1\big]\Big]+jt\mu_{0}\biggr\},~~~~\mathrm{{if}}~~\alpha \ne 1,\\ 
\exp\biggl\{-\left| \sigma t \right| \Big[1+j\beta~\mathrm{sgn}(t)\frac{2}{\pi}\log \left| t \sigma \right|\Big]+j t\mu_{0} \biggr\},~~~~~~~~~~~~~~~~~~~~~~~~~\mathrm{if}~\alpha= 1. 
\end{array} \right. 
\end{eqnarray} 
where $j^{2}$=-1 and $\mathrm{sgn}(.)$ is the well-known sign function. The family of stable distributions has four parameters: tail index $\alpha \in (0, 2]$, skewness $\beta \in [-1, 1]$, scale $\sigma \in \mathbb{R}^{+}$, and location $\mu_{0} \in \mathbb{R}$. If $\beta$=0, it would be the class of symmetric stable distributions. If $\beta$=1 and $\alpha< 1$, we have the class of the positive stable distributions. In this case, $Y$ varies over the positive semi-axis of real line. The chf of $Y$ in $S_{1}$ parameterization is given by
\begin{eqnarray} 
\label{chf1} 
\varphi_{Y}{(t)}=\left\{\begin{array}{*{20}c} 
\exp\biggl\{-\left| \sigma t \right|^\alpha\Big[1-j\beta~\mathrm{sgn}(t)\tan\left(\frac{\pi \alpha}{2}\right)\Big]+jt\mu_{1} \biggr\},~~~~\mathrm{{if}}~\alpha \ne 1,\\ 
\exp\biggl\{-\left| \sigma t \right| \Big[1+j\beta~\mathrm{sgn}(t)\frac{2}{\pi}\log \left| t \right|\Big]+j t\mu_{1} \biggr\},~~~~~~~\mathrm{if}~\alpha= 1, 
\end{array} \right. 
\end{eqnarray} 
where $\mu_{1} \in \mathbb{R}$ is the location parameter. The pdf and chf in $S_{0}$ parameterization are continuous functions of all four parameters while the pdf and chf in $S_{1}$ parameterization both have a discontinuity at $\alpha=1$. It can be seen that the location parameters in $S_{0}$ and $S_{1}$ parameterizations are related as
\begin{eqnarray}
\mu_{1}=\left\{\begin{array}{*{20}c} 
\mu_{0}-\beta \sigma \tan\left(\frac{\pi \alpha}{2}\right),~~~~~\mathrm{{if}}~\alpha \neq1,\\ 
\mu_{0}-\beta\frac{2}{\pi} \sigma \log \sigma,~~~~~~~~~\mathrm{if}~ \alpha= 1.\\ 
\end{array} \right. 
\end{eqnarray} 
It is clear that in symmetric case, both parameterizations are equal. Hereafter, we write $S_{0}(\alpha,\beta,\sigma,\mu)$ and $S_{1}(\alpha,\beta,\sigma,\mu)$ to denote the class of stable distributions in $S_{0}$ and $S_{1}$ parameterizations, respectively. Also, we assume that $Y$ follows stable distribution.
\par 
Frequency curve of phenomena such as asset return \cite{nolan2003modeling}, flood \cite{bernardara2008flood}, medical expenditure \cite{venturini2008gamma}, rainfall \cite{carreau2009statistical}, and precipitation \cite{neykov2014stochastic} exhibits heavy tails, multimodality, or both heavy tails and multimodality. Mixture models are very useful tool for modelling data whose frequency curve is multimodal. Also, the mixture models are widely used in model-based clustering. The normal mixture models are frequently used for model-based clustering, but they show poor performance in the presence of outliers. So, the robust mixture models which, aim to tackle tail-weight (\cite{andrews2012model, peel2000robust}); deal with skewness \cite{lin2007finite}; or account for both (\cite{basso2010robust, browne2015mixture, lee2013mixtures, lee2014finite, subedi2014variational, vrbik2014parsimonious}) are frequently used in model-based clustering. Since stable distributions have heavy tails, the symmetric stable mixture models (SSMM) can be considered as the best choice for robust mixture modelling. The pdf of a $K$-component SSMM has the form 
\begin{equation}\label{mix} 
g(y|\Theta)=\sum_{j=1}^{K}\omega_j f_{Y_{j}}(y|\alpha_j,0, \sigma_j, \mu_j), 
\end{equation} 
where $\Theta=(\boldsymbol{\alpha}^T, \boldsymbol{\sigma}^T, \boldsymbol{\mu}^T, \boldsymbol{\omega}^T)^T$; for $\boldsymbol{\alpha}=(\alpha_1,\dots,\alpha_K)^T$, $\boldsymbol{\sigma}=(\sigma_1,\dots,\sigma_K)^T$, $\boldsymbol{\mu}=(\mu_1,\dots,\mu_K)^T$, and $\boldsymbol{\omega}=(\omega_1,\dots,\omega_K)^T$ is the vector of parameters and $f_{Y_{j}}(.|\alpha_j, 0, \sigma_j, \mu_j)$ denotes the pdf of $j$th component which, follows a symmetric stable distribution. The mixing parameters $\omega_j$s are non-negative and sum to one, i.e., $\sum_{j=1}^{K} \omega_j=1$. Using some type of the EM algorithm, namely the expectation conditional maximization either algorithm (ECME), the parameters of SSMM can be estimated, see \cite{teimouri2018algorithm}. The performance and robustness of the proposed EM algorithm versus model violations and presence of outliers were proved via comprehensive simulations for different scenarios. For example, when data follow $t$ mixture model, the SSMM outperforms other mixture models including generalized hyperbolic (GH), normal, skew normal, and skew $t$. Also, when data follow a mixture of exponential power distribution, the SSMM performs better than a mixture of GH, normal, skew normal, and $t$ distributions.
\par Truncated stable random variables arise in many applied fields such as characterization of impulsive network traffic \cite{tang2009characterizing} and GARCH models and option pricing \cite{menn2009smoothly}. Some algorithms proposed in \cite{nadarajah2007programs}, \cite{soltani2010truncated}, \cite{shirvani2013characterization}, \cite{teimouri2013simulating}, and \cite{teimouri2017simulating} to simulate truncated stable random variables. In what follows, we give two definitions for stable random vector, see \cite[p.~57; p.~65]{samorodnitsky1994stable}.
\begin{definition}\label{def1} A random vector ${\boldsymbol{X}}=(X_{1},\dots,X_{d})^{T}$ is said to be stable in $\mathbb{R}^{d}$ if for any positive numbers $A$ and $B$ there is a positive number $C$ and a vector ${\boldsymbol{D}}\in \mathbb{R}^{d}$ such that
\begin{equation}\label{edef1}
A{\boldsymbol{X}}_{1}+B{\boldsymbol{X}}_{2}\mathop=\limits^d C{\boldsymbol{X}}+{\boldsymbol{D}},
\end{equation}
where ${\boldsymbol{X}}_{1}$ and ${\boldsymbol{X}}_{2}$ are independent and identical copies of ${\boldsymbol{X}}$ and $C=(A^{\alpha}+B^{\alpha})^{1/\alpha}$.
\end{definition}
\begin{definition} \label{def2}
Let $0<\alpha<2$. Then ${\boldsymbol{X}}$ is a non-Gaussian stable random vector in $\mathbb{R}^{d}$ if there exists a finite measure $\Gamma$ on the unit sphere
${\mathbb{S}}^d=\{\boldsymbol{x}=(x_{1},\dots,x_{d})^{T}\in \mathbb{R}^d|\langle{\boldsymbol{x}},{\boldsymbol{x}^T}\rangle=1\}$ and a vector $\boldsymbol{\mu}=(\mu_{1},\dots,\mu_{d})^{T}\in \mathbb{R}^{d}$ such that
\begin{align}
\label{chf}
\varphi _{\boldsymbol{X}}(\boldsymbol{t})&=\log E\left(\exp (i \boldsymbol{t} \boldsymbol{X})\right)\nonumber\\
&=\left\{ {\begin{array}{*{20}c}
-\int_{\boldsymbol{S}^d}{{\left|{\left\langle{\boldsymbol{t},\boldsymbol{s}}\right\rangle}\right|}^\alpha}\left[1-i~\mathrm{sgn}\left\langle{\boldsymbol{t},\boldsymbol{s}}\right\rangle \tan \left(\frac{\pi \alpha}{2}\right)\right]\,\Gamma (d{\boldsymbol{s}})
+i\left\langle {\boldsymbol{t},\boldsymbol{\mu}}\right\rangle,~~~~~\alpha \ne 1,\\
-\int_{\boldsymbol{S}}^d\left|{\left\langle{\boldsymbol{t},\boldsymbol{s}}\right\rangle}\right|\,\left[1
+i~\mathrm{sgn}\left\langle {\boldsymbol{t},\boldsymbol{s}} \right\rangle \frac{2}{\pi} \log \left|{\left\langle {\boldsymbol{t},\boldsymbol{s}}\right\rangle}\right| \right]\,\Gamma (d{\boldsymbol{s}})
+i\left\langle {\boldsymbol{t},\boldsymbol{\mu}}\right\rangle,~~~\alpha =1,\\
\end{array}} \right.
\end{align}
where $\left\langle \boldsymbol{t},\boldsymbol{s}\right\rangle=\sum_{i=1}^{d}t_{i}s_{i}$ for $\boldsymbol{t}=(t_{1},\dots,t_{d})^T$, $\boldsymbol{s}=(s_1,\dots,s_d)^{T}$, $i^2=-1$, and $\mathrm{sgn}(.)$ denotes the well-known sign function. The pair $\left(\Gamma,\boldsymbol{\mu}\right)$ is unique.
\end{definition}

Each $d$-dimensional multivariate stable distribution is fully described by the triple $(\alpha, \Gamma, \boldsymbol{\mu})$. A random vector $\boldsymbol{X}$ is said to be strictly stable in $\mathbb{R}^{d}$ if $\boldsymbol{\mu}=\boldsymbol{0}$ for $\alpha \neq1$. We note that $\boldsymbol{X}$ is strictly stable, in the sense of Definition \ref{def1}, if $\boldsymbol{D}=\boldsymbol{0}$. Throughout, we assume that $\boldsymbol{X}$ follows a multivariate strictly stable distribution for $\alpha \neq 1$. 
\par
The class of multivariate stable distributions are intractable for dimensions more than two, see \cite{nolan2013multivariate}. As a computationally tractable subclass, the pdf a $d$-dimensional elliptically contoured stable distribution can be computed numerically for any $d \geq 2$. Suppose $\boldsymbol{Z}$ follows a $d$-dimensional elliptically contoured stable distribution with dispersion matrix $\Sigma$ and location vector $\boldsymbol{\mu}\in \mathbb{R}^{d}$, then the chf of $\boldsymbol{Z}$ is given by
\begin{equation}\label{chf2}
\varphi_{\boldsymbol{Z}}(\boldsymbol{t})=E\,\exp (i\boldsymbol{t}^T\boldsymbol{Z}) = 
\exp\left\{-\left(\boldsymbol{t}^T\Sigma\boldsymbol{t}\right)^{\alpha/2}+i \boldsymbol{t}^T\boldsymbol{\mu}\right\}, 
\end{equation} 
where $\Sigma$ is a $d \times d$ positive definite matrix, see \cite{nolan2013multivariate}. Throughout, we assume that random vector $\boldsymbol{Z}$ has chf given by (\ref{chf2}).
Teimouri et al. (2018) proposed the EM algorithm to estimate the parameters of a $d$-dimensional elliptically contoured stable distribution. Since finding the full maximum likelihood (ML) estimations of the parameters of a multivariate elliptically contoured stable distribution is computationally expensive, the projecting approach is suggested by which $\boldsymbol{Z}$ is projected into real line. Under this approach each projected random variable follows a univariate symmetric stable distribution and so the CF, ML, and SQ methods can be applied to estimate the parameters, see \cite{nolan2013multivariate}. The results obtained in such a way are known as projected estimations and corresponding methods are called projected CF, ML, and SQ, respectively. It should be noted that the estimated dispersion matrix using the projection approach is not necessarily positive definite and is computationally expensive when $d$ gets large.
\par
Free packages developed in the literature are {\bf{stabledist}} \cite{stabledist} and {\bf{libstableR}} \cite{libstableR}. The {\bf{stabledist}} generates stable random variables, computes pdf(s), cumulative distribution function(s) (cdf) and quantiles of stable distribution for given admissible values of the parameters. The {\bf{libstableR}} is developed for fast and accurate evaluation, generation, and parameter estimation of stable distributions. It should be noted that 
packages are not applicable for a mixture of univariate stable and multivariate stable distributions.
\section{Methodology}
In this section, we briefly provide some explanations about the methodology used in {\bf{alphastable}}.
\subsection{Simulation} 
\subsubsection{Random variable generation from univariate stable distributions}
Based on formulas given in \cite{chambers1976method}, \cite{nolan2010stable}, and \cite{weron1996chambers}, we simulate stable random variables in both $S_0$ and $S_1$ parameterizations.
\subsubsection{Random variable generation from truncated stable distributions}
Let $F^{\langle a,b\rangle}_{Y}(.)$ denote the cdf of a truncated stable random variable over $(a,b)$. Then
\begin{align*}
F^{\langle a,b\rangle}_{Y}(y)=\frac{F_{Y}(y)-F_{Y}(a)}{F_{Y}(b)-F_{Y}(a)},
\end{align*}
where $-\infty <a<y<b<\infty$. Using the characterization theorems given in \cite{soltani2010truncated} and \cite{shirvani2013characterization}, we simulate truncated stable random variables.
\subsubsection{Random vector generation from elliptically contoured stable distribution}
Using the definition of an elliptically contoured stable distribution given in \cite{samorodnitsky1994stable}, we simulate $d$-dimensional elliptically contoured stable random vector. 
\subsubsection{Random vector generation from multivariate stable distributions}
An algorithm for random vector generation from multivariate stable distribution was given in \cite{modarres1994method}. Based on this algorithm, we simulate bivariate stable random vector. 
\subsection{Computing the pdf and cdf of univariate stable distributions}
Using series given in \cite{teimouri2008anovel}, we compute the cdf and pdf of stable distribution in both $S_0$ and $S_1$ parameterizations for $\alpha \neq 1$. These series provide a very accurate approximation of the cdf and pdf within their convergence regions. For those points that series are not convergent the {R} package {\bf{stabledist}} is used to compute the pdf and cdf. Fortunately these series approximate accurately the pdf (or cdf) and have no computational problem in boundary cases including 
$\alpha \downarrow 0$, $|\alpha-1| \rightarrow 0$, and $|\beta| \rightarrow 1$, see \cite{teimouri2008anovel}. Let $k=150$, $\lambda=\bigl(1+\beta^2\tan^2(\pi\alpha/2)\bigr)^{\frac{1}{2\alpha}}$, $\eta=\frac{2}{\pi}\arctan(\beta\tan(\pi\alpha/2))\mathrm{sgn}(y-\mu-\xi)$, and

\begin{eqnarray*} 
\xi=\left\{\begin{array}{*{20}c} 
0~~~~~~~~~~~~~~~~~~~~~\mathrm{{for}}~S_1~\mathrm{{form}},\\ 
-\sigma \beta \tan (\pi \alpha/2)~~\mathrm{for}~ S_0~\mathrm{{form}}.\\ 
\end{array} \right. 
\end{eqnarray*} 
For any admissible quadruple $(\alpha, \beta, \sigma, \mu)$ we have the following approximations. 
\begin{itemize}
\item If $|y-\mu-\xi|\geq\sigma\lambda \Bigl(\alpha \frac{\Gamma(k\alpha+\alpha)}{\Gamma(k\alpha+1)}\Bigr)^{\frac{1}{\alpha}}+\alpha$, then
\begin{align*}
f_Y(y)\approx&\frac{1}{\pi|x-\mu-\xi|}\sum_{i=1}^{k}(-1)^{i-1}\frac{\Gamma(i \alpha+1)}{\Gamma(i+1)}\Big|\frac{x-\mu-\xi}{\sigma \lambda}\Big|^{-i\alpha}\sin\Bigl(\frac{i\pi(\alpha+\eta)}{2}\Bigr), \nonumber\\
F_Y(y)\approx& \frac{1+\mathrm{sgn}(y-\mu-\xi)}{2}\nonumber\\&+\frac{\mathrm{sgn}(y-\mu-\xi)}{\pi}
\sum_{i=1}^{k}(-1)^{i}\frac{\Gamma(i \alpha+1)}{i \alpha \Gamma(i+1)}\Big|\frac{y-\mu-\xi}{\sigma \lambda}\Big|^{-i\alpha}
\sin\Bigl(\frac{i\pi(\alpha+\eta)}{2}\Bigr).\nonumber
\end{align*}
\end{itemize}
Also,
\begin{itemize}\item If $|y-\mu-\xi|\leq\sigma\lambda \alpha \frac{\Gamma(k/\alpha+1)}{\Gamma(k/\alpha+1/\alpha)}-\alpha$, then
\begin{align*}
f_Y(y)\approx&\frac{1}{\pi|x-\mu-\xi|}\sum_{i=1}^{k}(-1)^{i-1}\frac{\Gamma(i /\alpha+1)}{\Gamma(i+1)}\Big|\frac{x-\mu-\xi}{\sigma \lambda}\Big|^{i}\sin\Bigl(\frac{i\pi(\alpha+\eta)}{2\alpha}\Bigr),\nonumber\\
F_Y(y)\approx&\frac{1}{2}-\frac{\arctan(\beta\tan(\pi\alpha/2))}{\pi \alpha}\nonumber\\&-\frac{\mathrm{sgn}(y-\mu-\xi)}{\pi}
\sum_{i=1}^{k}(-1)^{i}\frac{\Gamma(i /\alpha+1)}{i\Gamma(i+1)}\Big|\frac{x-\mu-\xi}{\sigma \lambda}\Big|^{i}
\sin\Bigl(\frac{i\pi(\alpha+\eta)}{2\alpha}\Bigr).\nonumber
\end{align*}
\end{itemize}

\subsection{Parameter estimation for Cauchy and mixture of Cauchy distributions}
\begin{figure}[h!]
\resizebox{\textwidth}{!}
{\begin{tabular}{ccc}
\includegraphics[width=40mm,height=40mm]{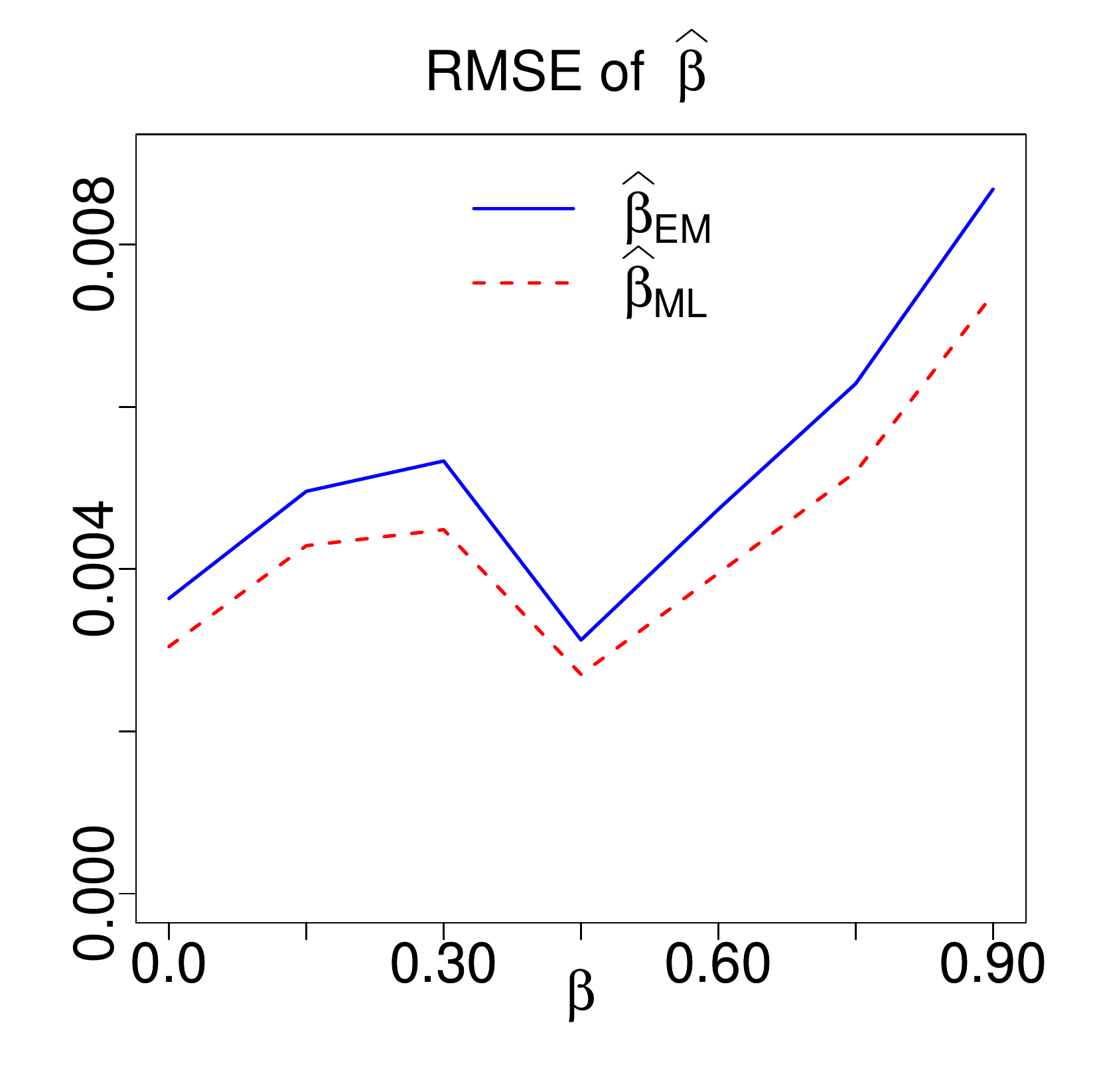}&
\includegraphics[width=40mm,height=40mm]{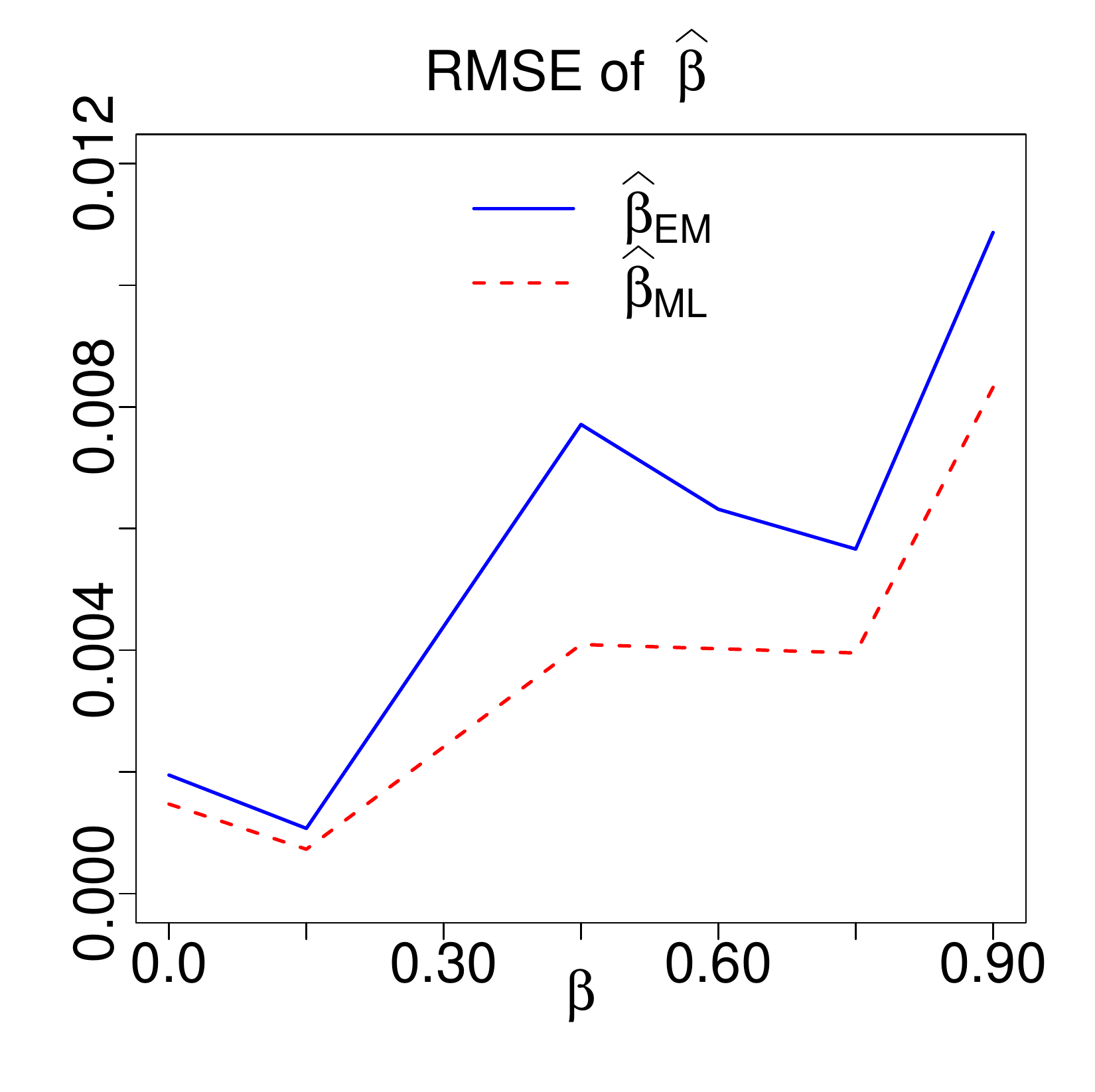}&
\includegraphics[width=40mm,height=40mm]{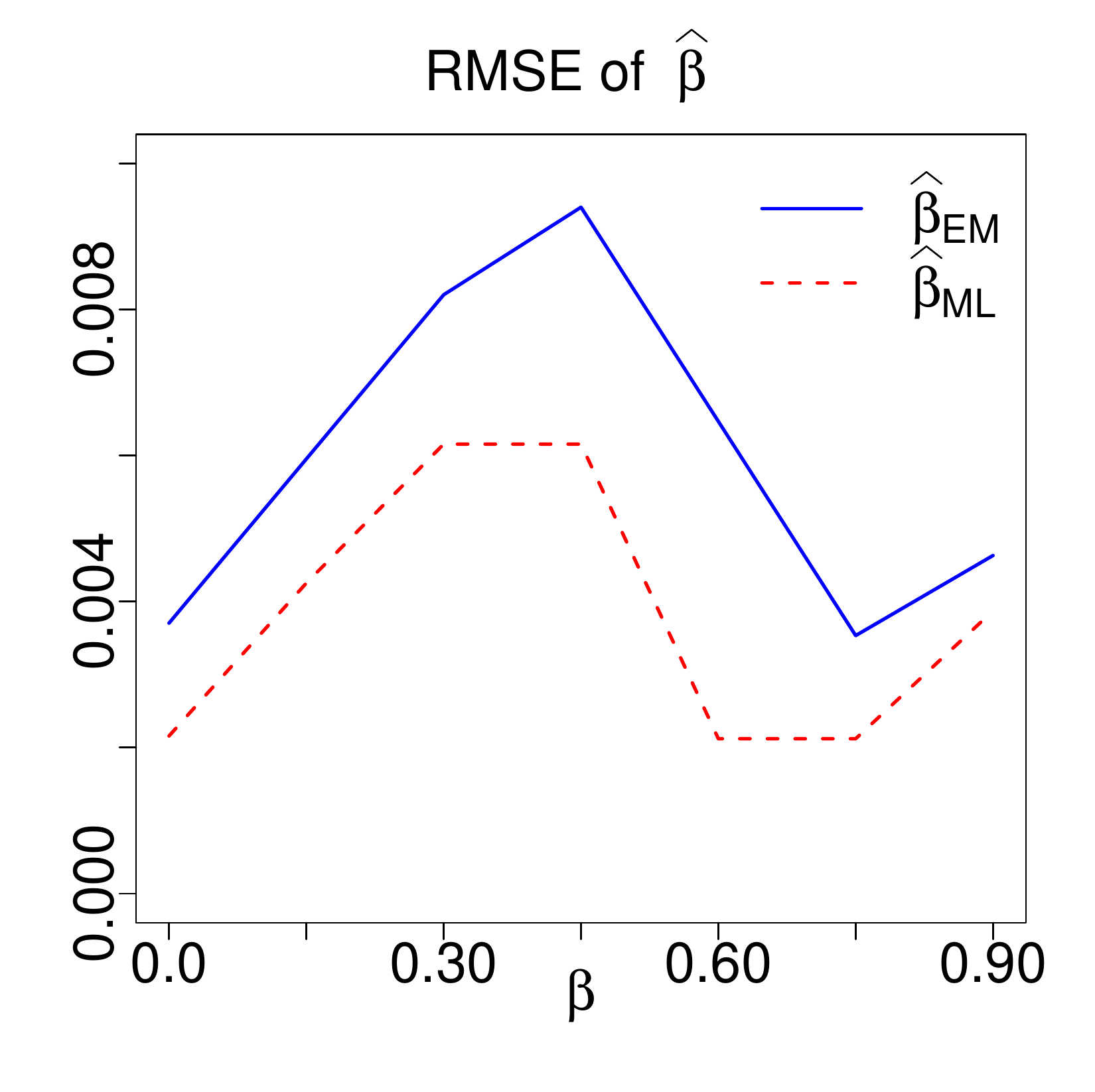}\\
\includegraphics[width=40mm,height=40mm]{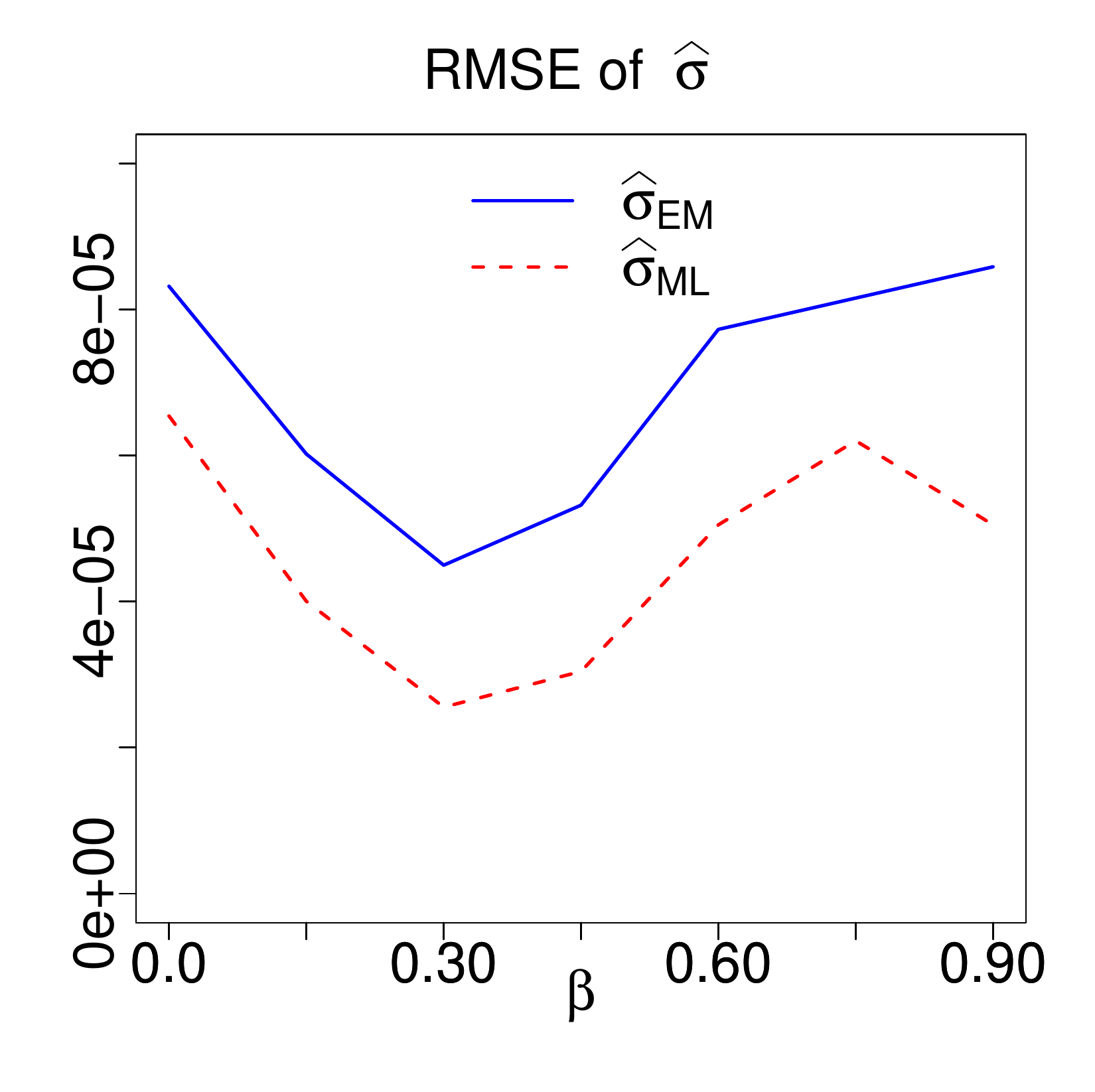}&
\includegraphics[width=40mm,height=40mm]{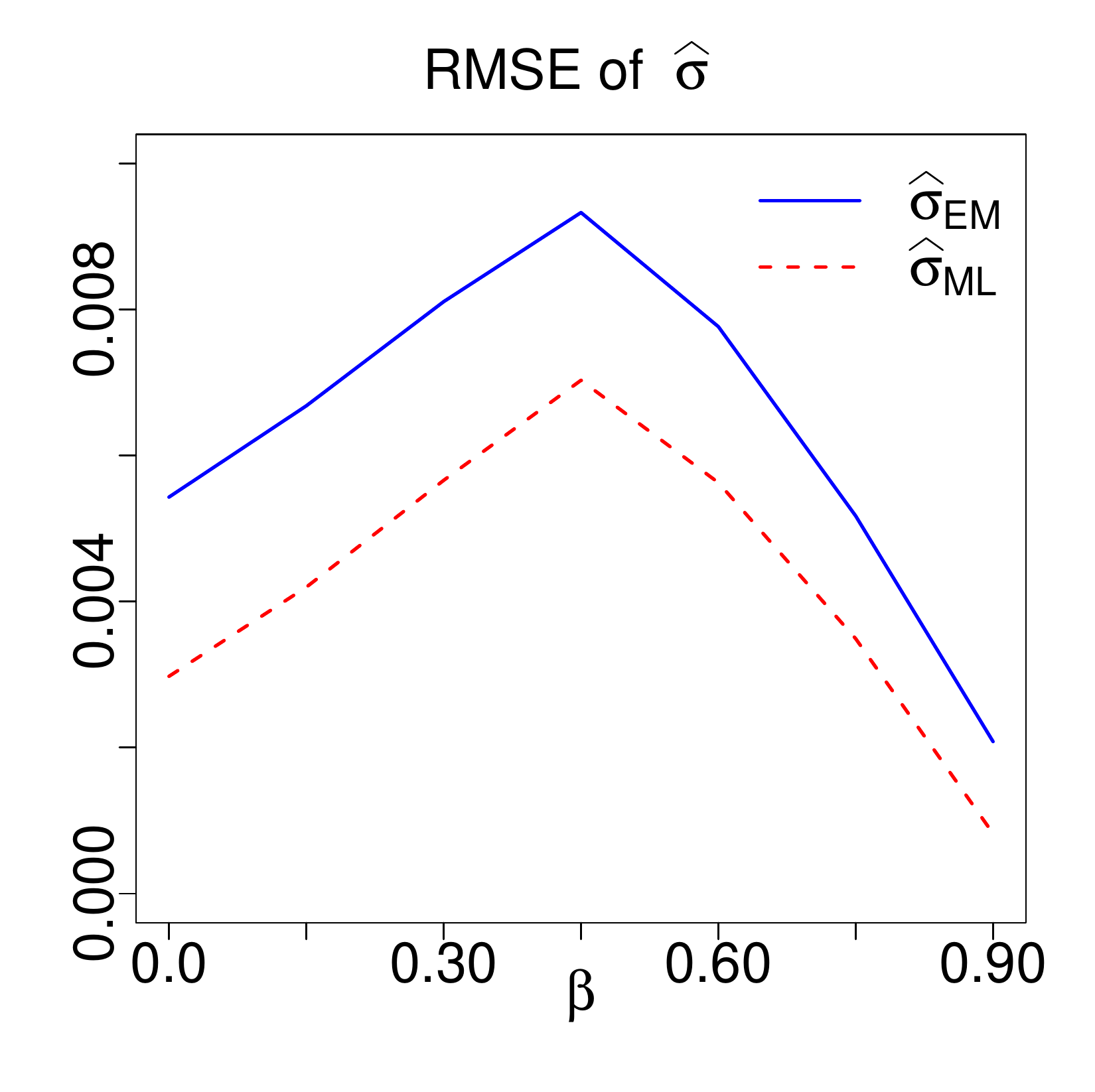}&
\includegraphics[width=40mm,height=40mm]{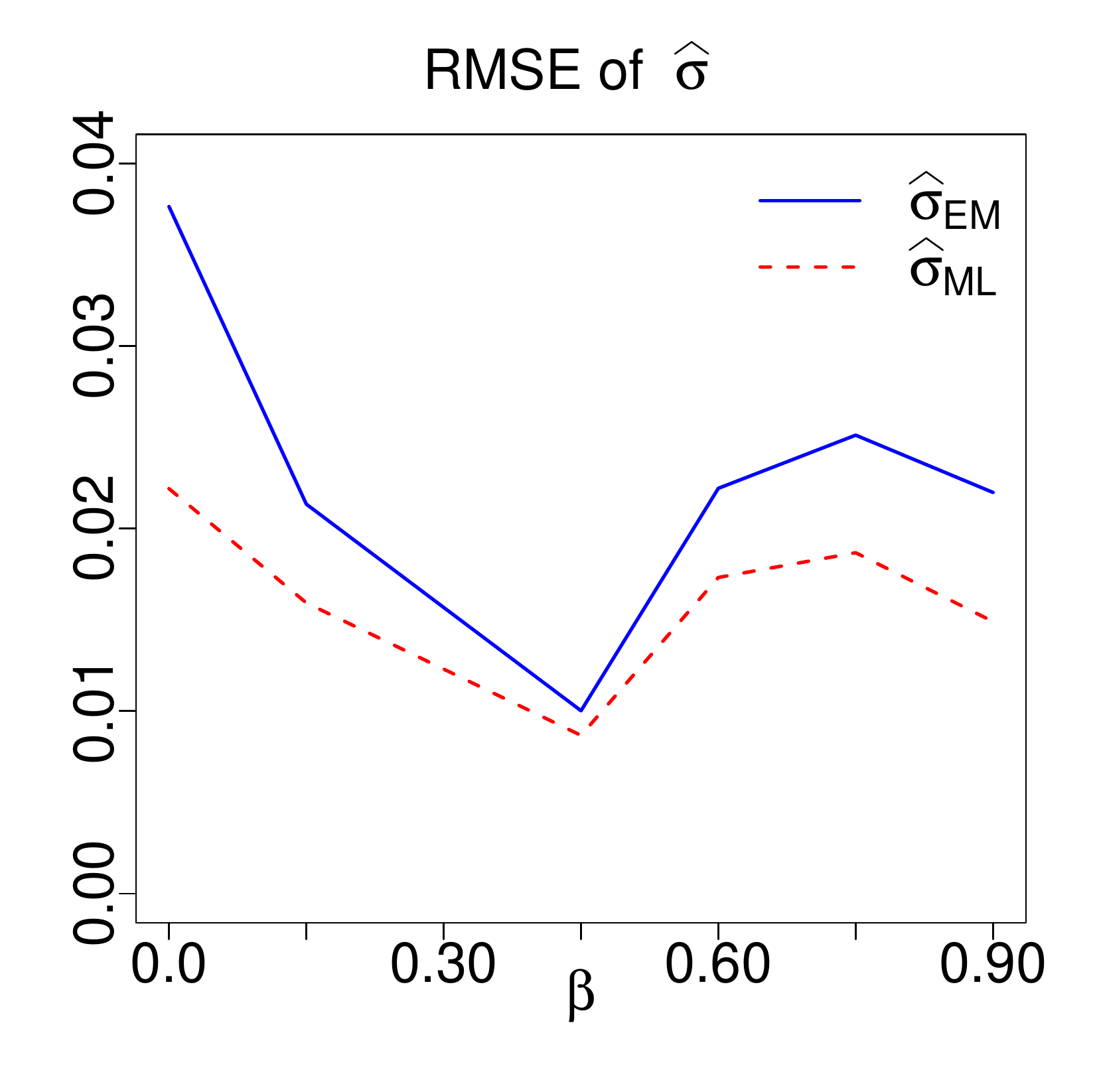}\\
\includegraphics[width=40mm,height=40mm]{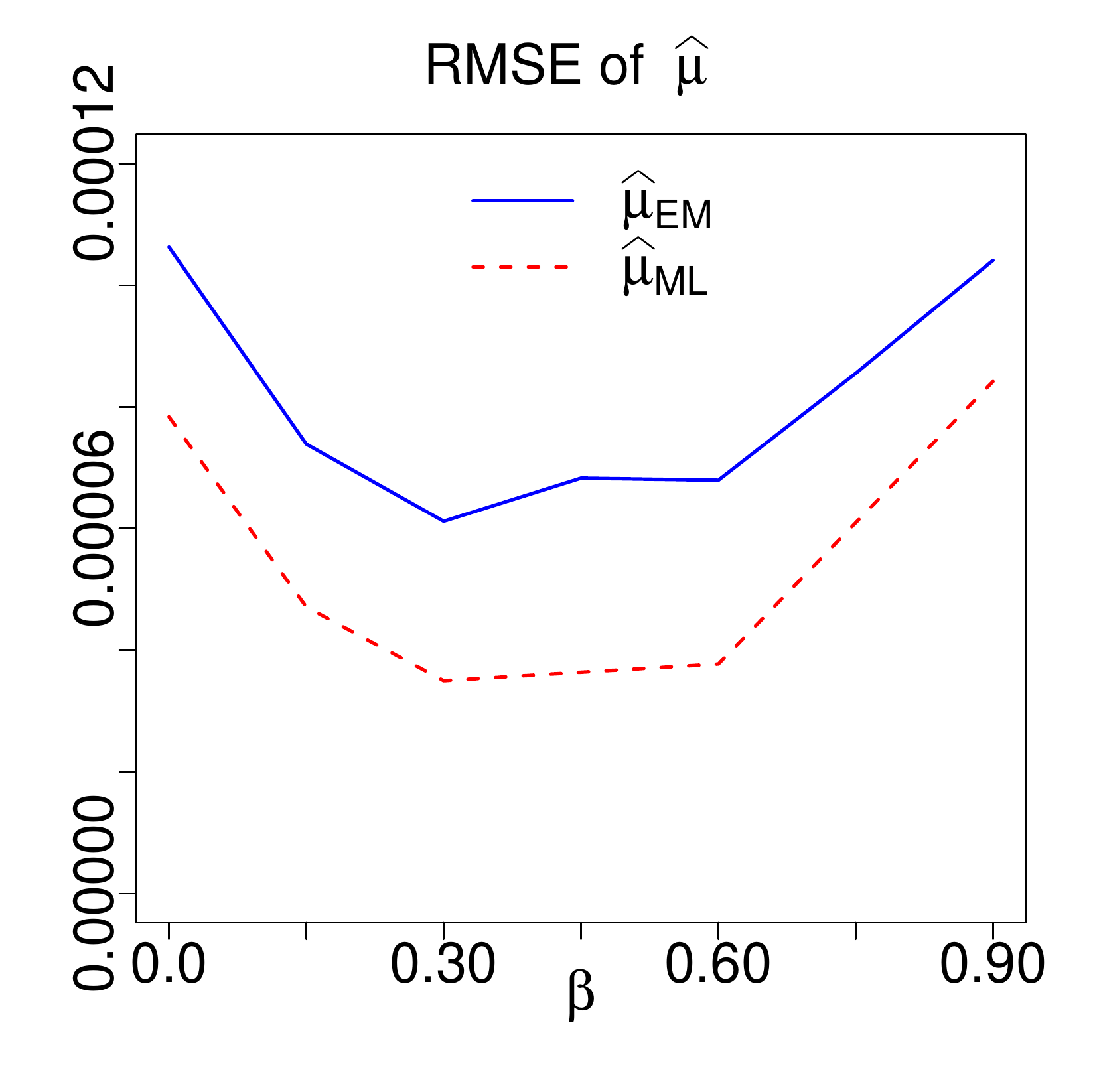}&
\includegraphics[width=40mm,height=40mm]{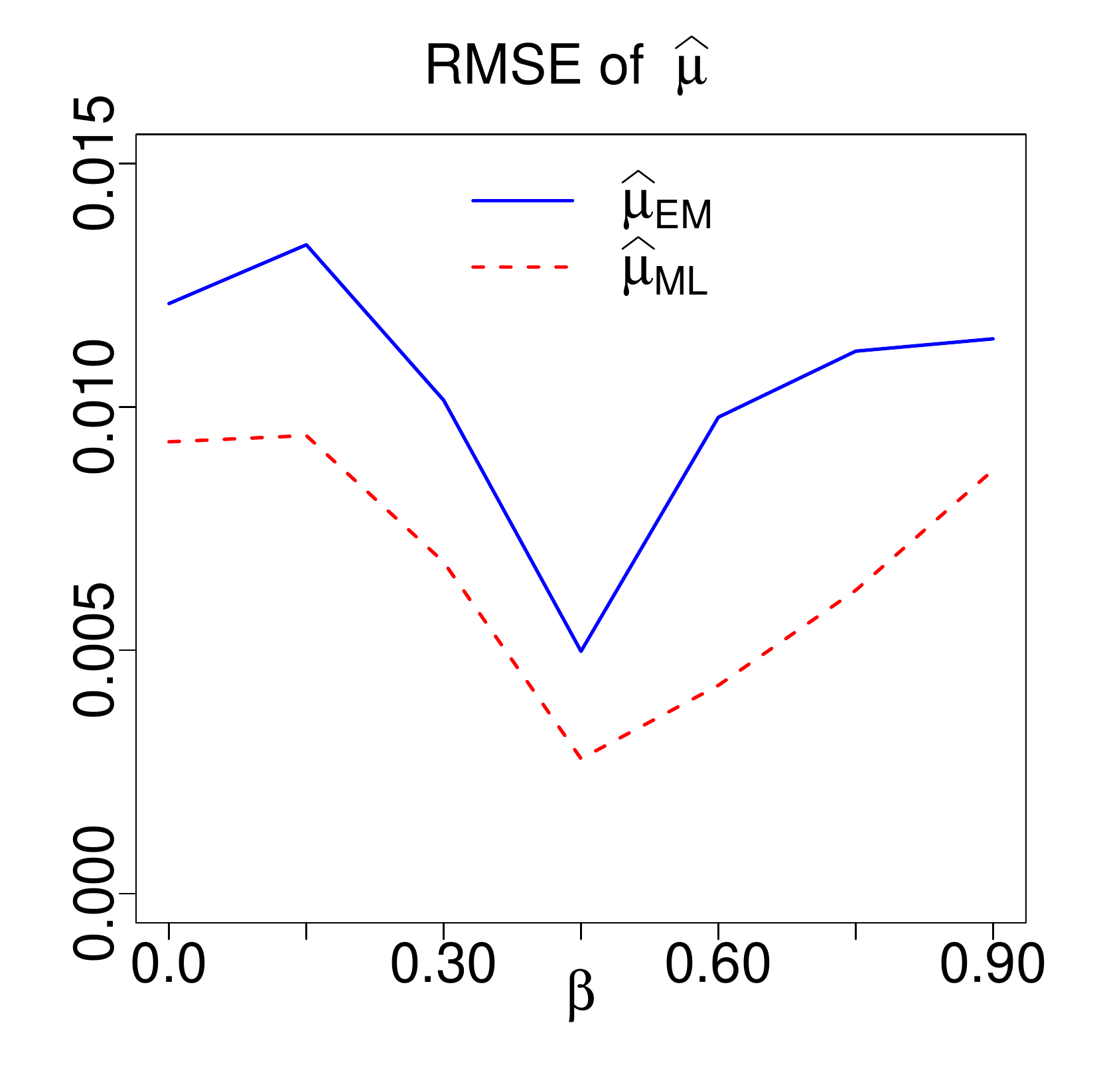}&
\includegraphics[width=40mm,height=40mm]{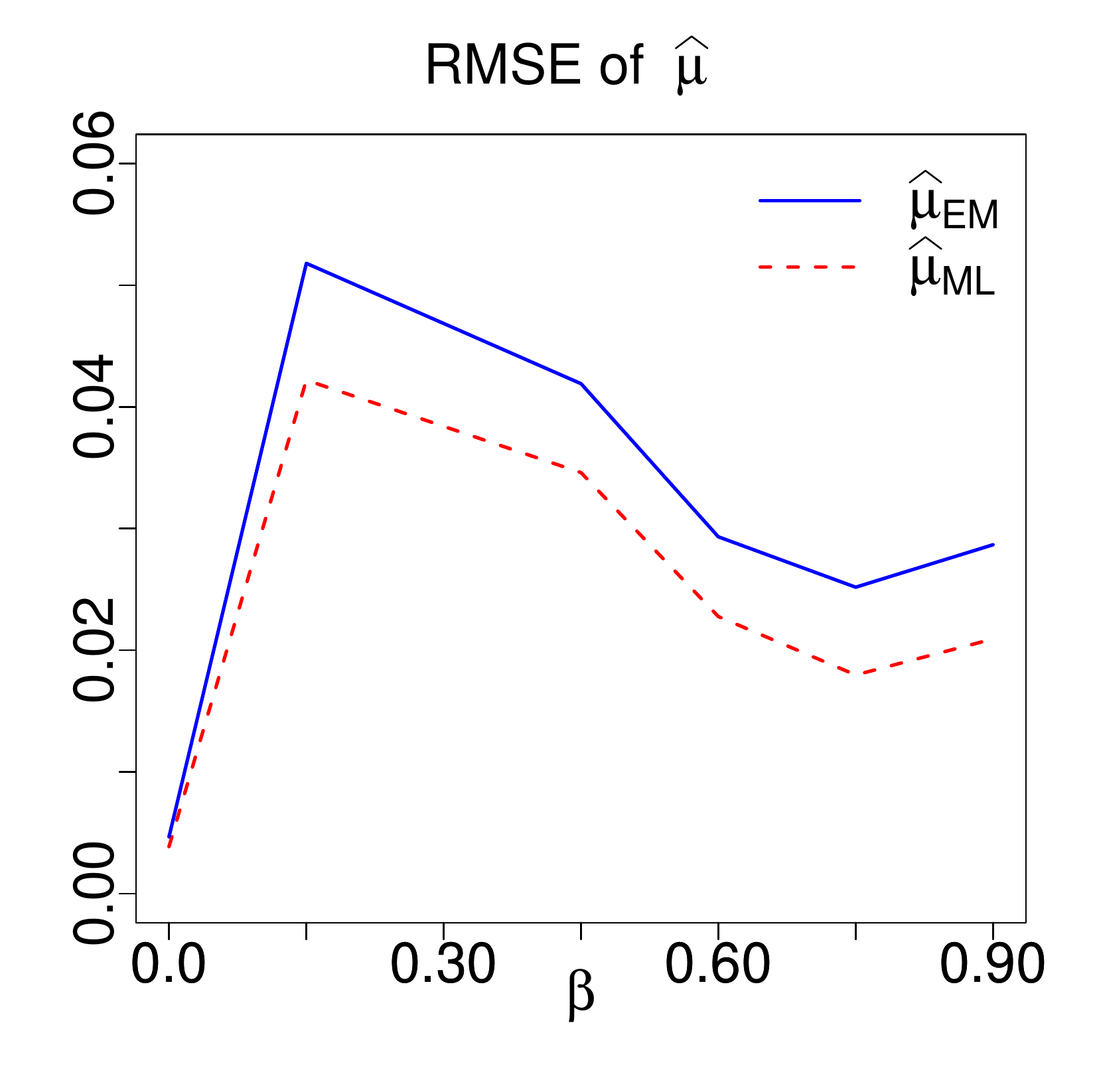}\\
\end{tabular}}
\caption{The RMSE of estimators obtained through the EM and ML approaches. The estimators are obtained when 300 realizations are generated from Cauchy distribution. In each sub-figure, the subscripts ML and EM indicate that the estimators $\hat{\beta}$, $\hat{\sigma}$, and $\hat{\mu}$ are obtained using the EM algorithm (blue solid line) or the ML approach (red dashed line). Note that, the levels of the skewness parameter on the horizontal axis are 0.0,0.15,0.30,0.45,0.60,0.75, and 0.90. The sub-figures in the first, second, and the third columns correspond to $\sigma=0.10$, $\sigma=2$, and $\sigma=5$, respectively.}
\label{cauchy}
\end{figure}

\begin{figure}[h!]
\resizebox{\textwidth}{!}
{\begin{tabular}{cc}
\includegraphics[width=20mm,height=20mm]{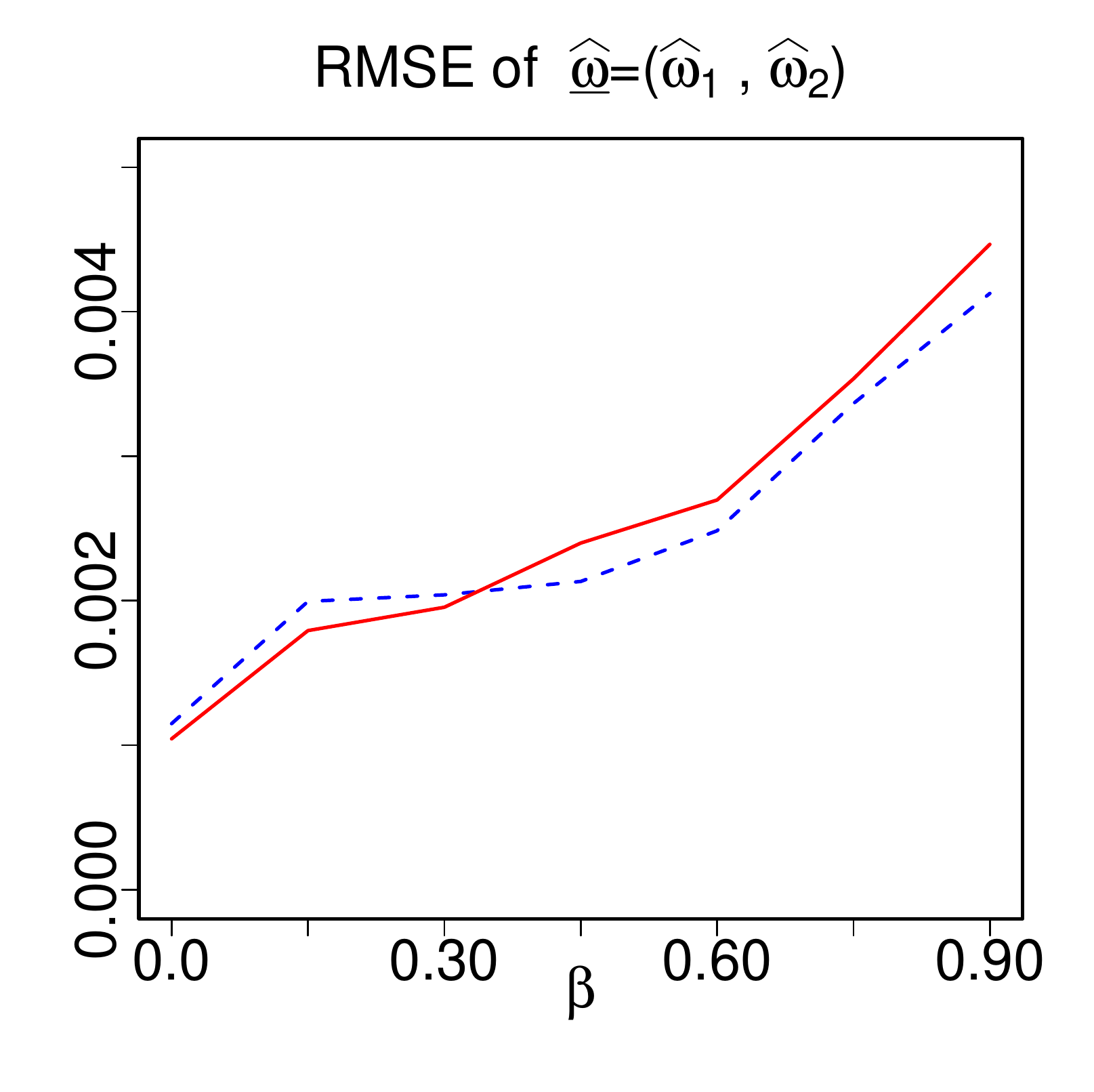}&
\includegraphics[width=20mm,height=20mm]{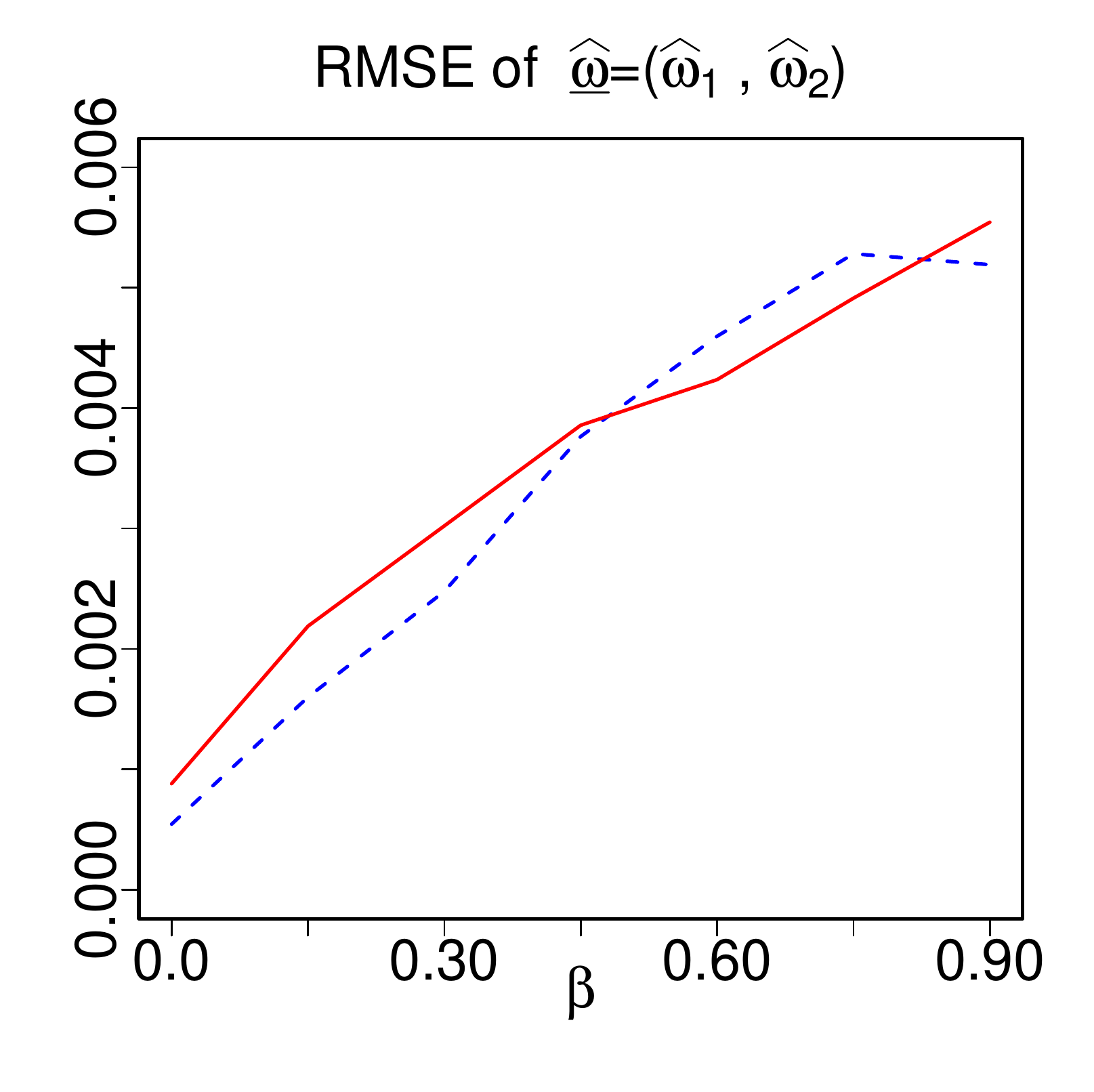}\\
\includegraphics[width=20mm,height=20mm]{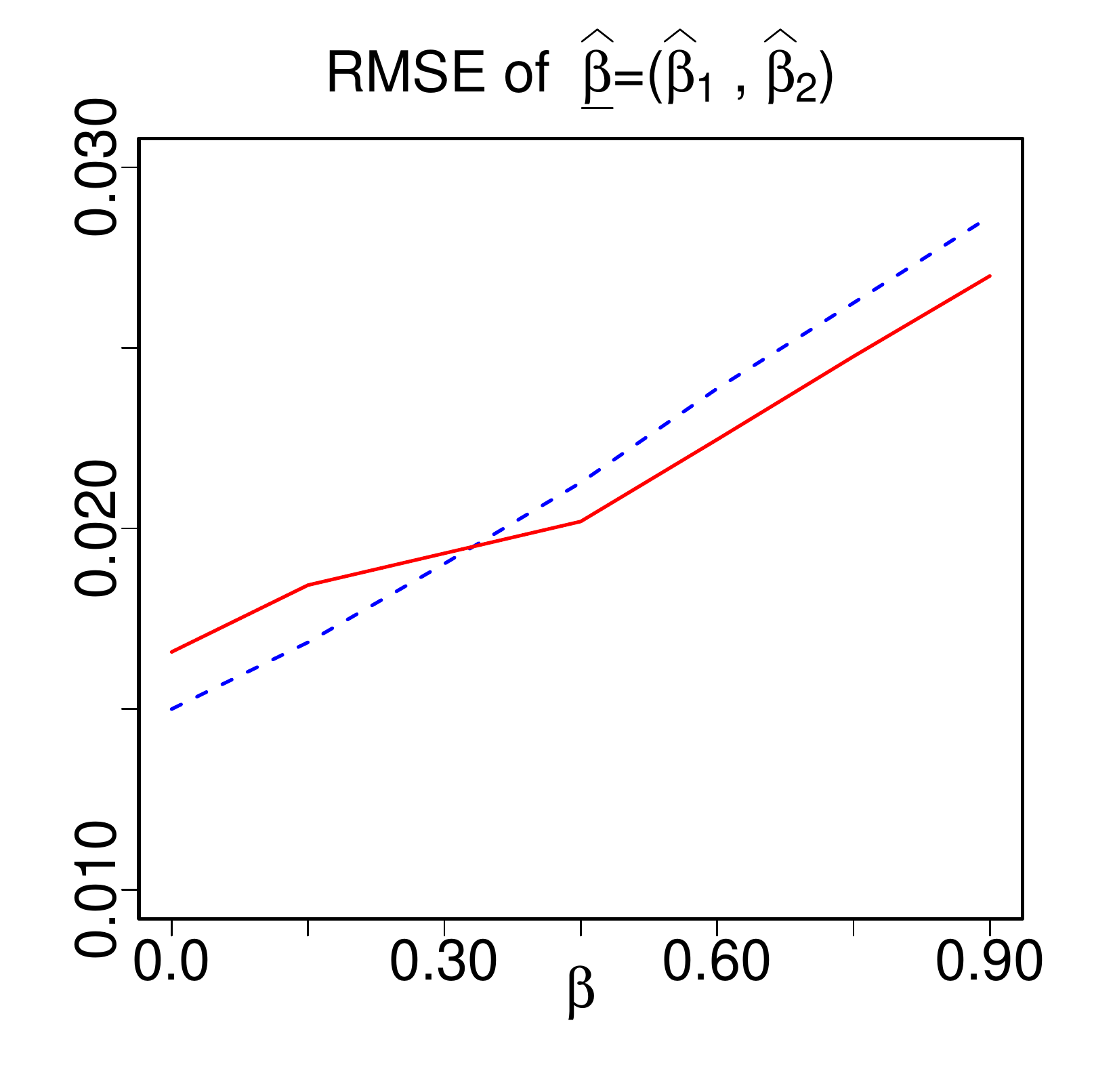}&
\includegraphics[width=20mm,height=20mm]{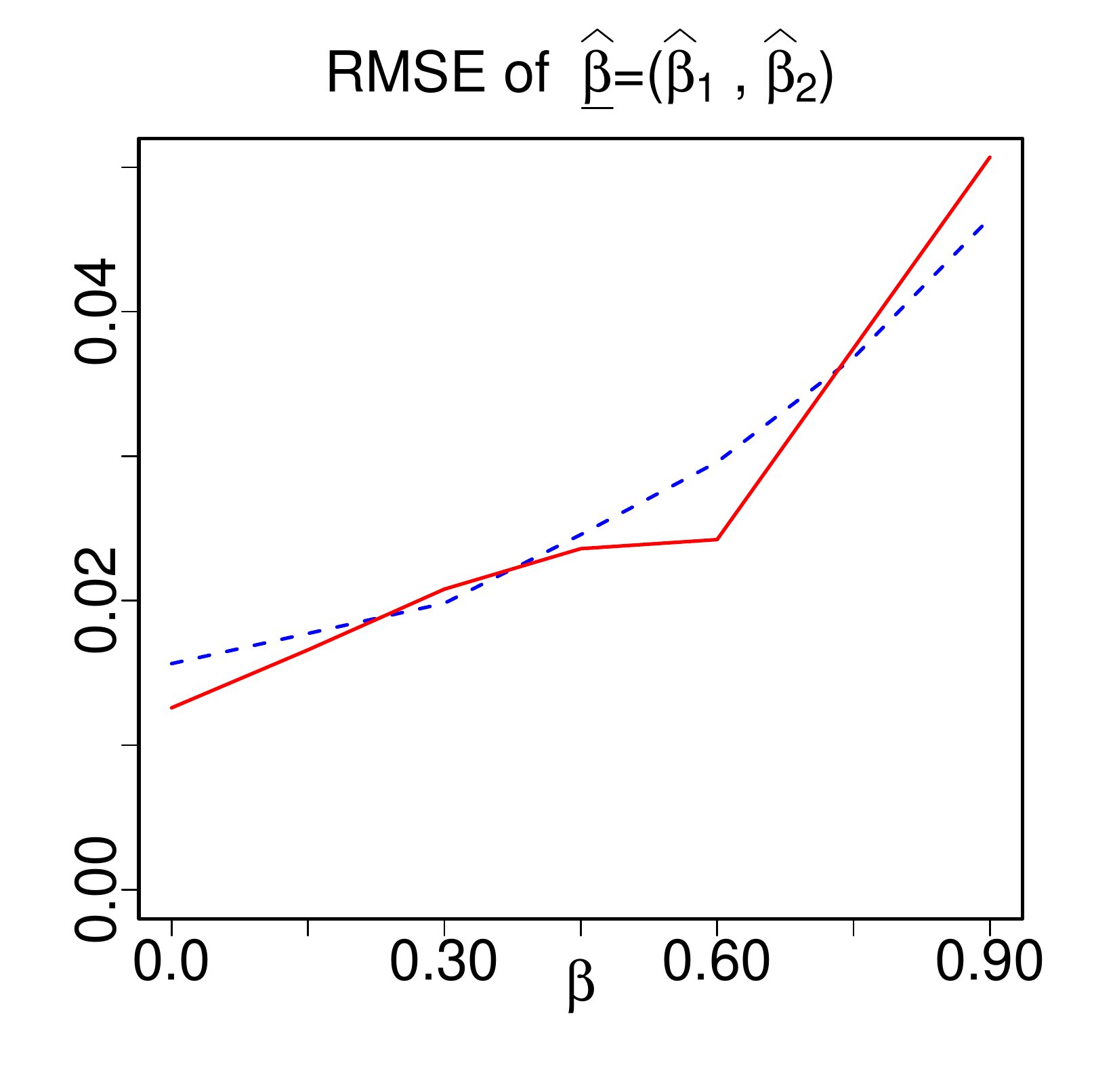}
\end{tabular}}
\caption{The RMSE of $\underline{\hat{\omega}}=(\widehat{\omega}_1,\widehat{\omega}_2)$ and $\underline{\hat{\beta}}=(\widehat{\beta}_1,\widehat{\beta}_2)$ when the EM algorithm is applied to the sample of size 1000 generated from two-component mixture of Cauchy distributions for 200 replications. In each sub-figure, dashed blue line and red solid line refer to the RMSE of estimator of the first and the second components, respectively. The levels of the skewness parameter on the horizontal axis are 0.0,0.15,0.30,0.45,0.60,0.75, and 0.90. The left-hand side and right-hand side sub-figures correspond to the scale vector $\underline{\sigma}=(0.25,0.25)$ and $\underline{\sigma}=(0.5,0.5)$, respectively. In simulations, we set $\underline{{\mu}}=(-2,2)$ and $\underline{{\beta}}=(\beta,\beta)$; for $\beta$=0.0,0.15,0.30,0.45,0.60,0.75, and 0.90.}
\label{mixcauchy1}
\end{figure}

\begin{figure}[h!]
\resizebox{\textwidth}{!}
{\begin{tabular}{cc}
\includegraphics[width=20mm,height=20mm]{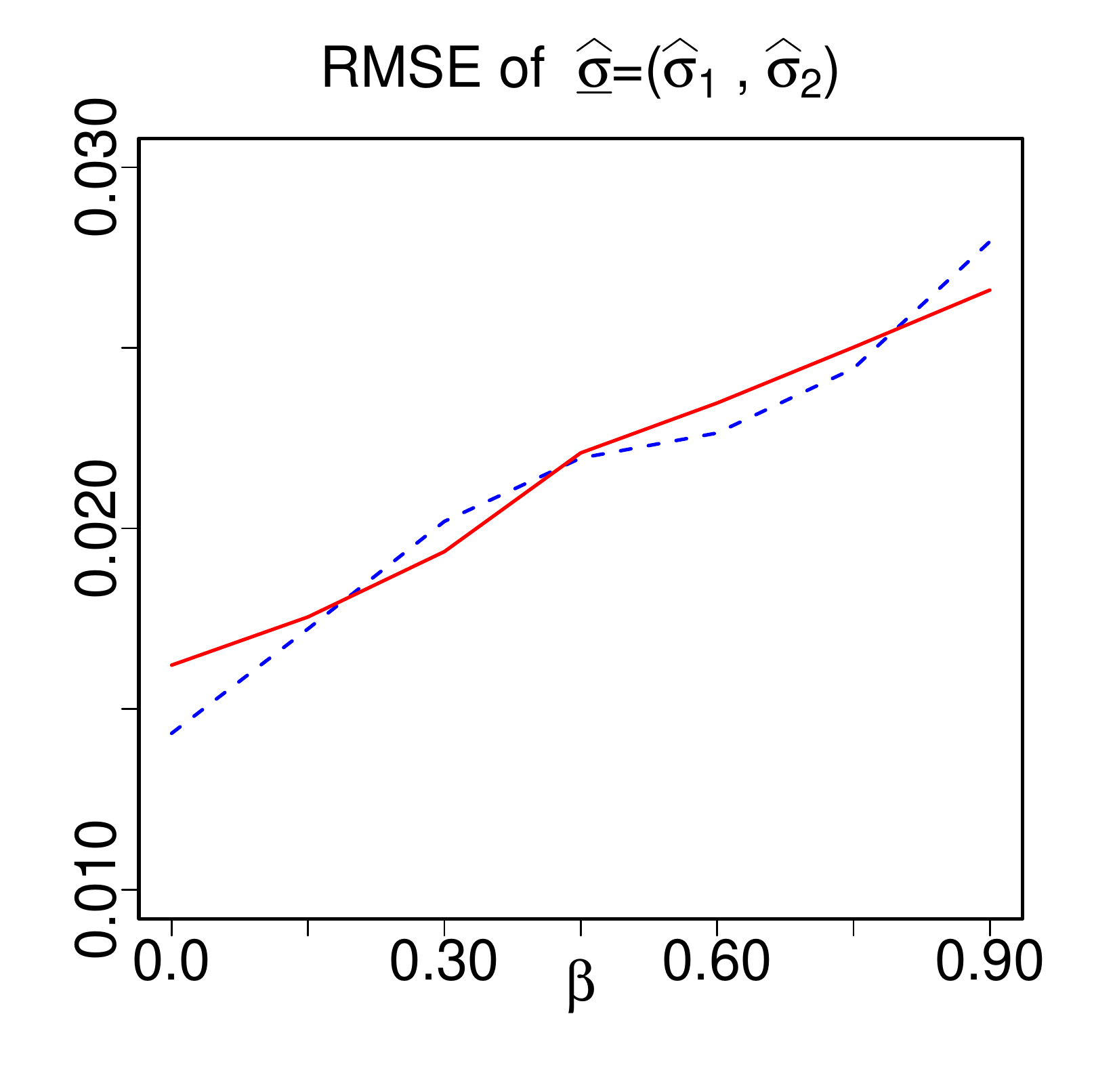}&
\includegraphics[width=20mm,height=20mm]{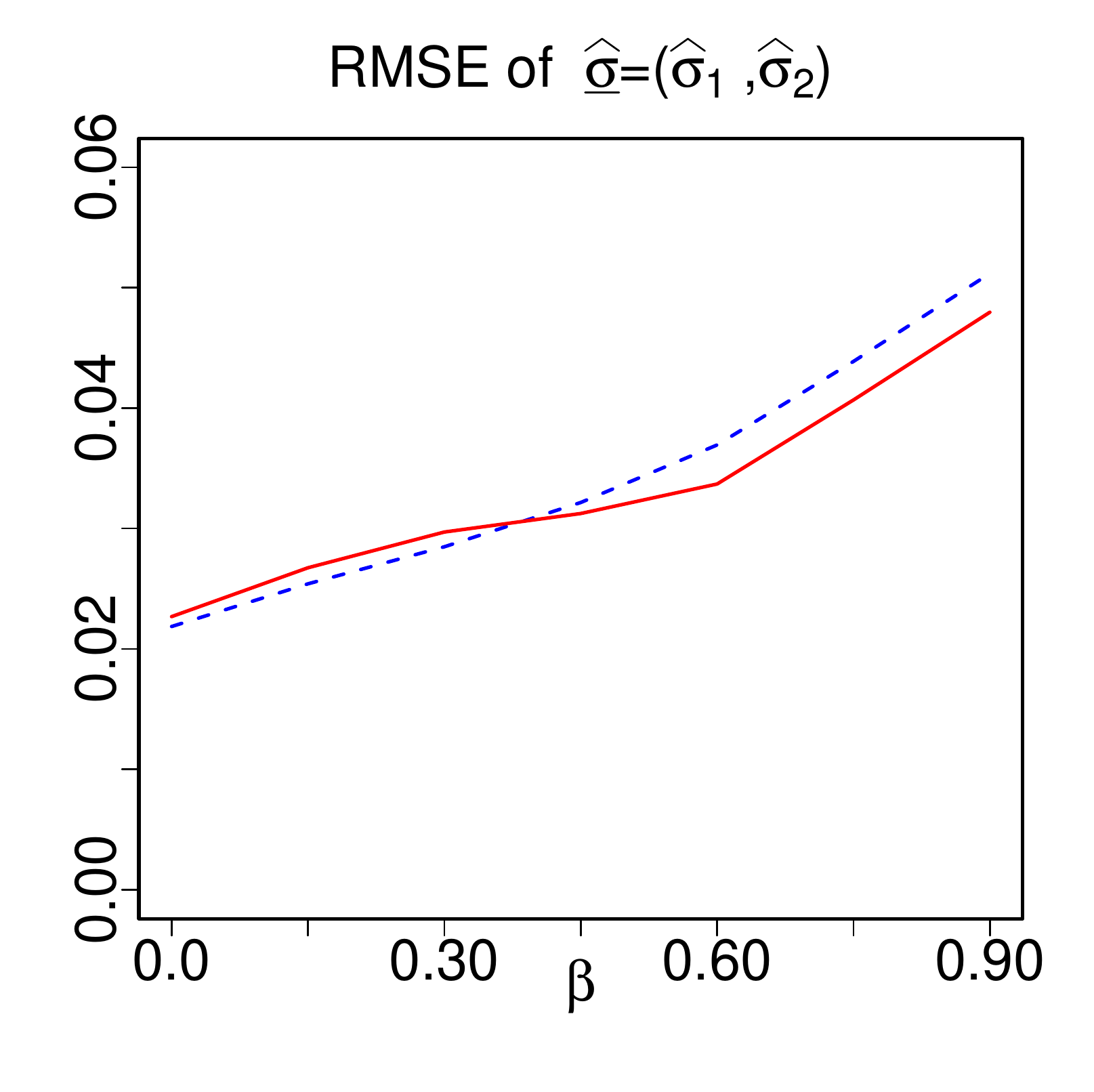}\\
\includegraphics[width=20mm,height=20mm]{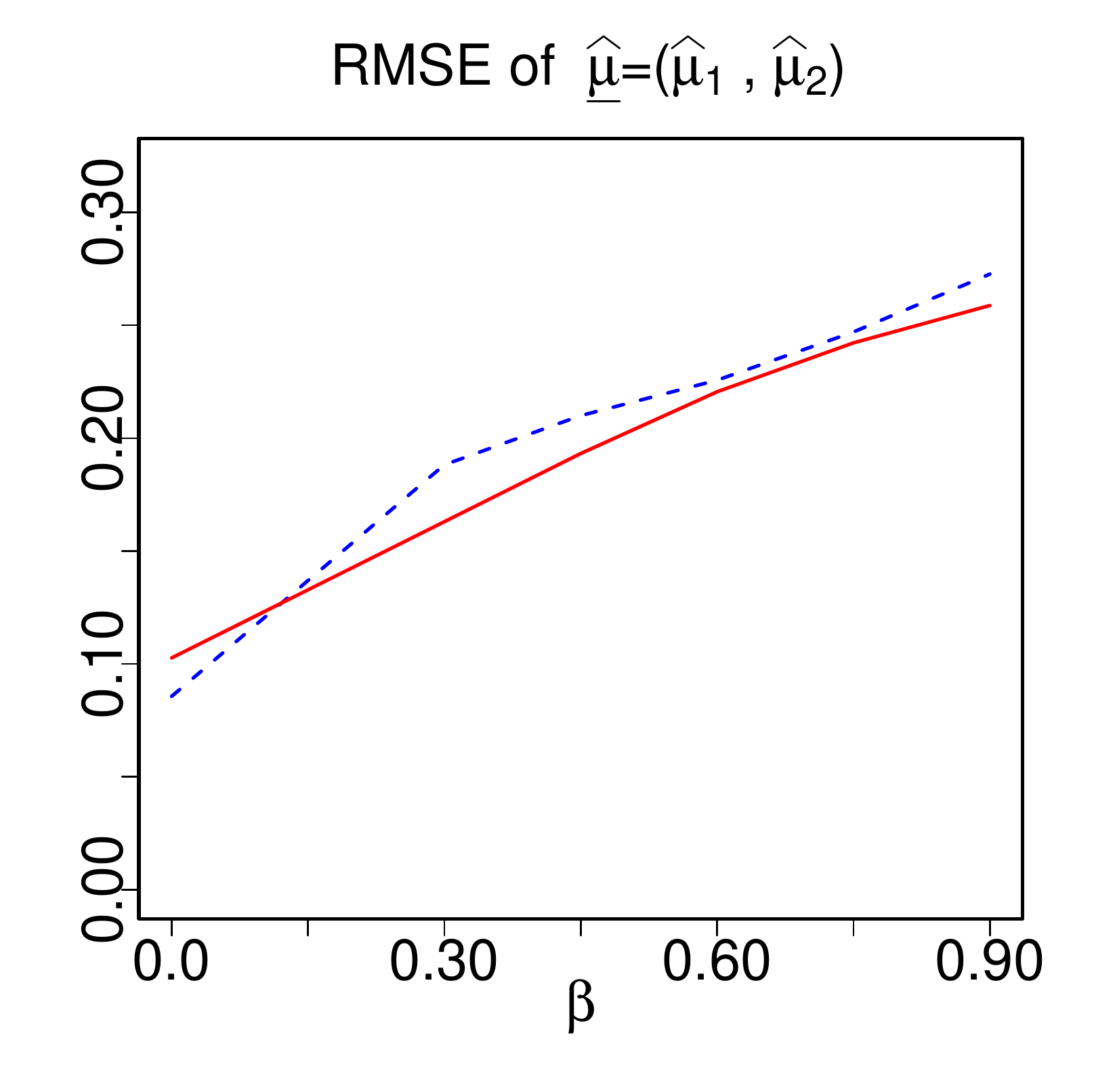}&
\includegraphics[width=20mm,height=20mm]{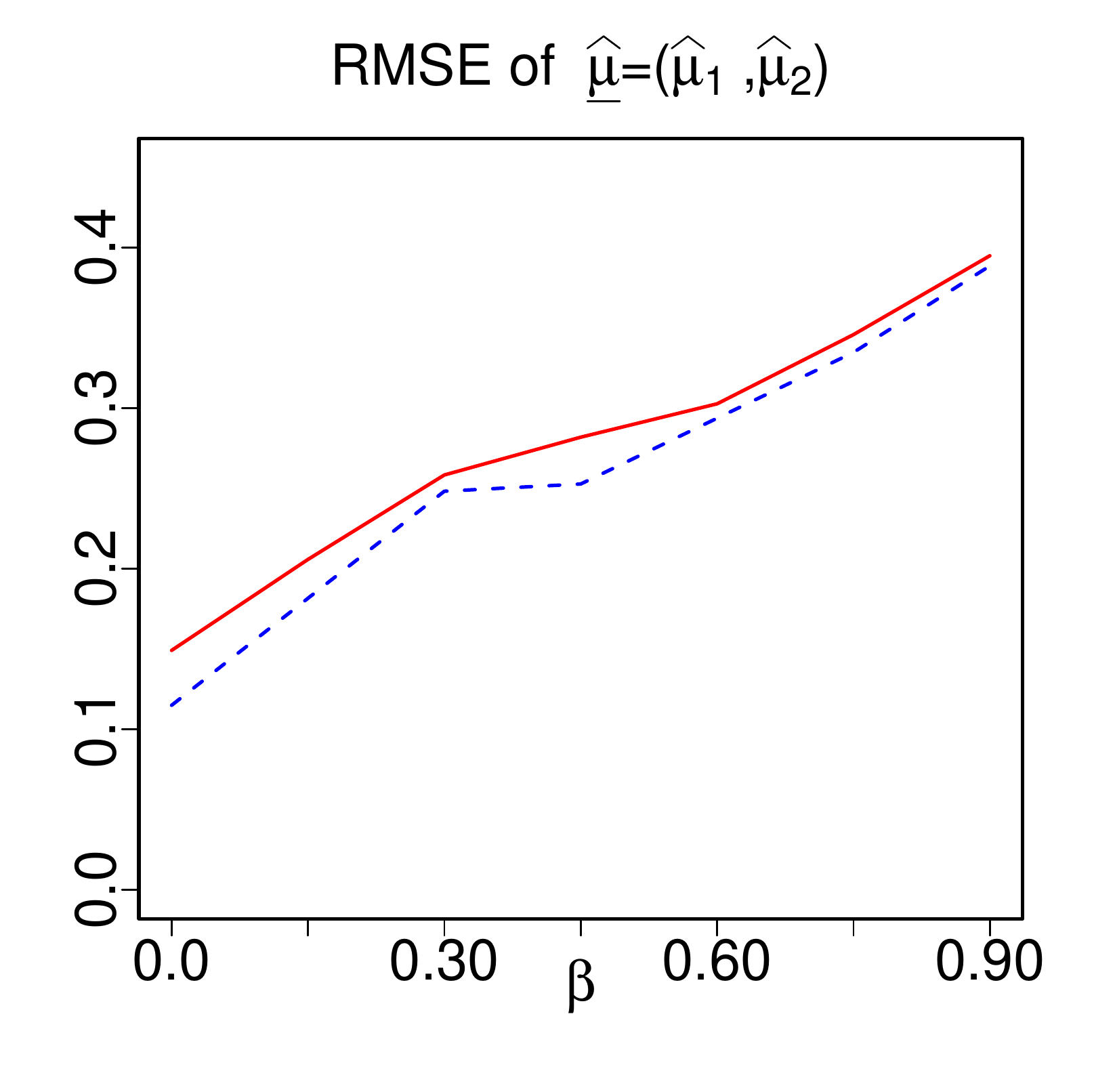}
\end{tabular}}
\caption{The RMSE of $\underline{\hat{\sigma}}=(\widehat{\sigma}_1,\widehat{\sigma}_2)$ and $\underline{\hat{\mu}}=(\widehat{\mu}_1,\widehat{\mu}_2)$ when the EM algorithm is applied to the sample of size 1000 generated from two-component mixture of Cauchy distributions for 200 replications. In each sub-figure, dashed blue line and red solid line refer to the RMSE of estimator of the first and the second components, respectively. The levels of the skewness parameter on the horizontal axis are 0.0,0.15,0.30,0.45,0.60,0.75, and 0.90. The left-hand side and right-hand side sub-figures correspond to the scale vector $\underline{\sigma}=(0.25,0.25)$ and $\underline{\sigma}=(0.5,0.5)$, respectively. In simulations, we set $\underline{{\mu}}=(-2,2)$ and $\underline{{\beta}}=(\beta,\beta)$; for $\beta$=0.0,0.15,0.30,0.45,0.60,0.75, and 0.90.}
\label{mixcauchy2}
\end{figure}

Due to a wide range of applications received for the Cauchy distribution, estimation of the parameters of Cauchy and mixture of Cauchy distributions are important tasks. To perform these tasks, we employing the EM algorithm. Generally, the EM algorithm is employed for computing the ML estimations when the log-likelihood function is not tractable mathematically. This is done by considering an extra missing (or latent) variable when the conditional expectation of the complete data log-likelihood given observed data and a guess of unknown parameter(s) is maximized. So, we look for a stochastic representation that involves the latent variable(s).
The representation given by the following proposition is valid for Cauchy distribution.
\begin{proposition}\label{prop1} Suppose $Y\sim S_{0}(1,\beta,\sigma,\mu)$ and $T \sim S_{1}(1,1,1,0)$. Then
\begin{eqnarray} \label{cauchyrep}
Y\mathop=\limits^d\sigma\left(1-|\beta|\right) \frac{N}{Z}+\sigma \beta T+\mu
\end{eqnarray}
where $\mathop=\limits^d$ denotes the equality in distribution and $N\sim Z\sim{\cal{N}}(0,1)$. The random variables $N$, $Z$, and $P$ are mutually independent.
\end{proposition}
Simulations reveal that the derived EM algorithm based on Proposition \ref{prop1} can be considered as a good competitor for the ML approach. We note that the EM algorithm outperforms the CF and SQ approaches and hence, two latter methods are removed by competitions, see Figure \ref{cauchy}. To show the performance of the EM algorithm for estimating the parameters of Cauchy mixture model, we consider a two-component ($K=2$) Cauchy mixture model in which the pdf of each component takes the form $f(.|1, \beta_j, \sigma_j, \mu_j)$; for $j=1, 2$. The results of simulations are depicted in Figures \ref{mixcauchy1}-\ref{mixcauchy2}. 
\subsection{Parameter estimation for symmetric and mixture of symmetric stable distributions}
\begin{figure}[h!]
\begin{tabular}{p{0.015cm} p{0.3cm}ccc}
&&\scriptsize{${\sigma=0.5}$}&\scriptsize{${\sigma=1}$}&\scriptsize{${\sigma=2}$}\\
&
\begin{rotate}{90}~~~~~~~~~{\scriptsize{n=200}}
\end{rotate}
&
\includegraphics[angle=90,width=40mm,height=40mm]{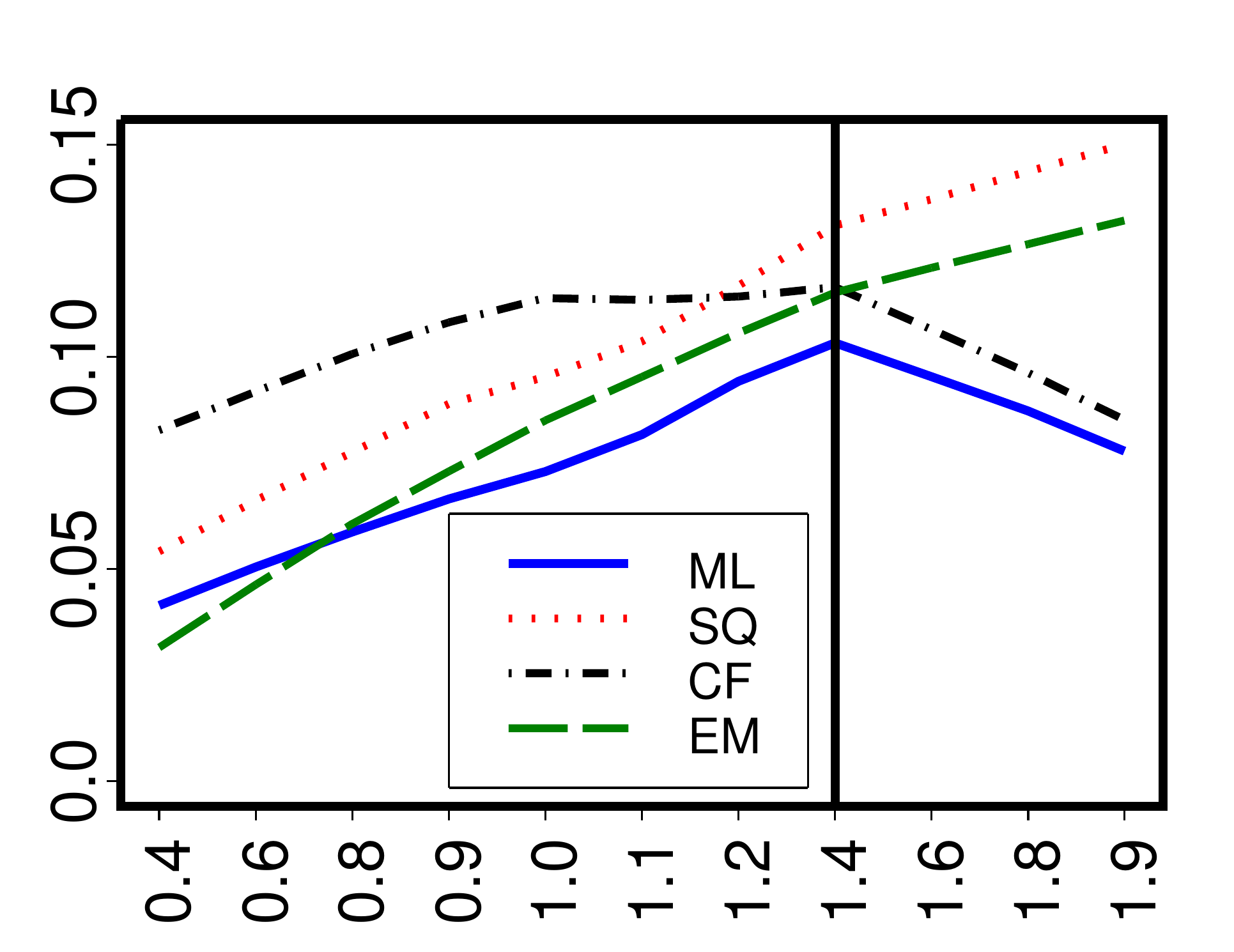}&
\includegraphics[angle=90,width=40mm,height=40mm]{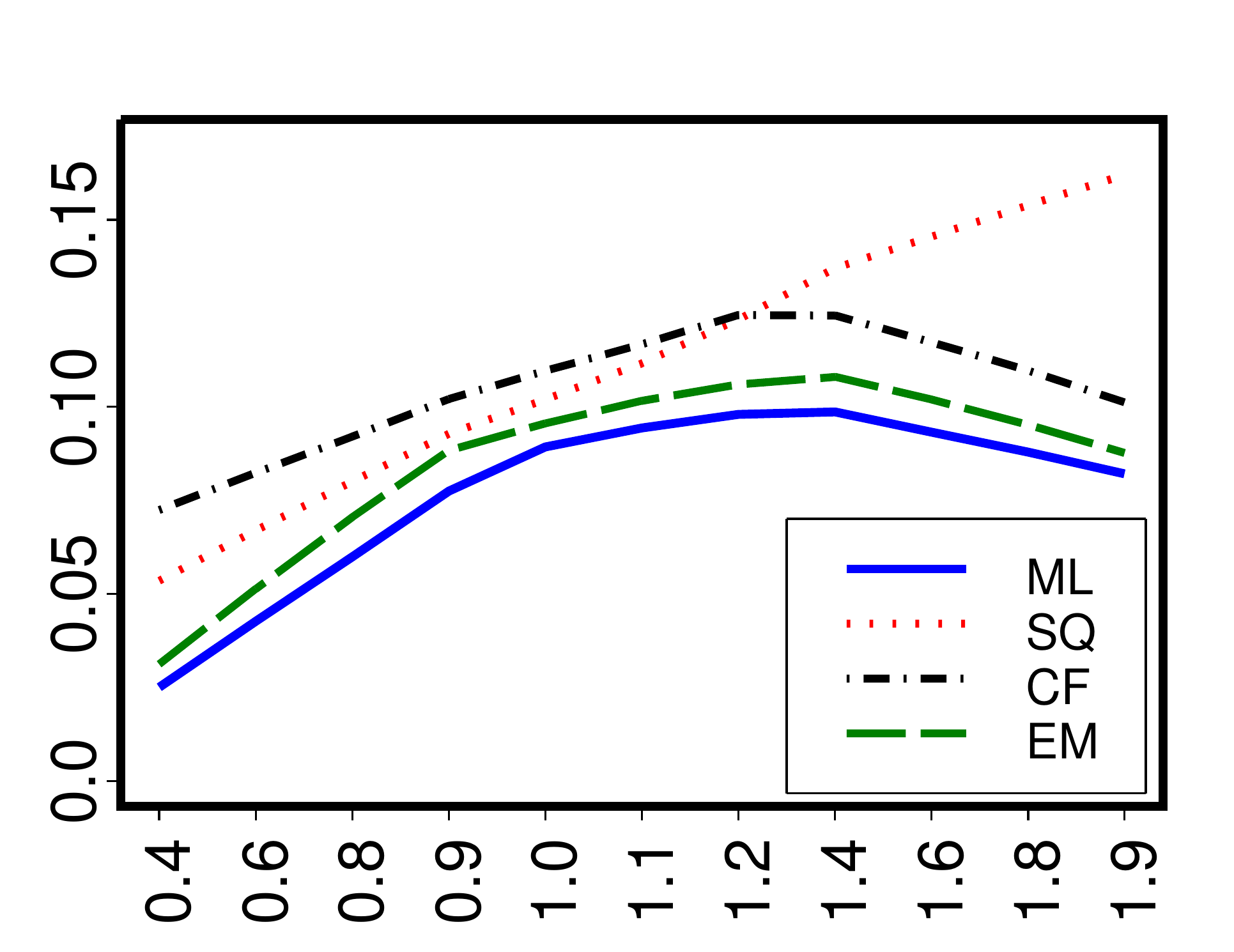}&
\includegraphics[angle=90,width=40mm,height=40mm]{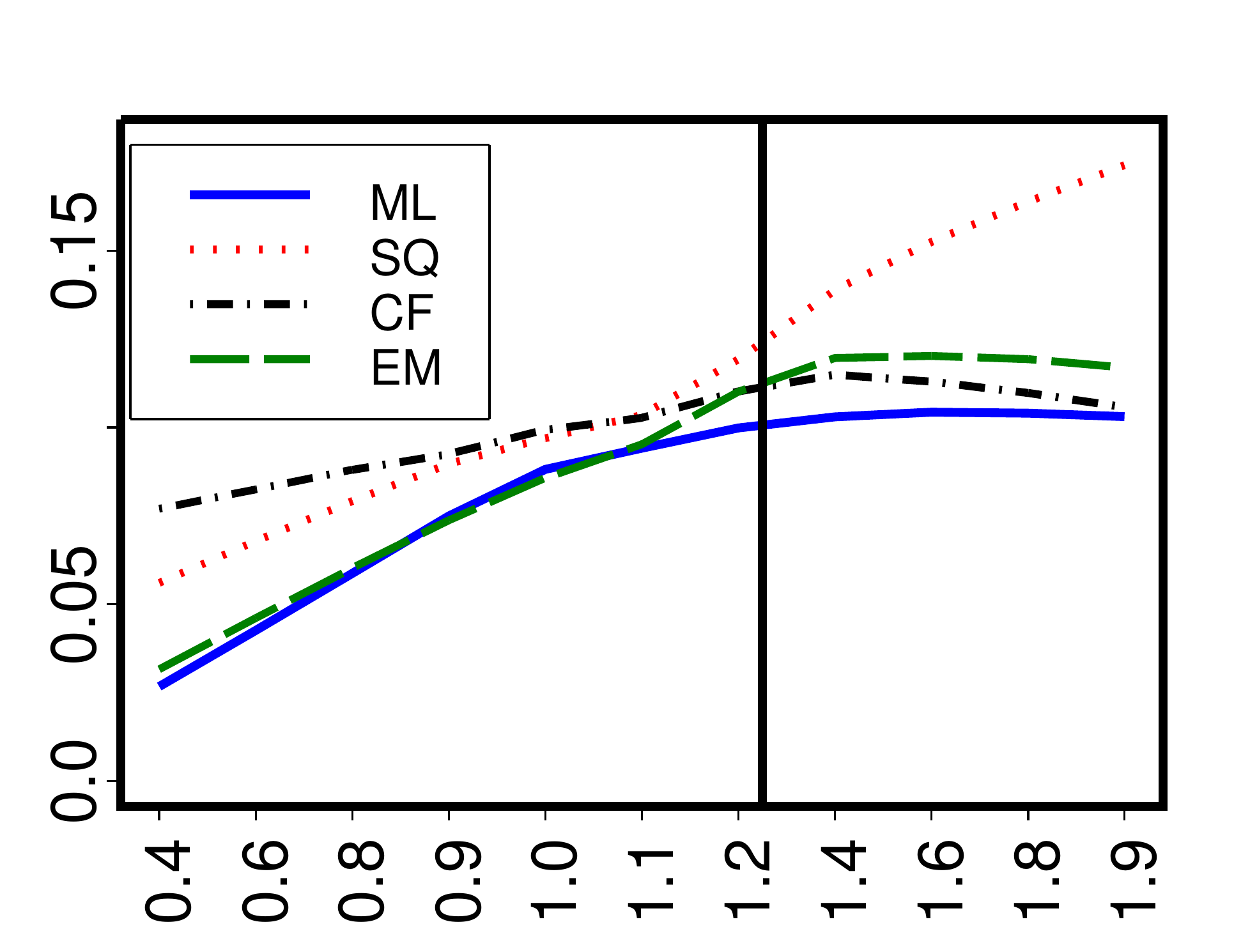}\\
&
\begin{rotate}{90}~~~~~~~~~{\scriptsize{n=500}}
\end{rotate}
&
\includegraphics[angle=90,width=40mm,height=40mm]{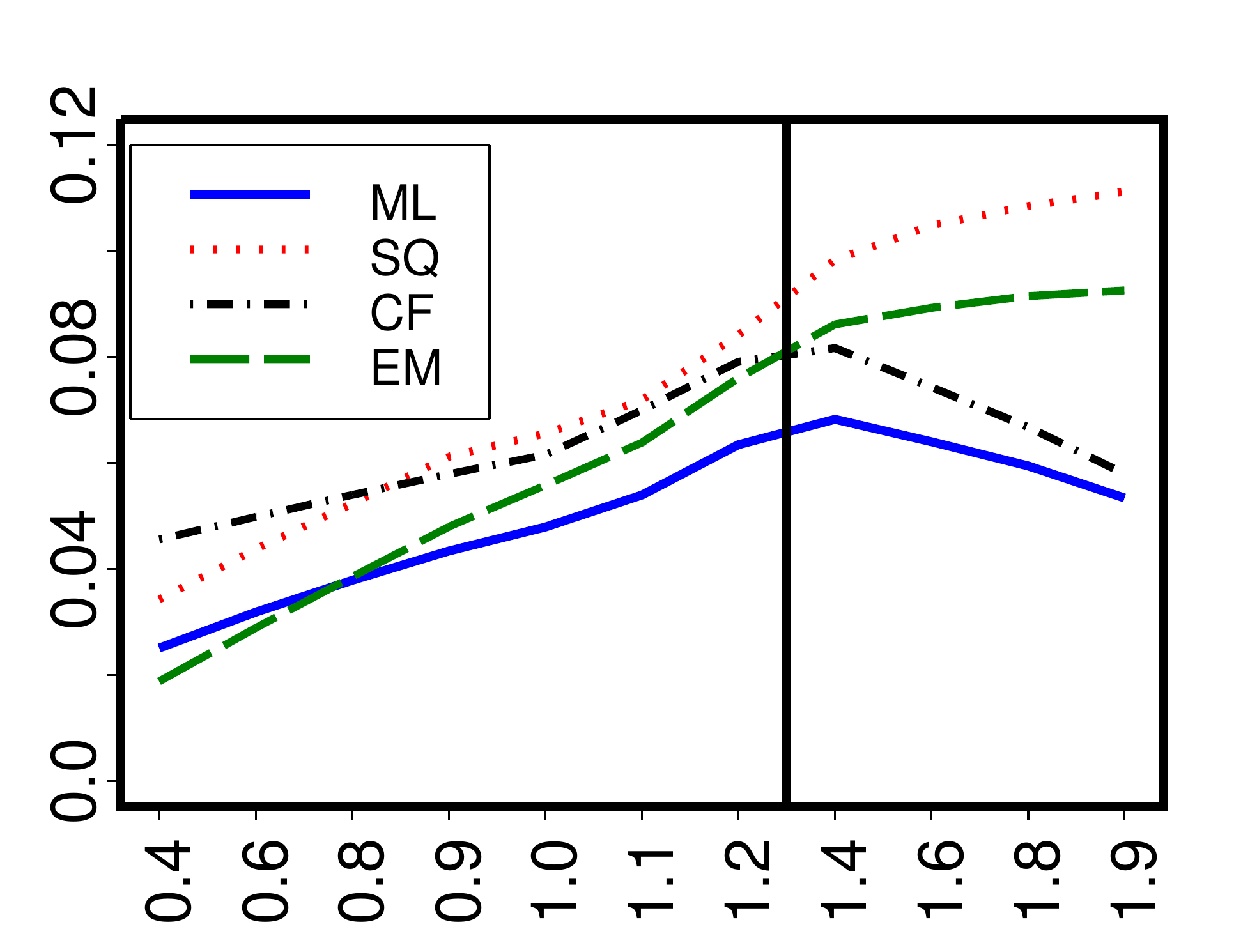}&
\includegraphics[angle=90,width=40mm,height=40mm]{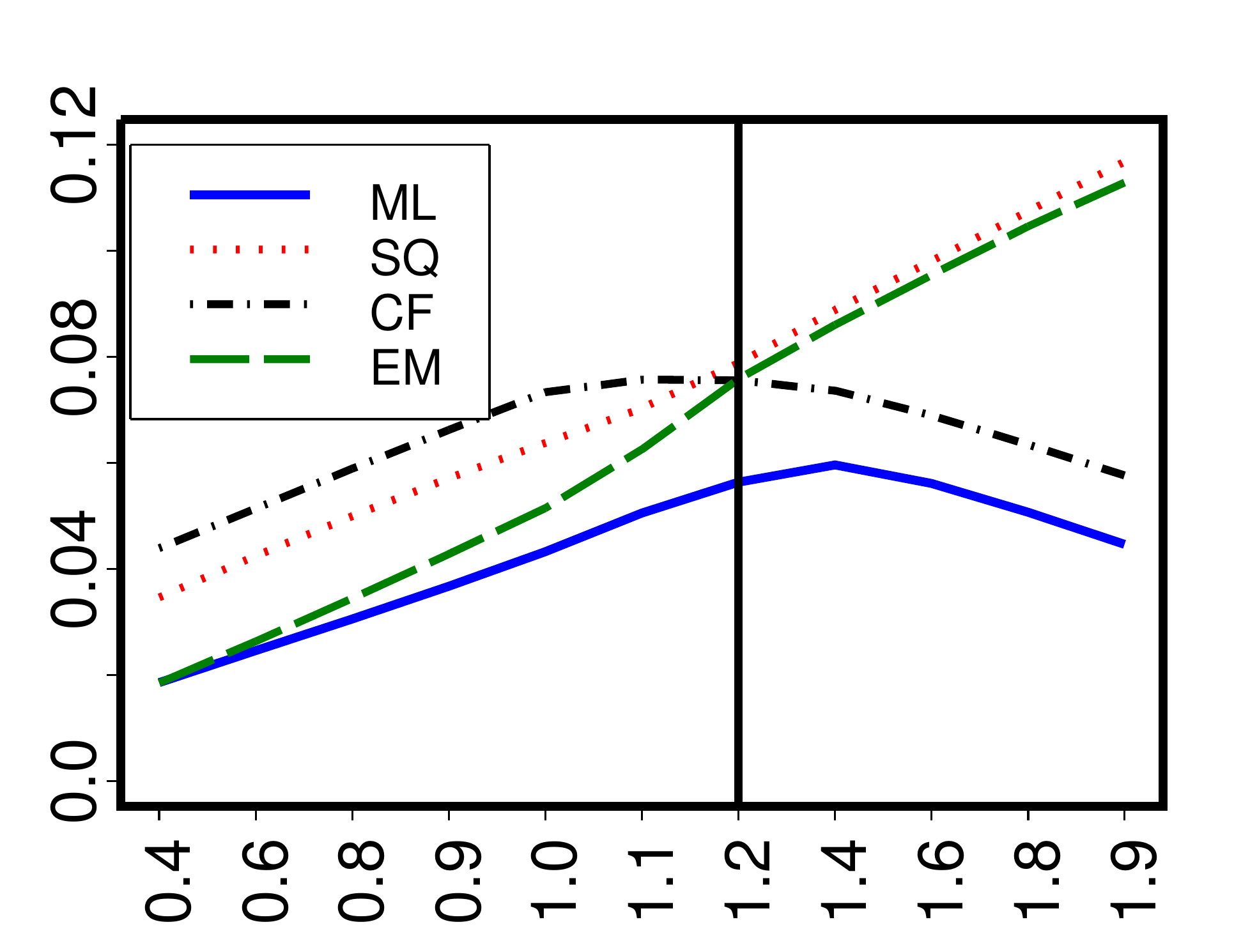}&
\includegraphics[angle=90,width=40mm,height=40mm]{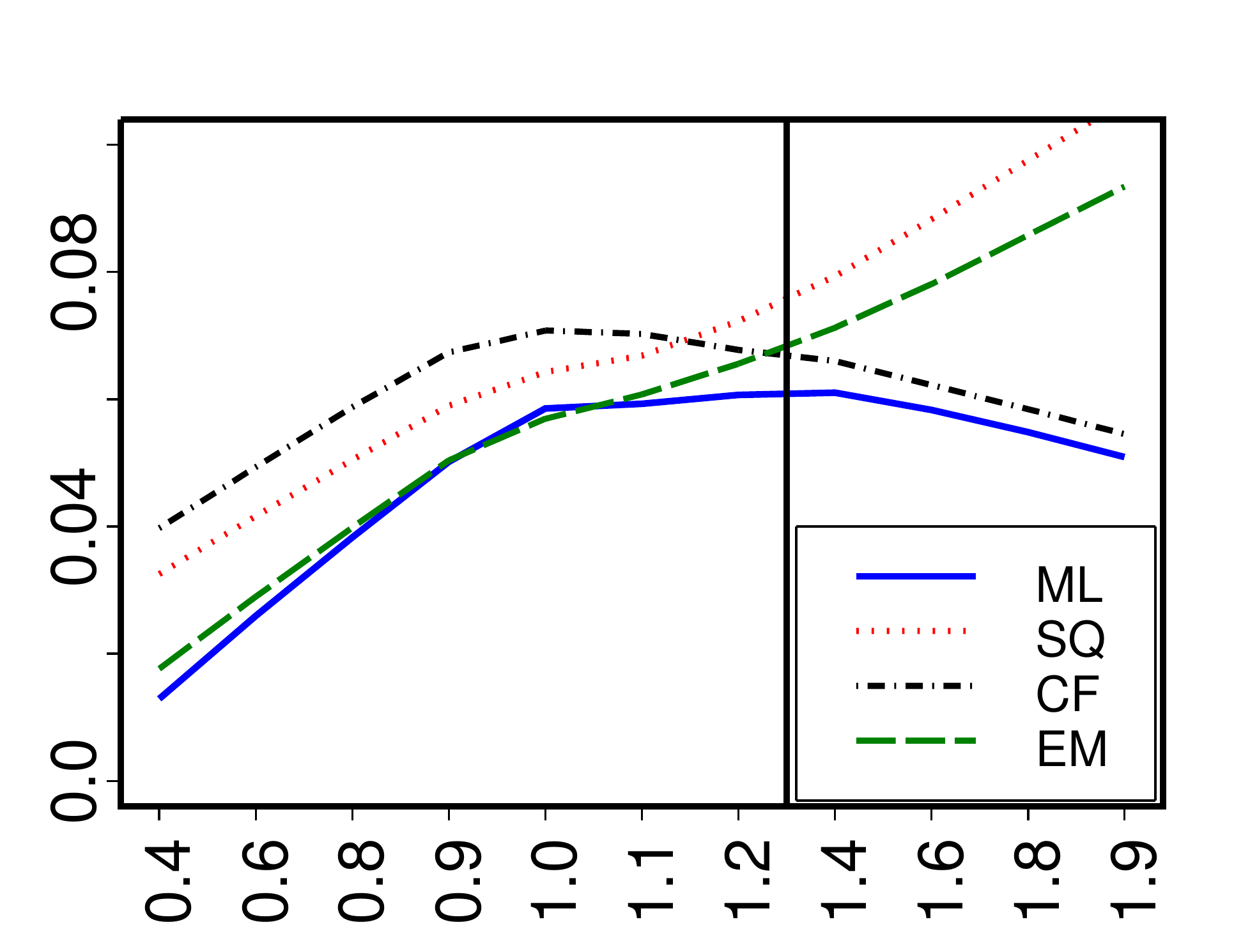}\\
&
\begin{rotate}{90}~~~~~~~~~{\scriptsize{n=1000}}
\end{rotate}
&
\includegraphics[angle=90,width=40mm,height=40mm]{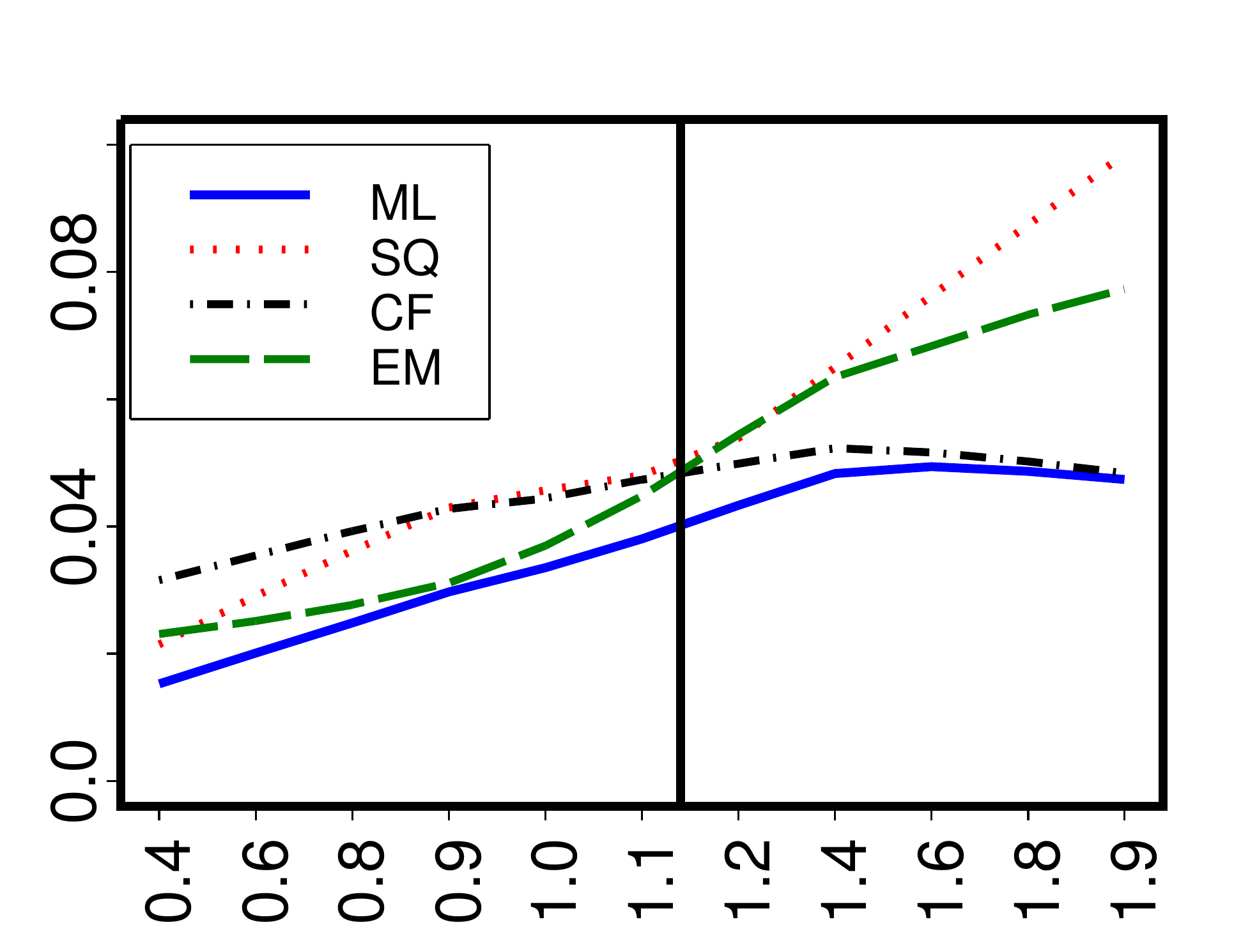}&
\includegraphics[angle=90,width=40mm,height=40mm]{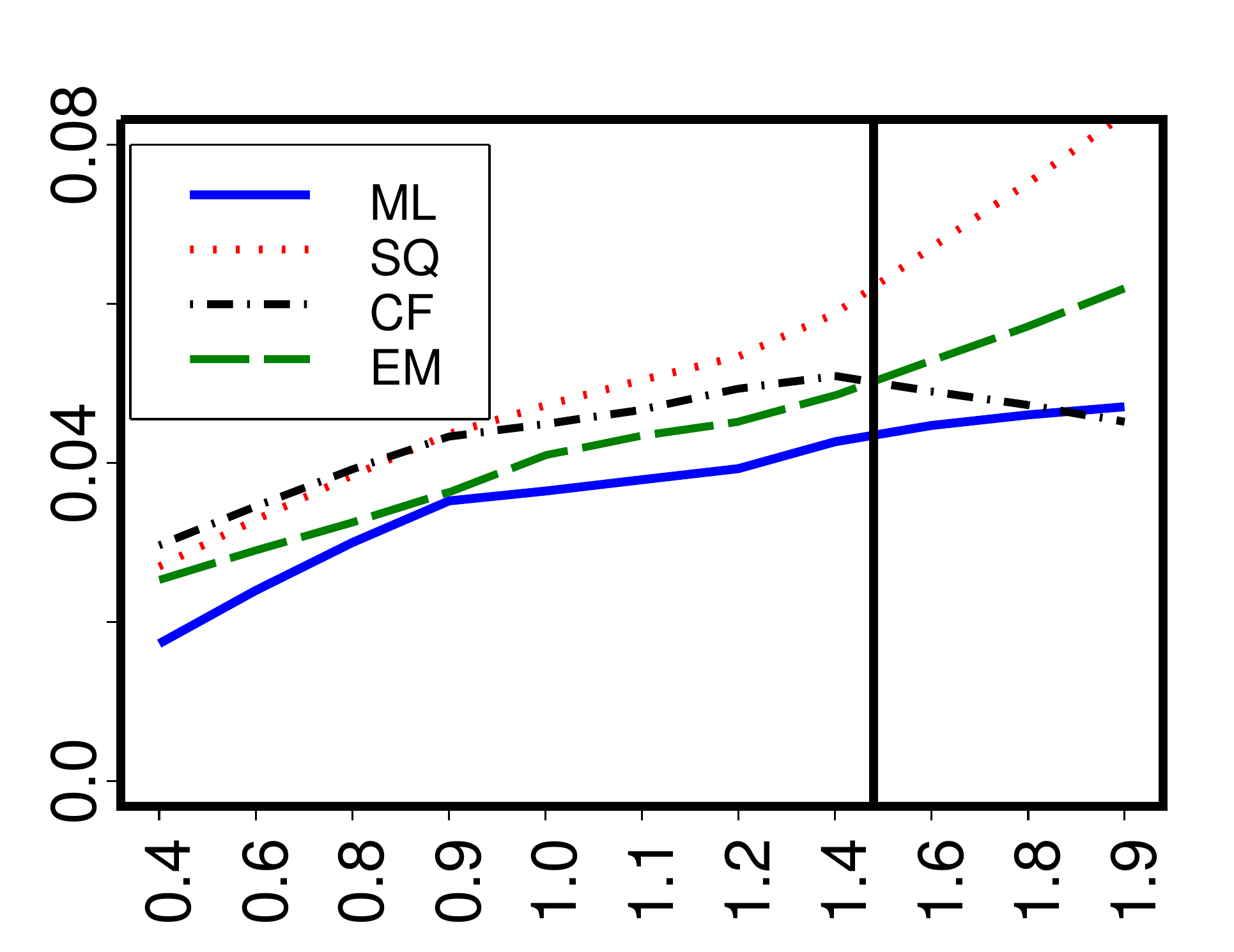}&
\includegraphics[angle=90,width=40mm,height=40mm]{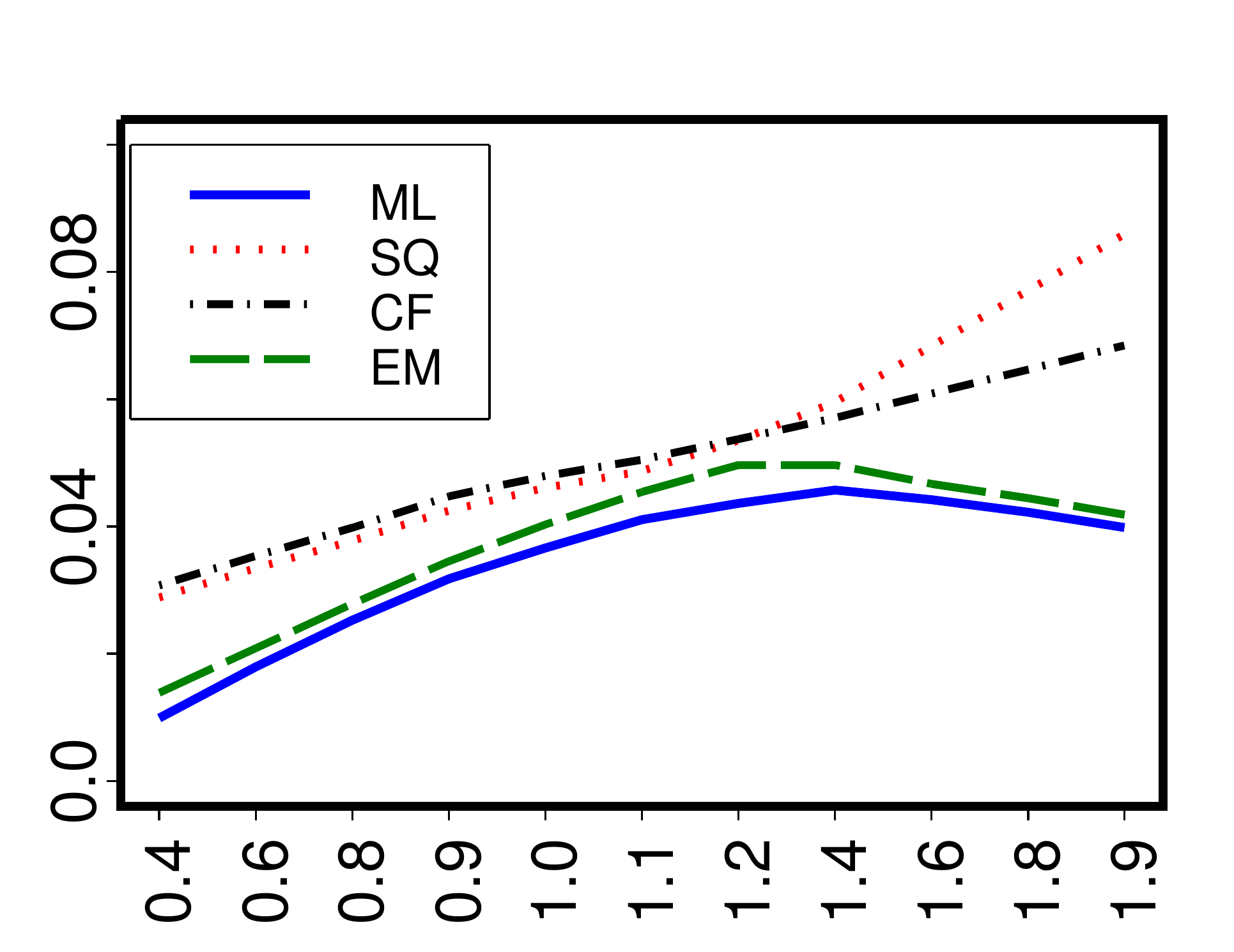}\\
\end{tabular}
\caption{RMSE of $\hat{\alpha}$ for different values of $\sigma=0.5, 1, 2$ and for $0.4 \leq \alpha \leq 1.9$. In all sub-figures {x-axis is values of} ${\alpha}$ {and y-axis is the RMSE of} ${\hat{\alpha}}$.}
\label{sas1}
\end{figure}

\begin{figure}[h!]
\begin{tabular}{p{0.01cm} p{0.1cm}ccc}
&&\scriptsize{${\sigma=0.5}$}&\scriptsize{${\sigma=1}$}&\scriptsize{${\sigma=2}$}\\
&
\begin{rotate}{90}~~~~~~~~~{\scriptsize{n=200}}
\end{rotate}
&
\includegraphics[angle=90,width=40mm,height=35mm]{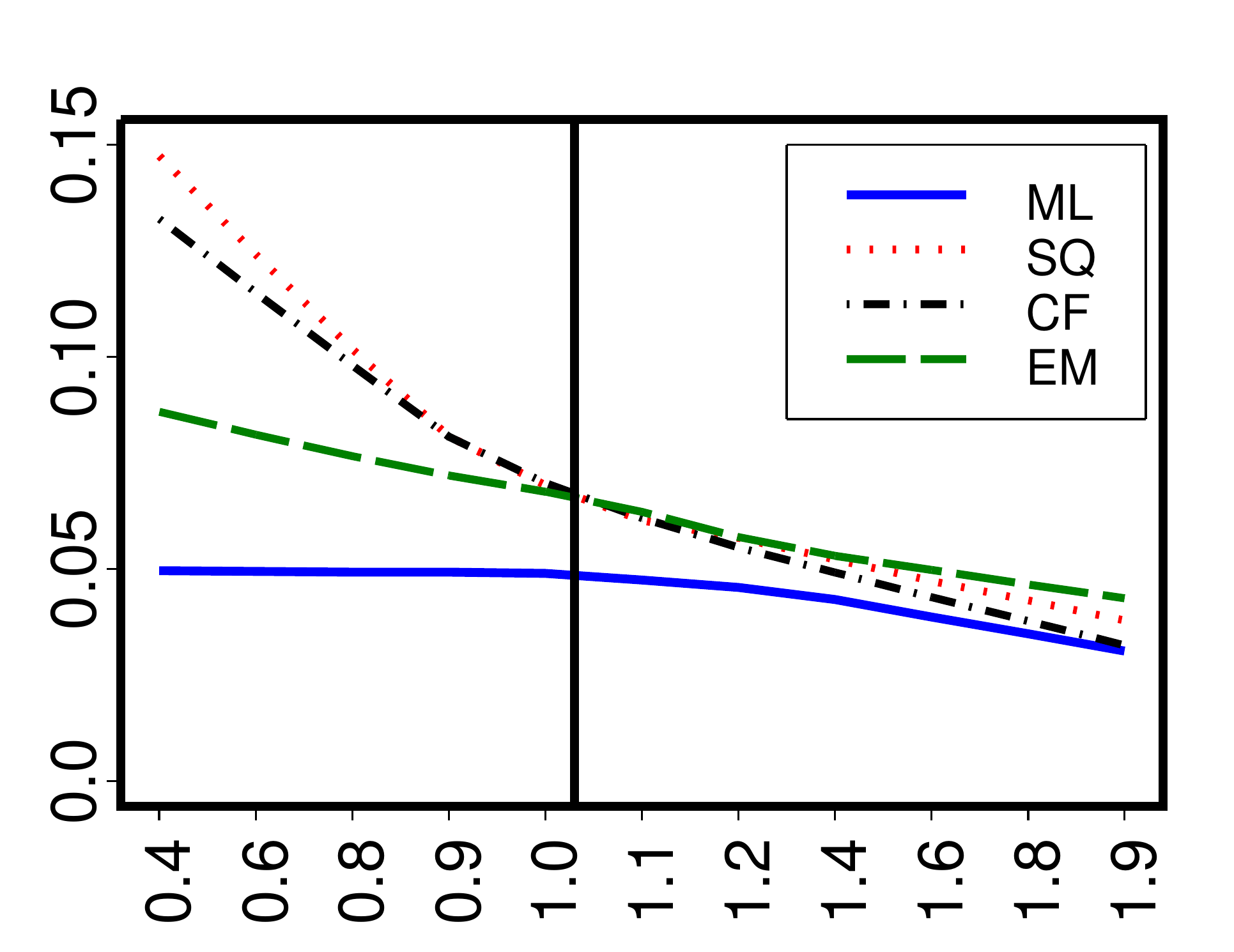}&
\includegraphics[angle=90,width=40mm,height=35mm]{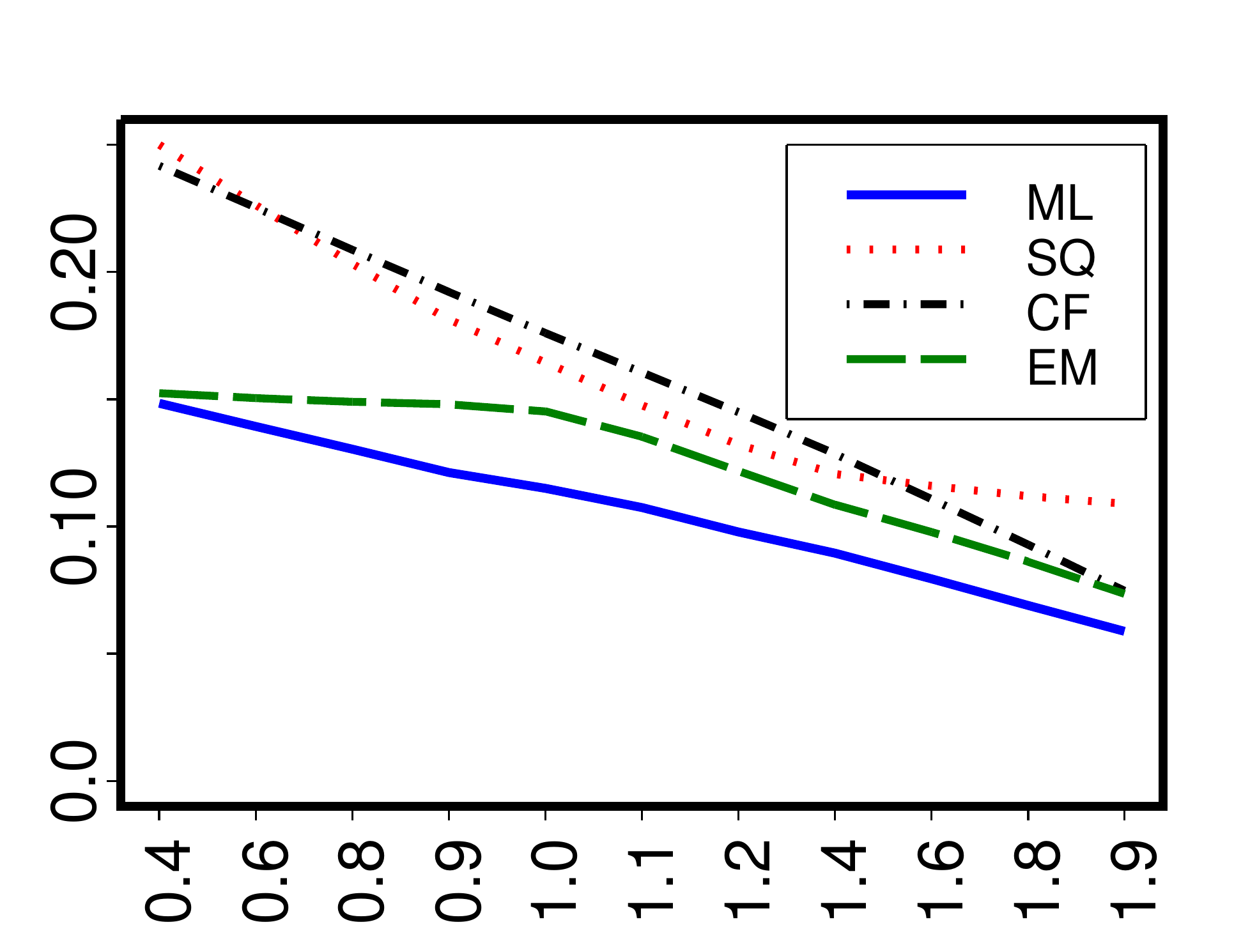}&
\includegraphics[angle=90,width=40mm,height=35mm]{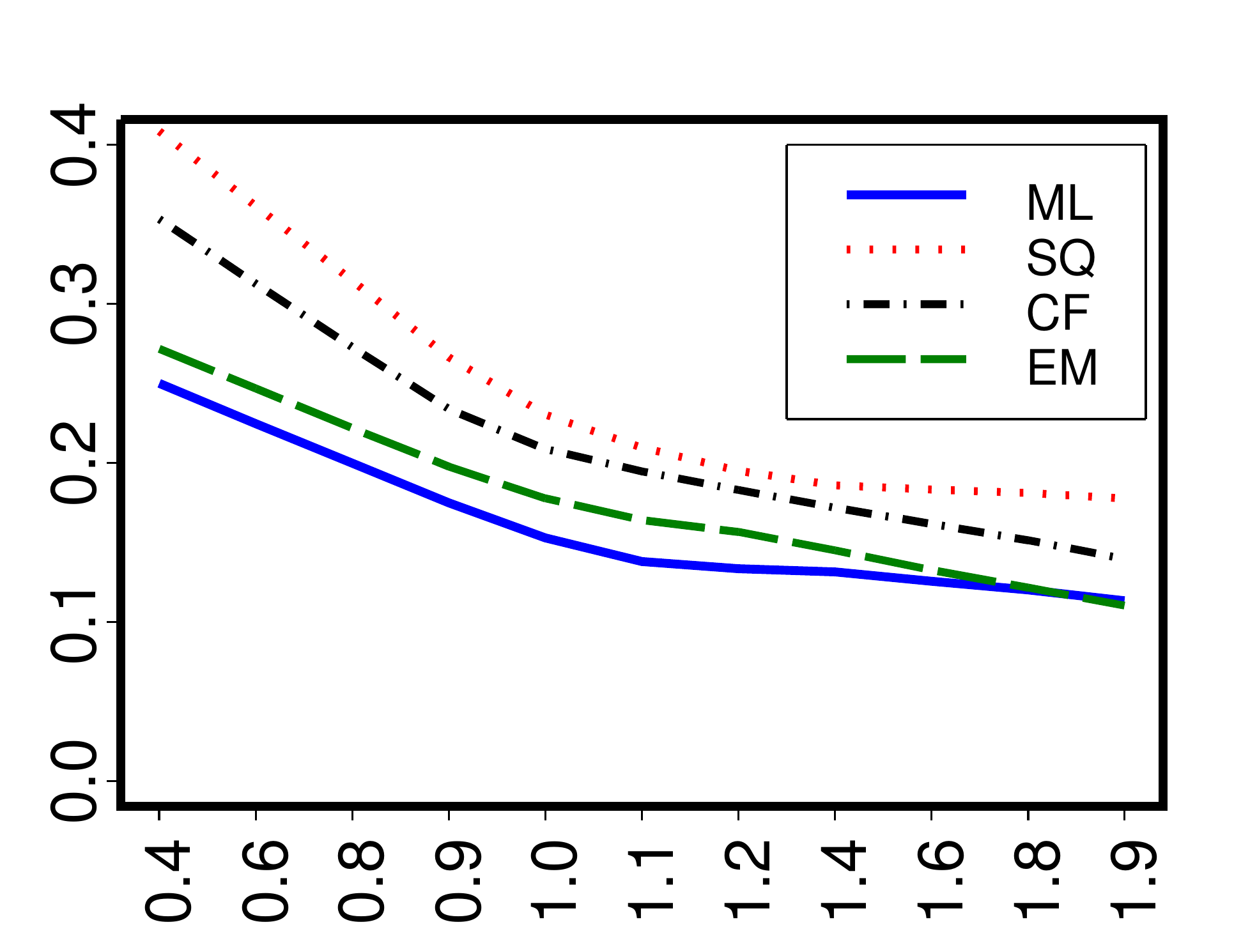}\\
&
\begin{rotate}{90}~~~~~~~~~{\scriptsize{n=500}}
\end{rotate}
&
\includegraphics[angle=90,width=40mm,height=35mm]{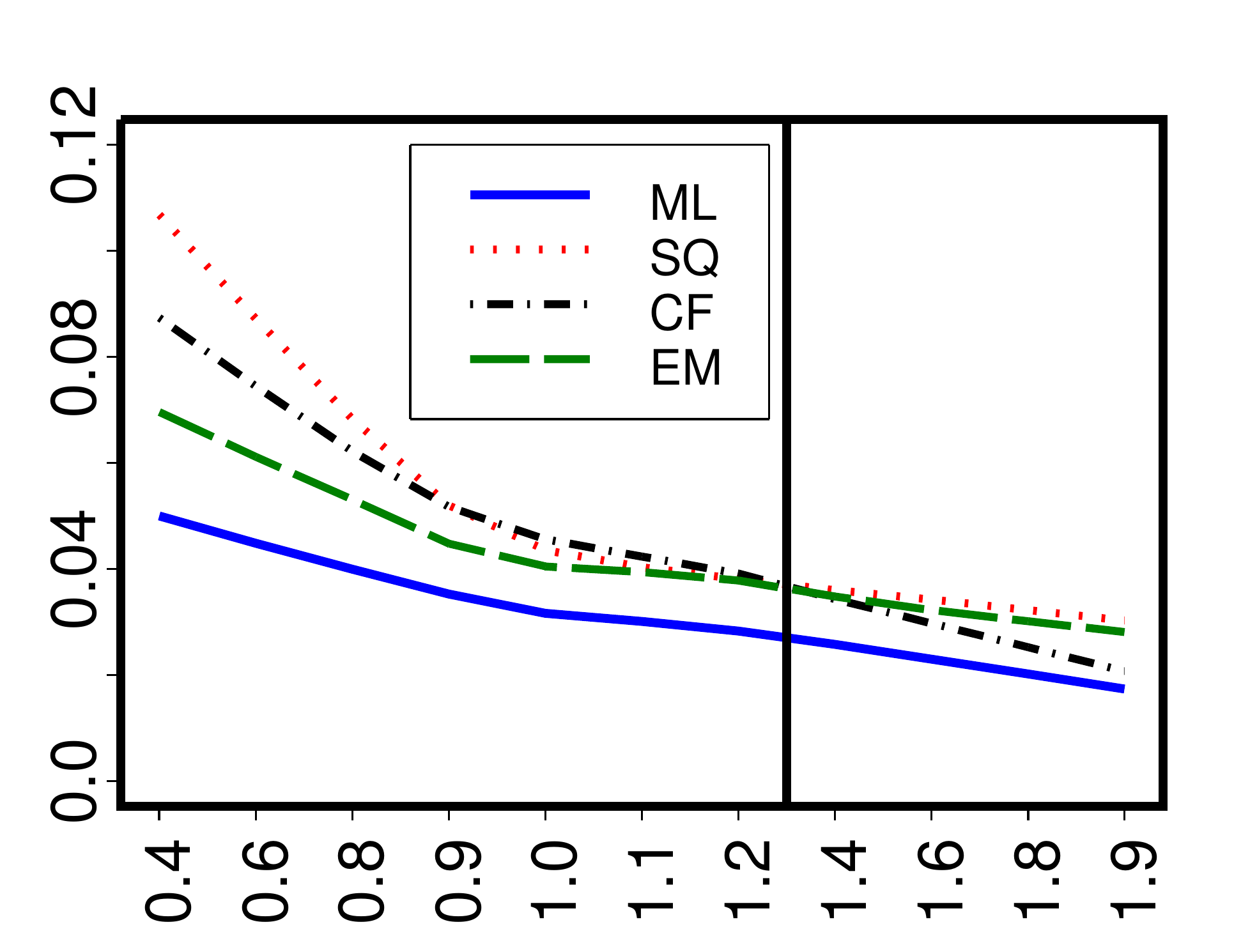}&
\includegraphics[angle=90,width=40mm,height=35mm]{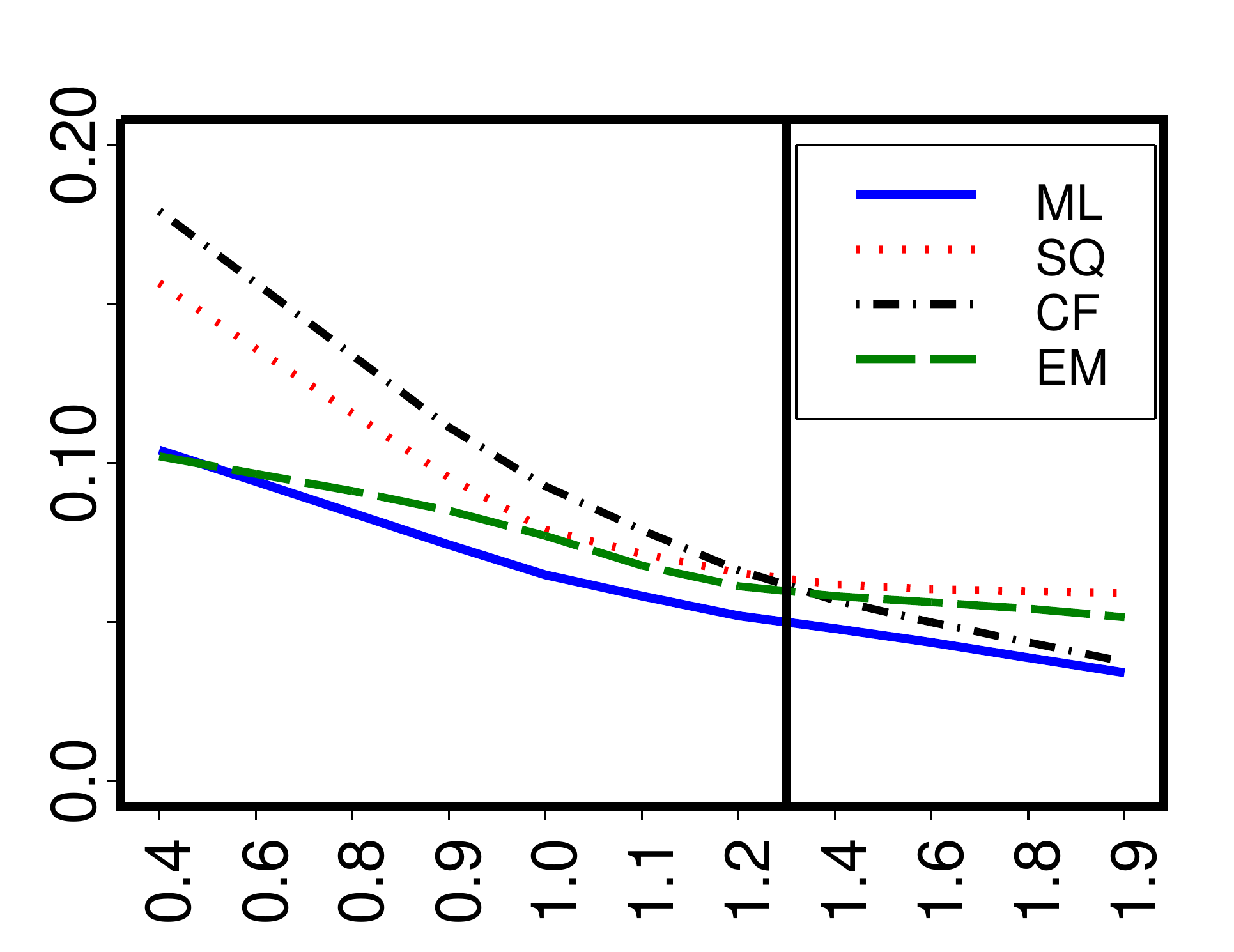}&
\includegraphics[angle=90,width=40mm,height=35mm]{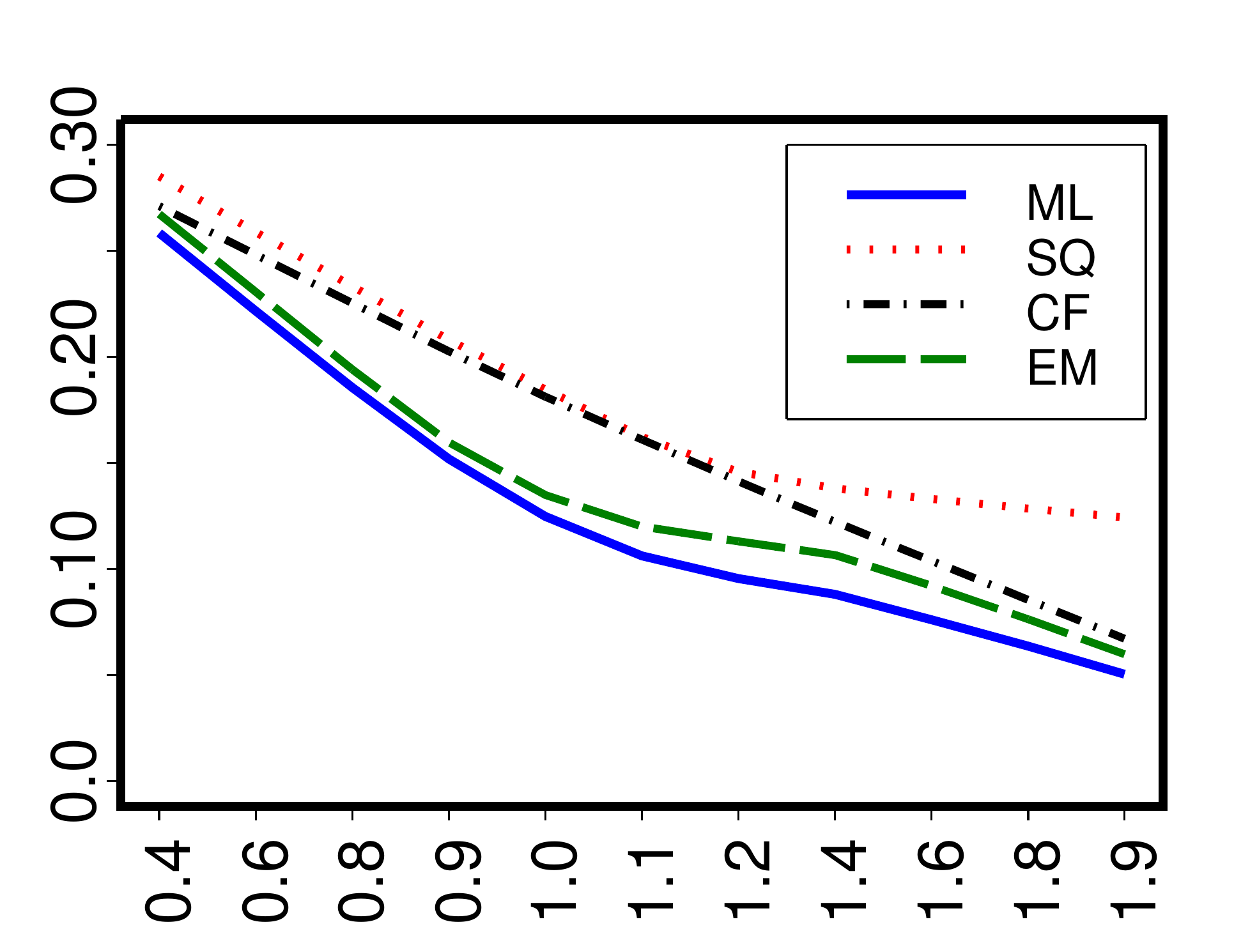}\\
&
\begin{rotate}{90}~~~~~~~~~{\scriptsize{n=1000}}
\end{rotate}
&
\includegraphics[angle=90,width=40mm,height=35mm]{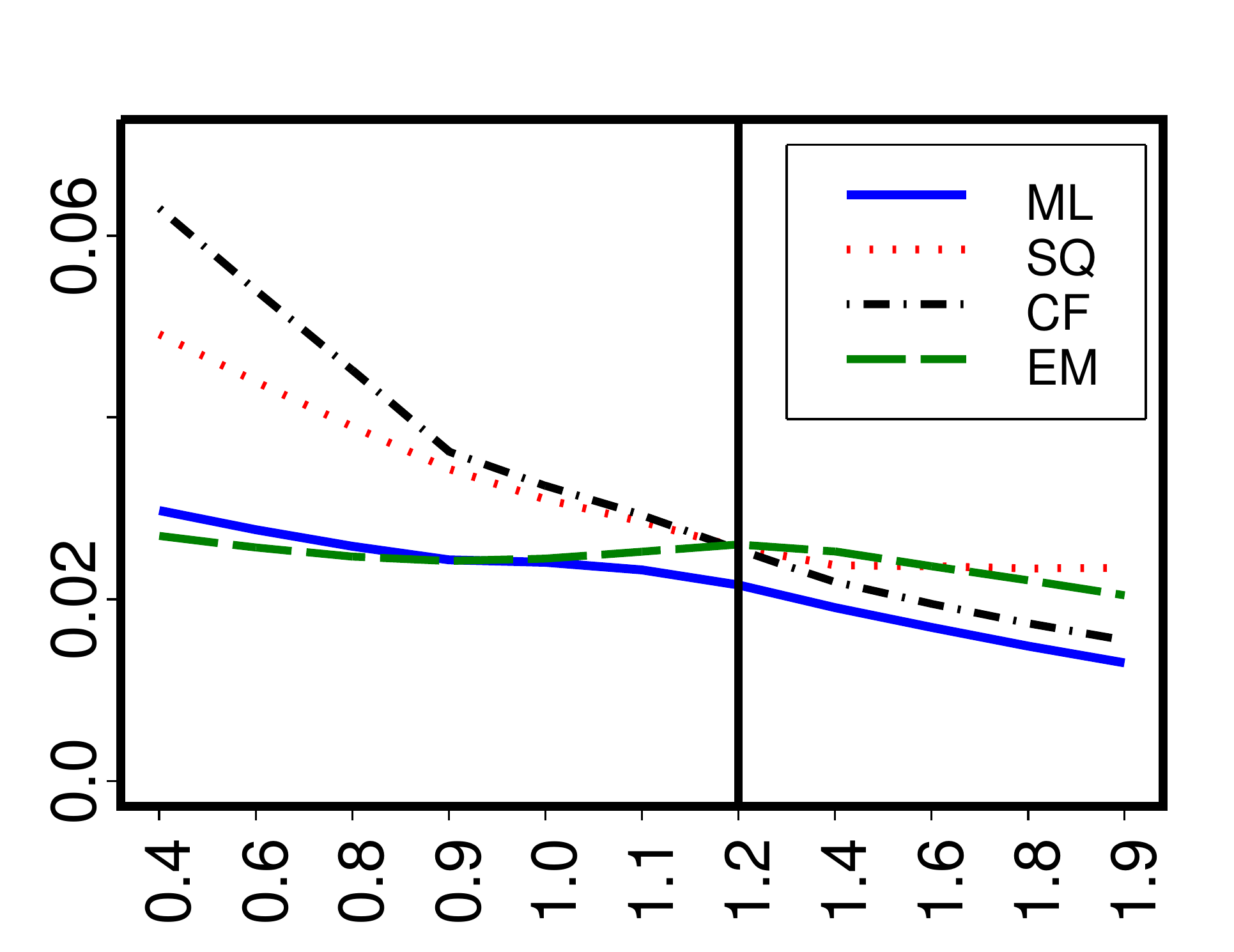}&
\includegraphics[angle=90,width=40mm,height=35mm]{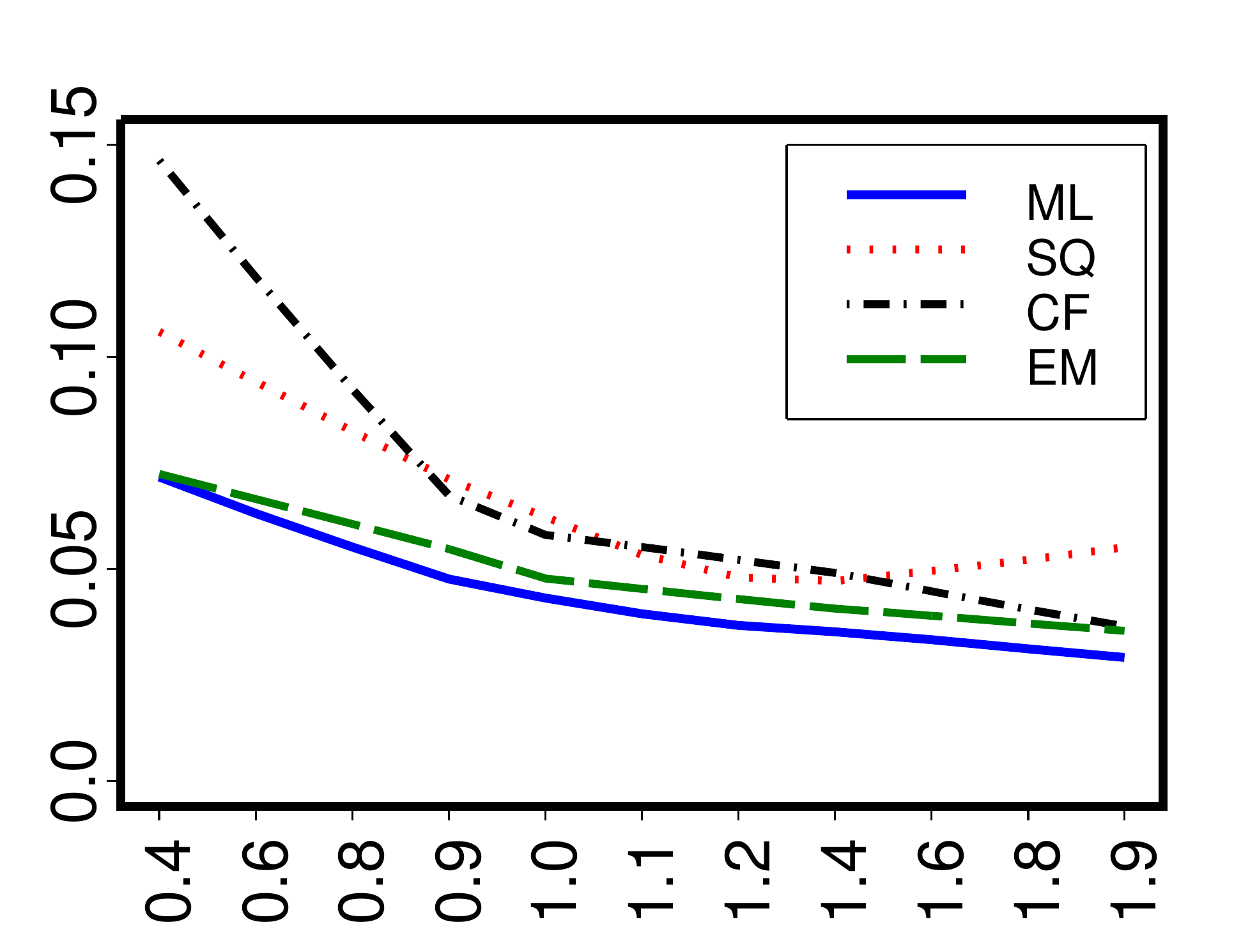}&
\includegraphics[angle=90,width=40mm,height=35mm]{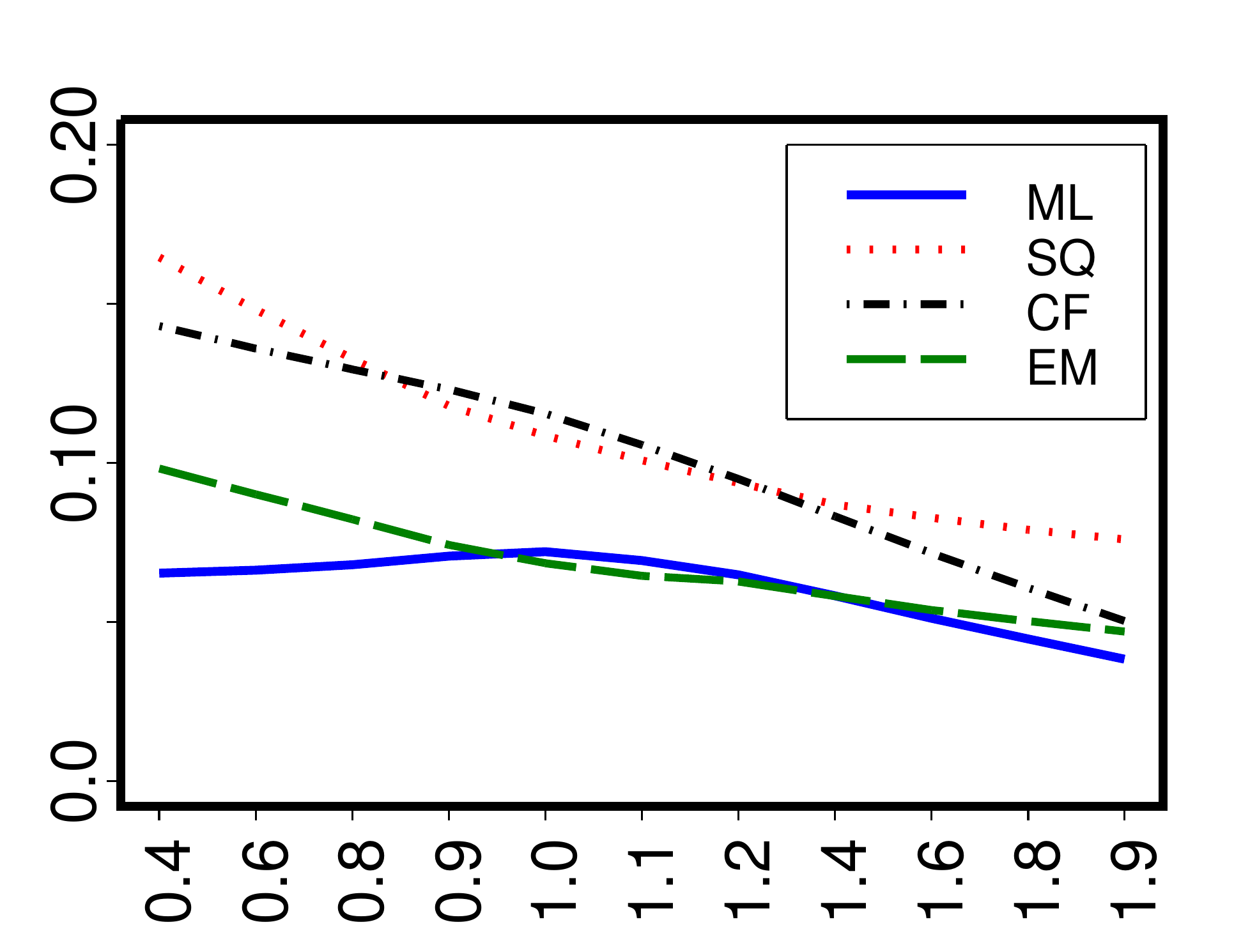}\\
\end{tabular}
\caption{RMSE of $\hat{\sigma}$ for different values of $\sigma=0.5, 1, 2$ and for $0.4 \leq \alpha \leq 1.9$. In all sub-figures {x-axis is values of} ${\alpha}$ {and y-axis is the RMSE of} ${\hat{\sigma}}$.}
\label{sas2}
\end{figure}

\begin{figure}[h!]
\begin{tabular}{p{0.01cm} p{0.1cm}ccc}
&&\scriptsize{${\sigma=0.5}$}&\scriptsize{${\sigma=1}$}&\scriptsize{${\sigma=2}$}\\
&
\begin{rotate}{90}~~~~~~~~~{\scriptsize{n=200 }}
\end{rotate}
&
\includegraphics[angle=90,width=40mm,height=35mm]{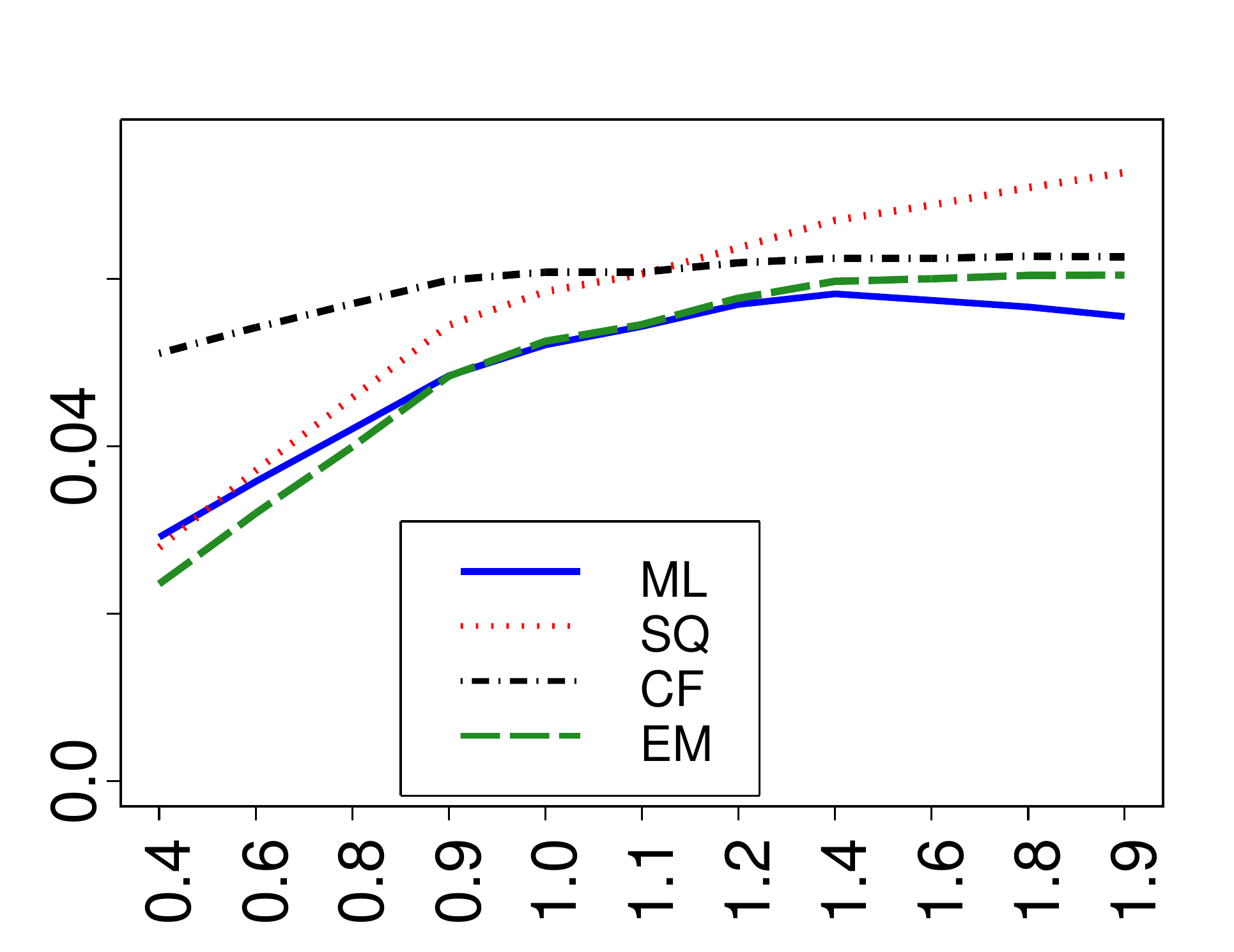}&
\includegraphics[angle=90,width=40mm,height=35mm]{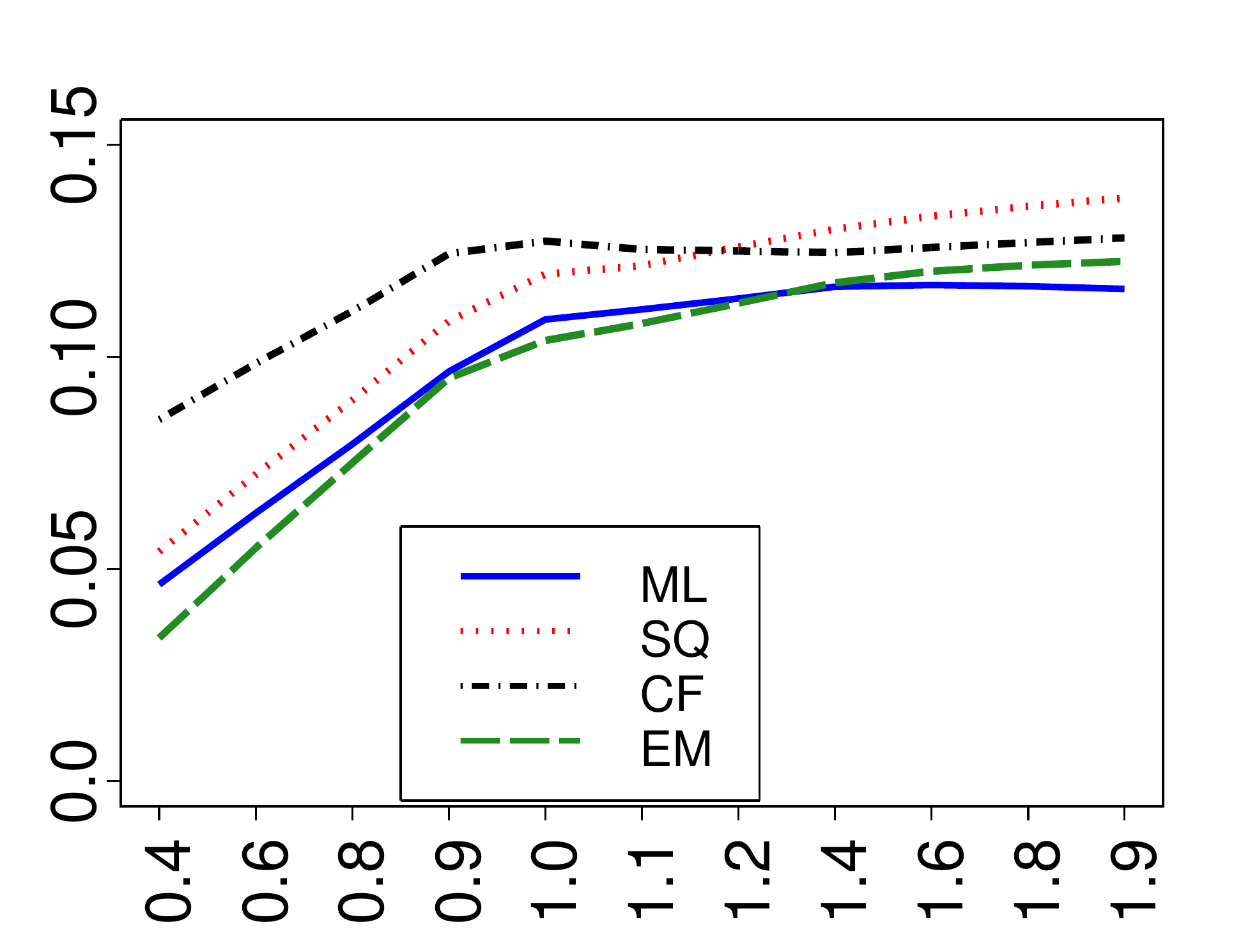}&
\includegraphics[angle=90,width=40mm,height=35mm]{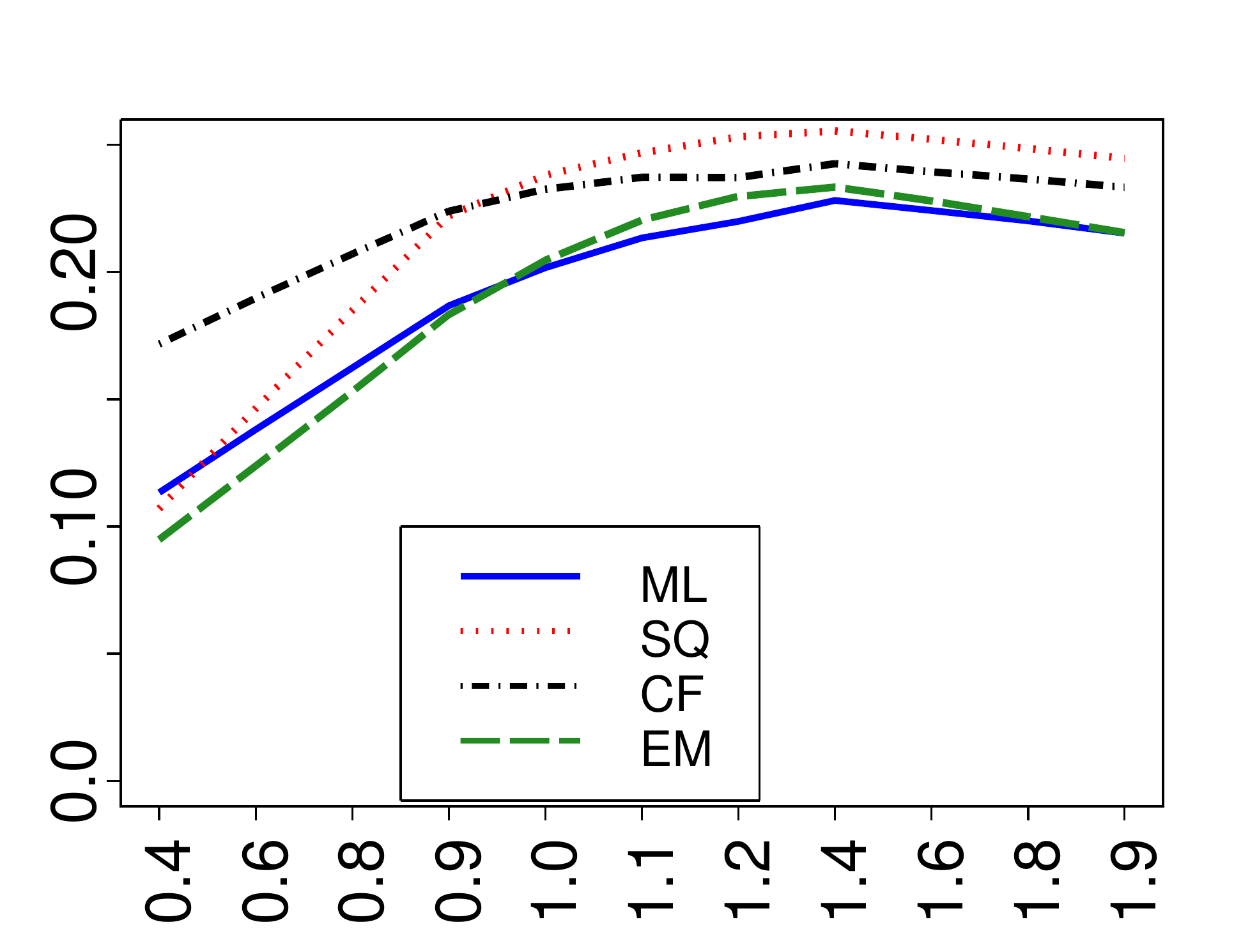}\\
&
\begin{rotate}{90}~~~~~~~~~{\scriptsize{n=500}}
\end{rotate}
&
\includegraphics[angle=90,width=40mm,height=35mm]{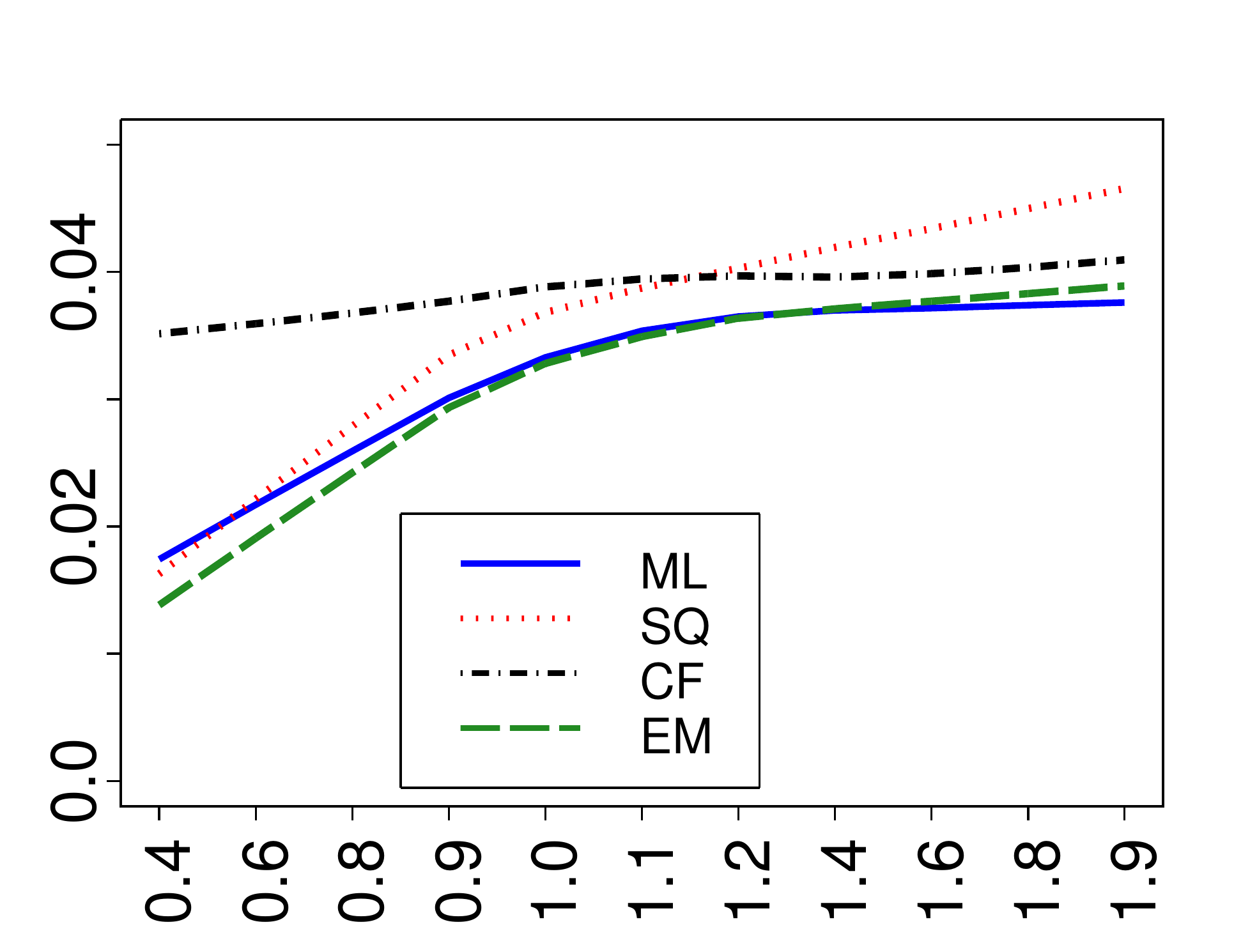}&
\includegraphics[angle=90,width=40mm,height=35mm]{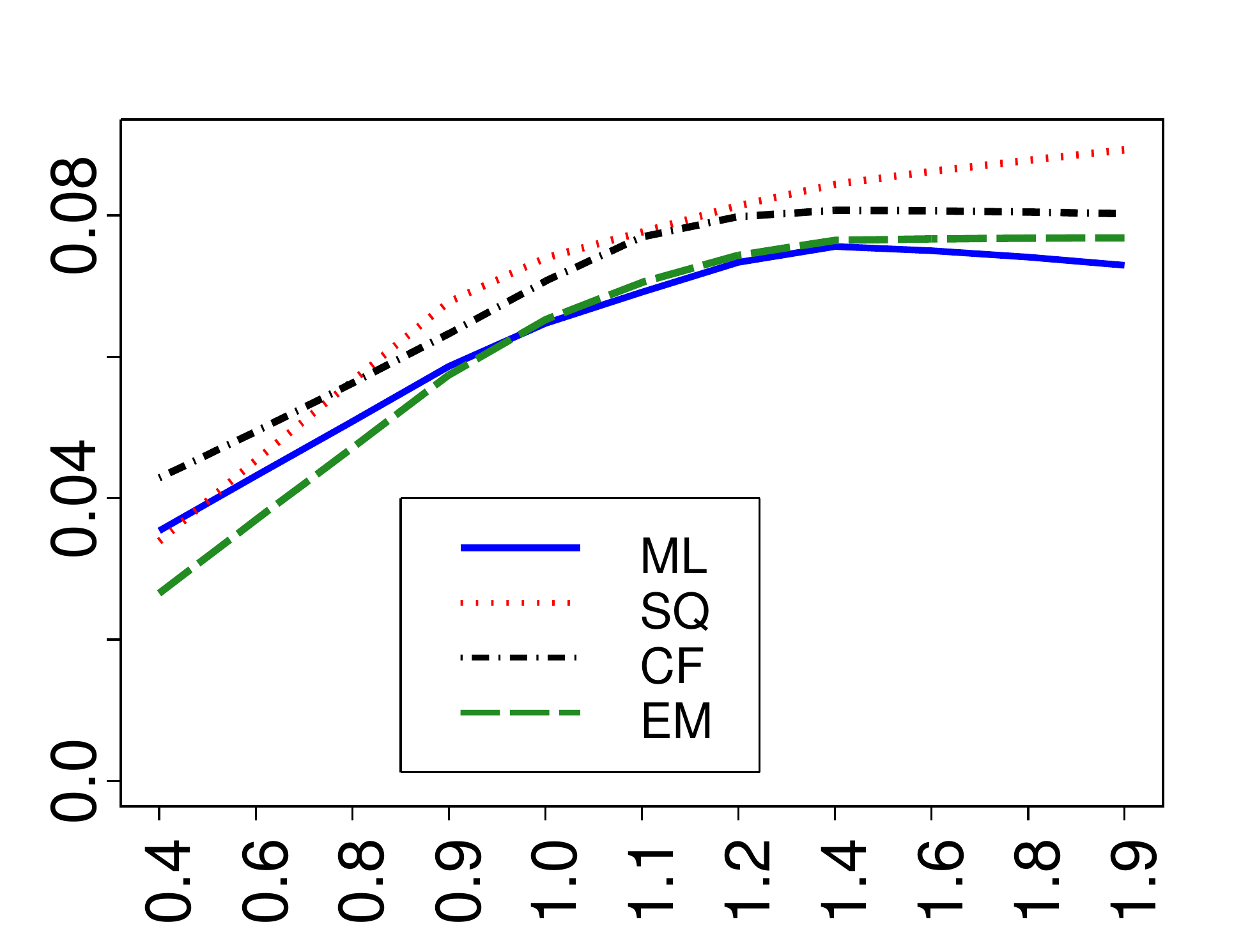}&
\includegraphics[angle=90,width=40mm,height=35mm]{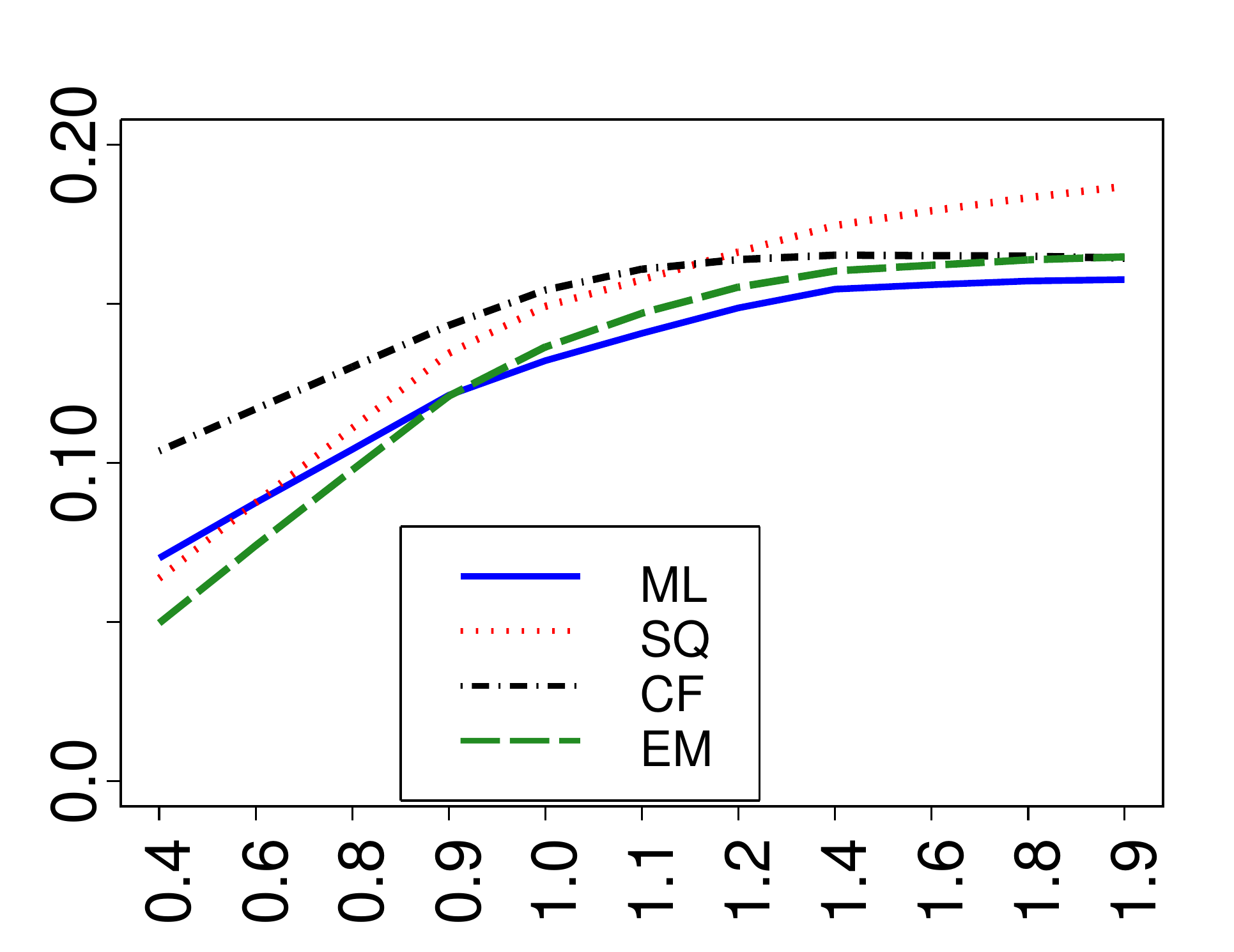}\\
&
\begin{rotate}{90}~~~~~~~~~{\scriptsize{n=1000}}
\end{rotate}
&
\includegraphics[angle=90,width=40mm,height=35mm]{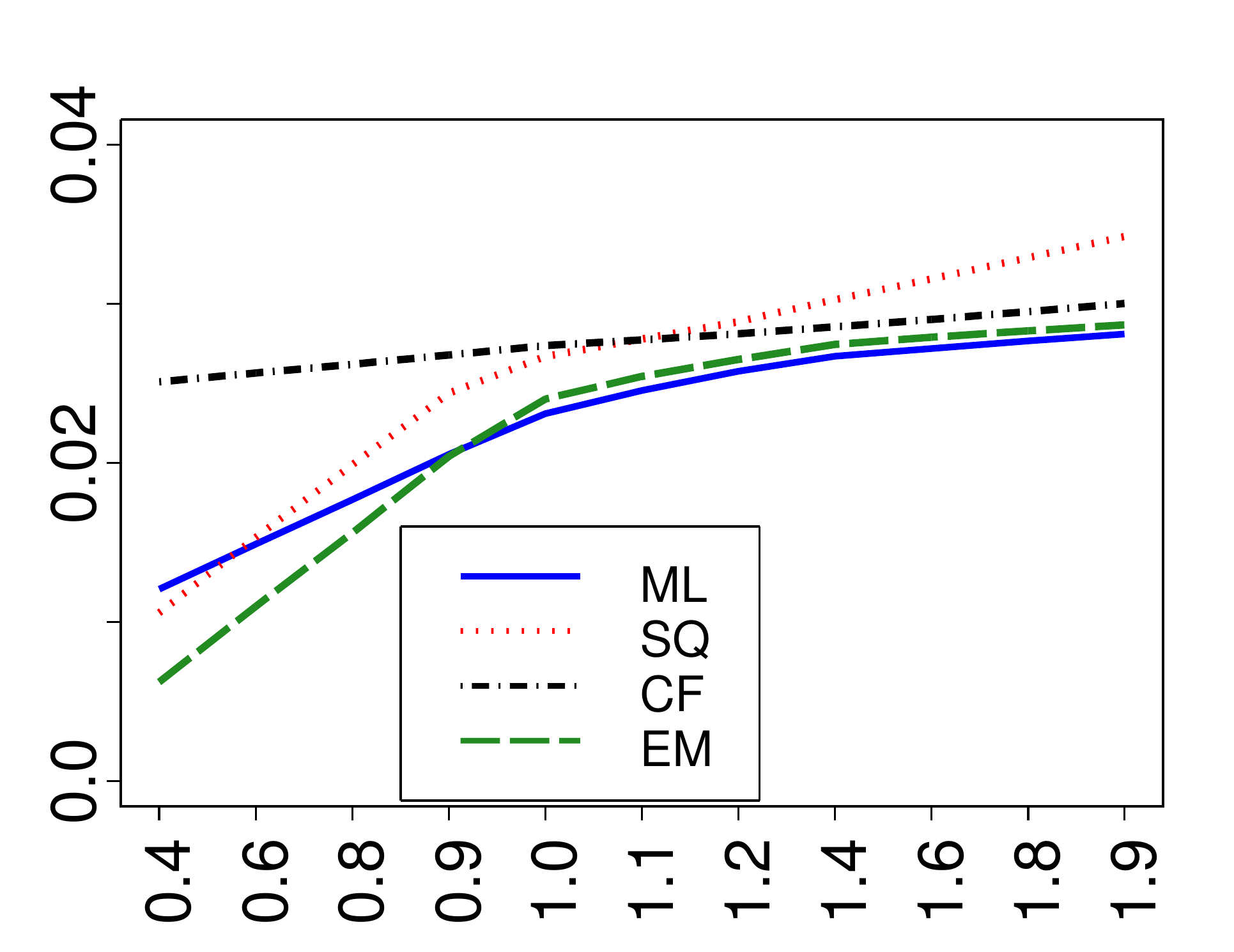}&
\includegraphics[angle=90,width=40mm,height=35mm]{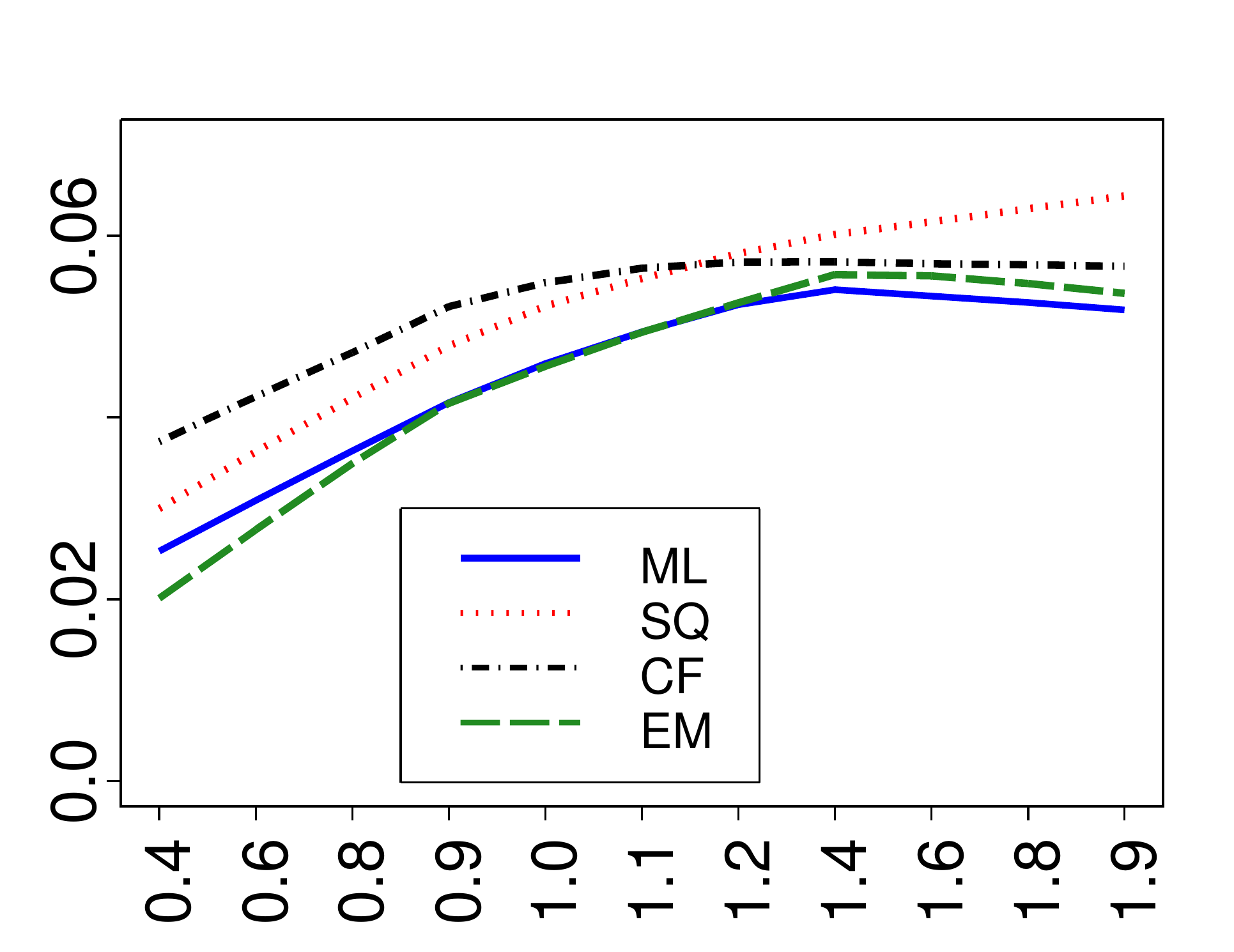}&
\includegraphics[angle=90,width=40mm,height=35mm]{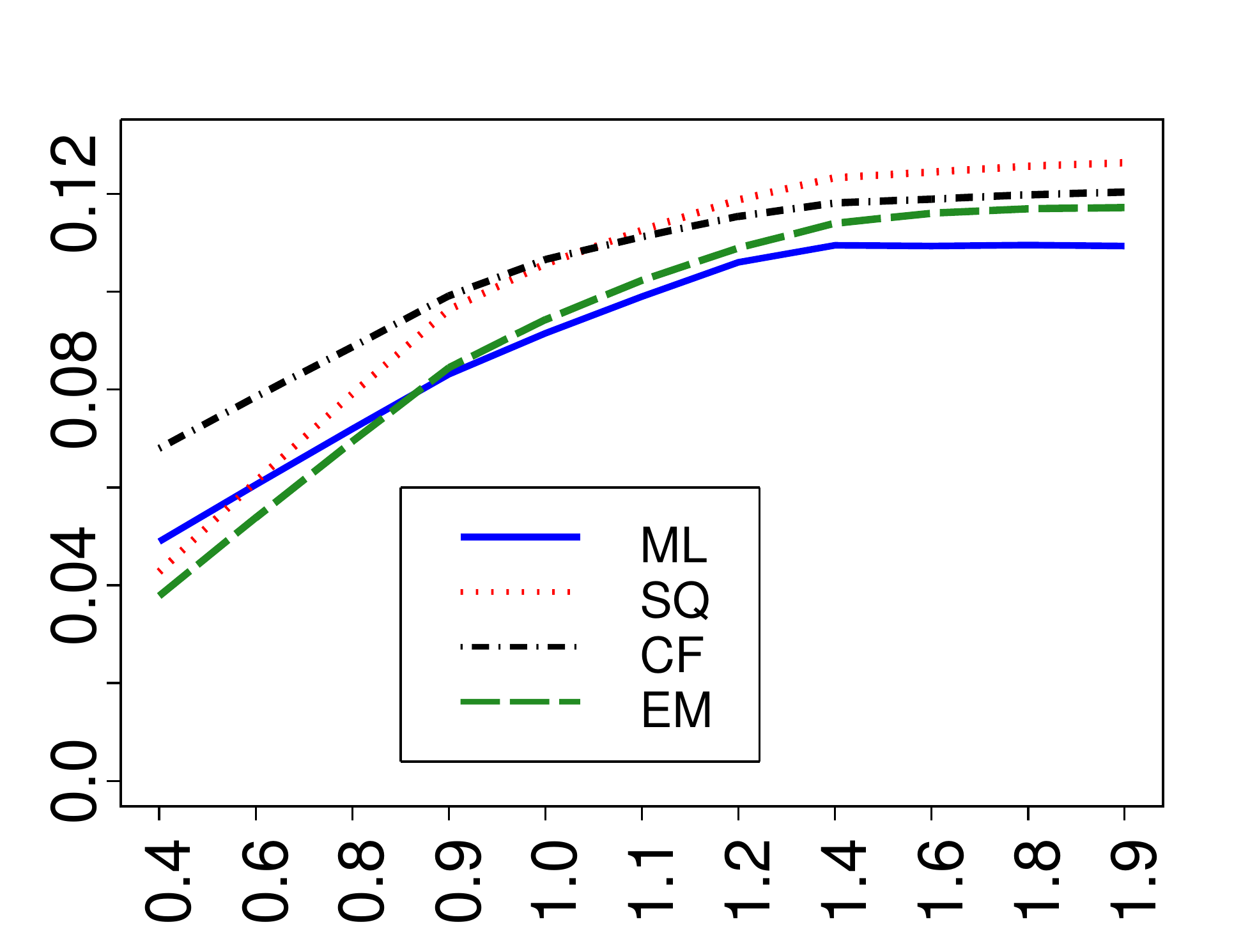}\\
\end{tabular}
\caption{RMSE of $\hat{\mu}$ for different values of $\sigma=0.5, 1, 2$ and for $0.4 \leq \alpha \leq 1.9$. In all sub-figures {x-axis is values of} ${\alpha}$ {and y-axis is the RMSE of} ${\hat{\mu}}$.}
\label{sas3}
\end{figure}

Similar to the previous subsection, by considering a latent variable such as $P$, we consider the following representation for each symmetric stable distribution, see \cite{samorodnitsky1994stable}.
\begin{proposition}\label{prop3} Suppose $Y\sim S_{1}(\alpha,0,\sigma,\mu)$ and $P \sim S_{1}\bigl(\alpha/2,1,\bigl(\cos(\pi \alpha/4)\bigr)^{2/\alpha},0\bigr)$. Then
\begin{eqnarray} \label{sym1}
Y\mathop=\limits^d \sigma \sqrt{2P}N+\mu
\end{eqnarray}
where $N\sim{\cal{N}}(0,1)$. The random variables $N$ and $P$ are independent.
\end{proposition}
Using representation (\ref{sym1}), the proposed EM algorithm is applied to estimate the parameters of symmetric stable distribution, see \cite{teimouri2018parameter}. We compare the performance of the CF, ML, SQ, and the proposed EM methods for estimating the parameters of the symmetric stable distribution through simulations. For applying the CF, ML, and SQ approaches we use {STABLE} software, see \cite{nolan2001maximum}. In simulations, the root of mean square error (RMSE) is computed for sample sizes $n=200, 500$, and 1000. The number of replications is $N=1000$ and parameter settings are: $\mu=0$, $\sigma=0.5,1,2$, and $\alpha=0.4,0.6,0.8,0.9,1.1,1.2,1.4,1.6,1.8,1.9$. The simulations results under four methods are shown in Figures \ref{sas1}-\ref{sas3}. It should be noted that the estimation results for the evaluated ML approach is not reliable for $\alpha<0.4$. The smoothed version of the actual values plotted in these figures are obtained using the {\bf{lowess}} package with the following settings: the smoothing span is considered to be 2/3, number of iterations is 3, and the speed of computations is determined by 0.01th of the range of the $\alpha$ values, see \cite{cleveland1981lowess}. 
\par In each sub-figure through Figures \ref{sas1}-\ref{sas3}, the vertical line indicates that the proposed EM algorithm works better than the CF and SQ methods for those $\alpha$s lie in the left side of the line. If the EM algorithm works as the best, such a line is disappeared. By considering the location parameter as zero, the following results are obtained from the simulations.
\begin{enumerate} 
\item EM-based $\hat{\alpha}$ and $\hat{\sigma}$ show better performance than both the CF and SQ for $\alpha<1.1$ in all cases, ref. Figures \ref{sas1}-\ref{sas2}, respectively. 
\item EM-based $\hat{\mu}$ outperforms the CF and SQ methods, ref. Figure \ref{sas3}. 
\item EM-based $\hat{\alpha}$ outperforms the SQ-based $\hat{\alpha}$. 
\item EM-based $\hat{\mu}$ shows superior performance than the CF- and SQ-based $\hat{\mu}$. 
\item We know that ML method gives the efficient estimation, particularly when sample size gets large. But, sometimes EM-based $\hat{\alpha}$ and $\hat{\sigma}$ outperform the ML counterparts. This occurs because, {{STABLE}} gives the approximated ML estimation not the exact ML estimation and also we are using {\bf{lowess}} package. 
\item RMSE of all estimators decreases by increasing the sample size $n$. 
\end{enumerate} 
Along with the simulations, we investigate the execution time for different sample sizes and scale parameter levels. The results are given in Table \ref{tab1}. For this purpose, all runs are performed on a machine with 3.5 GHz Core(TM) i7-2700K Intel(R) processor and 8 GB of RAM. 
\begin{table}
\caption{\small{Average of execution time for different sample sizes and scale parameter levels. Times are in seconds when we run the program written in {R} for implementing the EM algorithm. Note that values outside (inside) of parentheses are obtained for $\alpha=0.5$ ($\alpha=1.5$).}} 
\label{tab1} 
\centering 
\small\addtolength{\tabcolsep}{1pt} 
\begin{tabular}{cccc} 
\hline
&\multicolumn{3}{c}{$\sigma$}\\ 
\cline{2-4} 
\multicolumn{1}{c}{n} & 
\multicolumn{1}{c}{0.5} & 
\multicolumn{1}{c}{1} & 
\multicolumn{1}{c}{2}\\ 
\hline 
500&30(51)&26(55)&22(53)\\ 
1000&70(118)&55(110)&43(112)\\ 
5000&334(602)&283(611)&223(591)\\ 
\hline 
\end{tabular} 
\end{table}
The performance analysis of the proposed EM algorithm for modelling a SSMM has been reported by \cite{teimouri2018algorithm}.
\subsection{Parameter estimation for skewed stable distribution}
For skewed stable distribution, we give a new representation by the following.
\begin{proposition} Suppose $Y\sim S_{0}(\alpha,\beta,\sigma,\mu)$, $P \sim S_{1}\bigl(\alpha/2,1,\bigl(\cos(\pi \alpha/4)\bigr)^{2/\alpha},0\bigr)$, and $V\sim S_{1}(\alpha,1,1,0)$. We have
\begin{eqnarray} \label{skew}
Y\mathop=\limits^d\eta\sqrt{2P}N+ \theta V+\mu- \lambda
\end{eqnarray}
where $\eta=\sigma\left(1-|\beta|\right)^{\frac{1}{\alpha}}$, $\theta=\sigma\mathrm{sgn}(\beta)|\beta|^{\frac{1}{\alpha}}$, $\lambda=\sigma \beta \tan \bigl(\pi \alpha/2 \bigr)$, and $N\sim{\cal{N}}(0,1)$. All random variables $N$, $P$, and $V$ are mutually independent.
\end{proposition}

\begin{figure}[h!]
\resizebox{\textwidth}{!}
{\begin{tabular}{ccc}
\includegraphics[width=40mm,height=40mm]{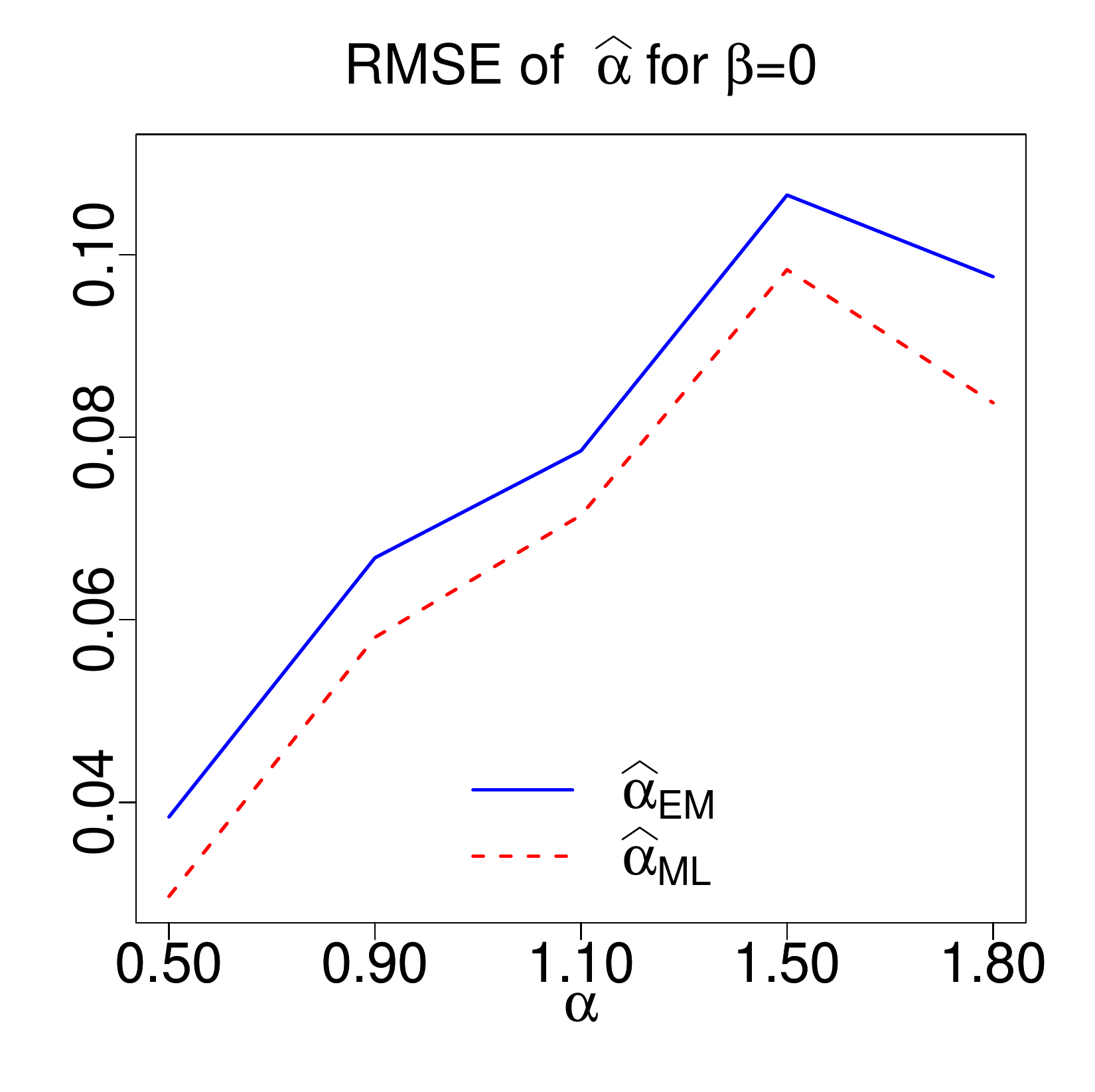}&
\includegraphics[width=40mm,height=40mm]{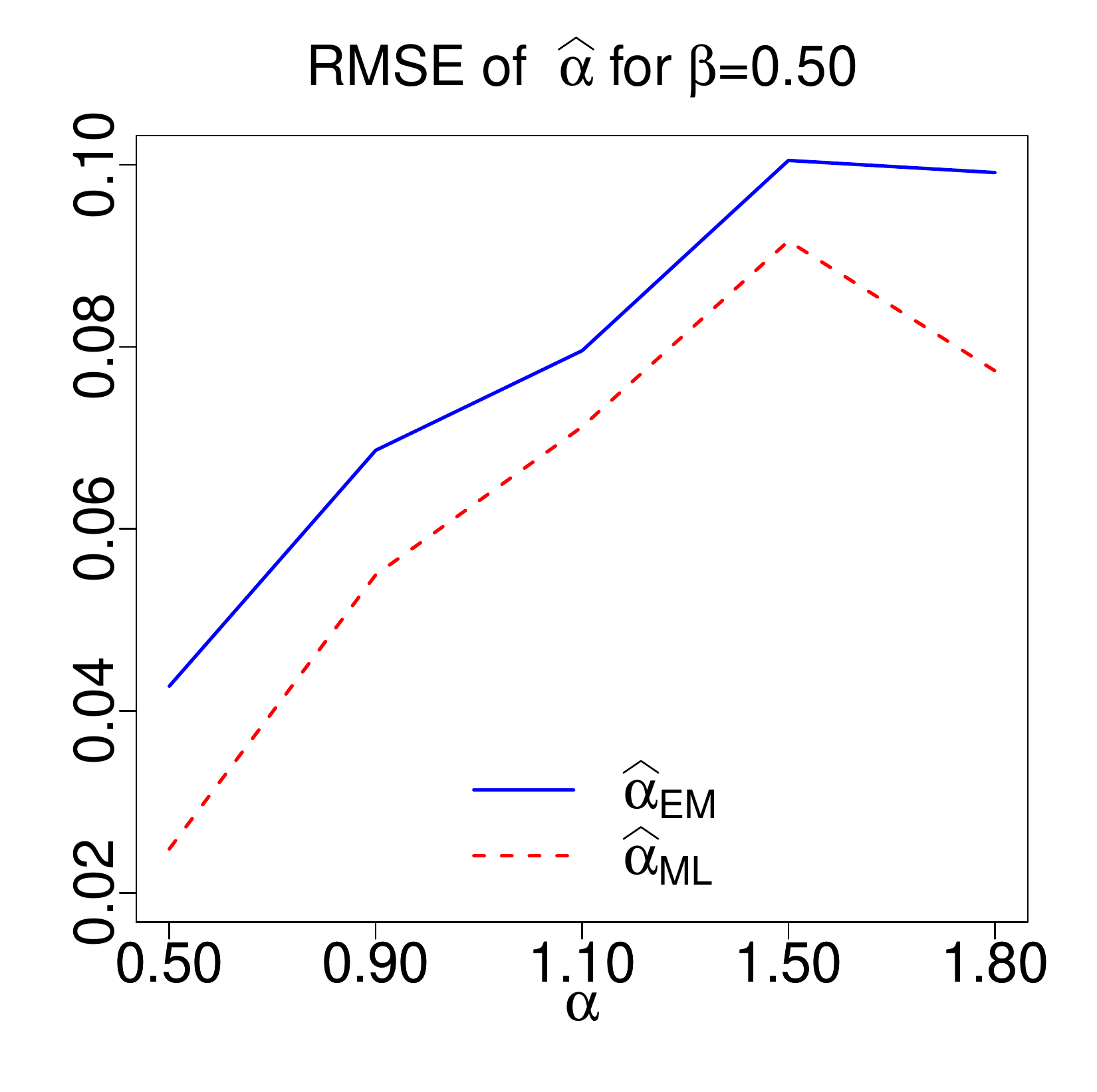}&
\includegraphics[width=40mm,height=40mm]{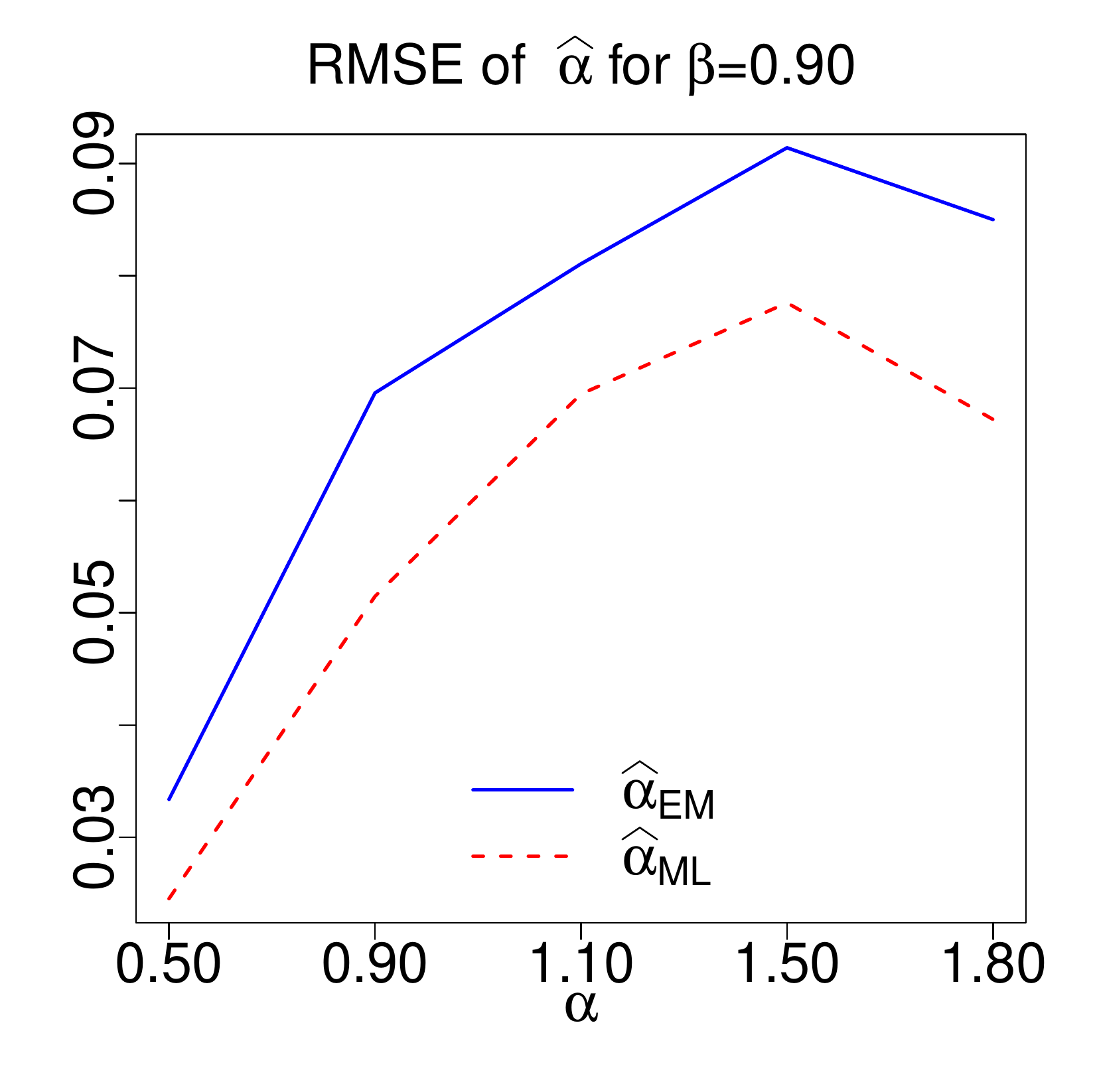}\\
\includegraphics[width=40mm,height=40mm]{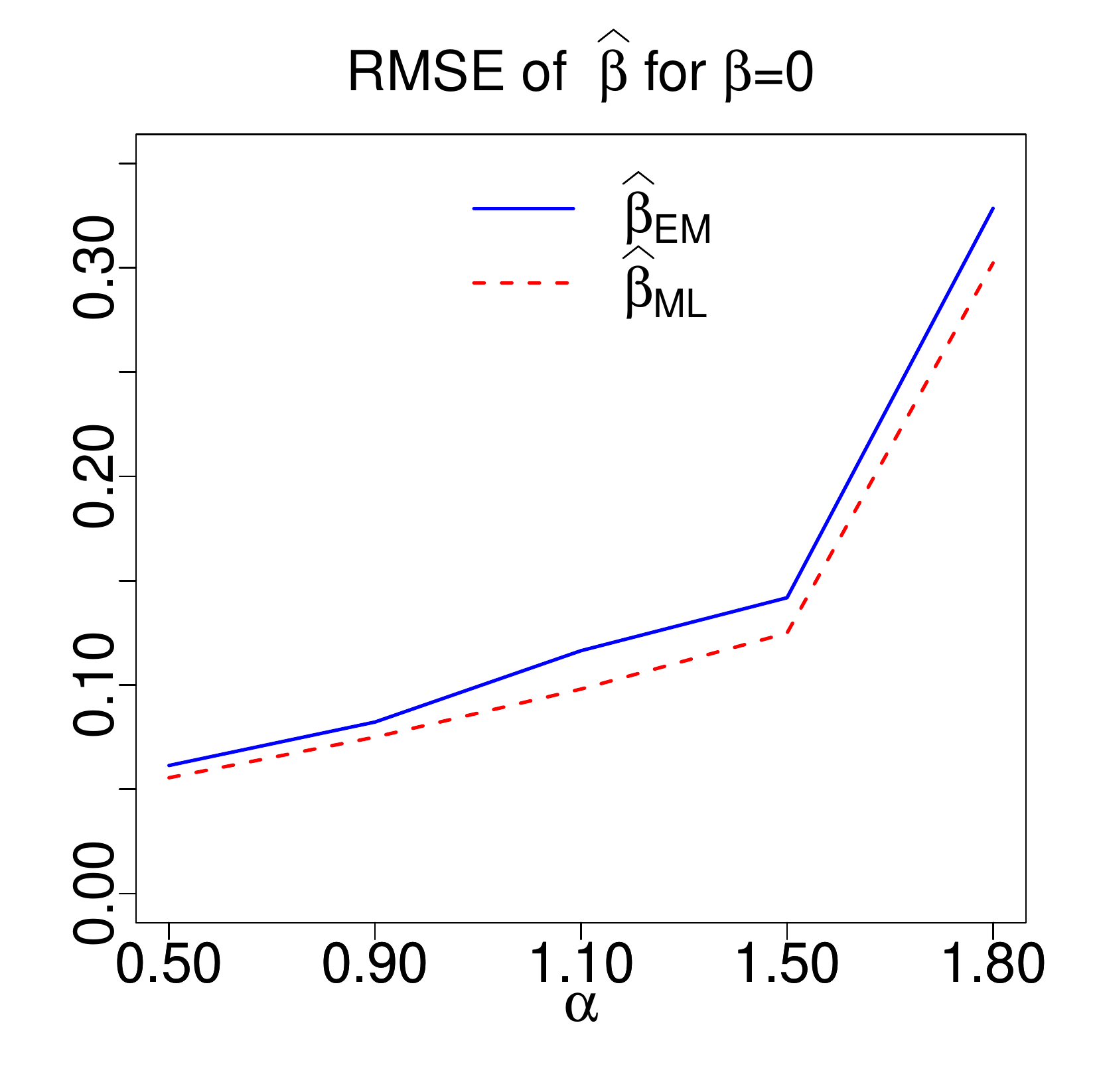}&
\includegraphics[width=40mm,height=40mm]{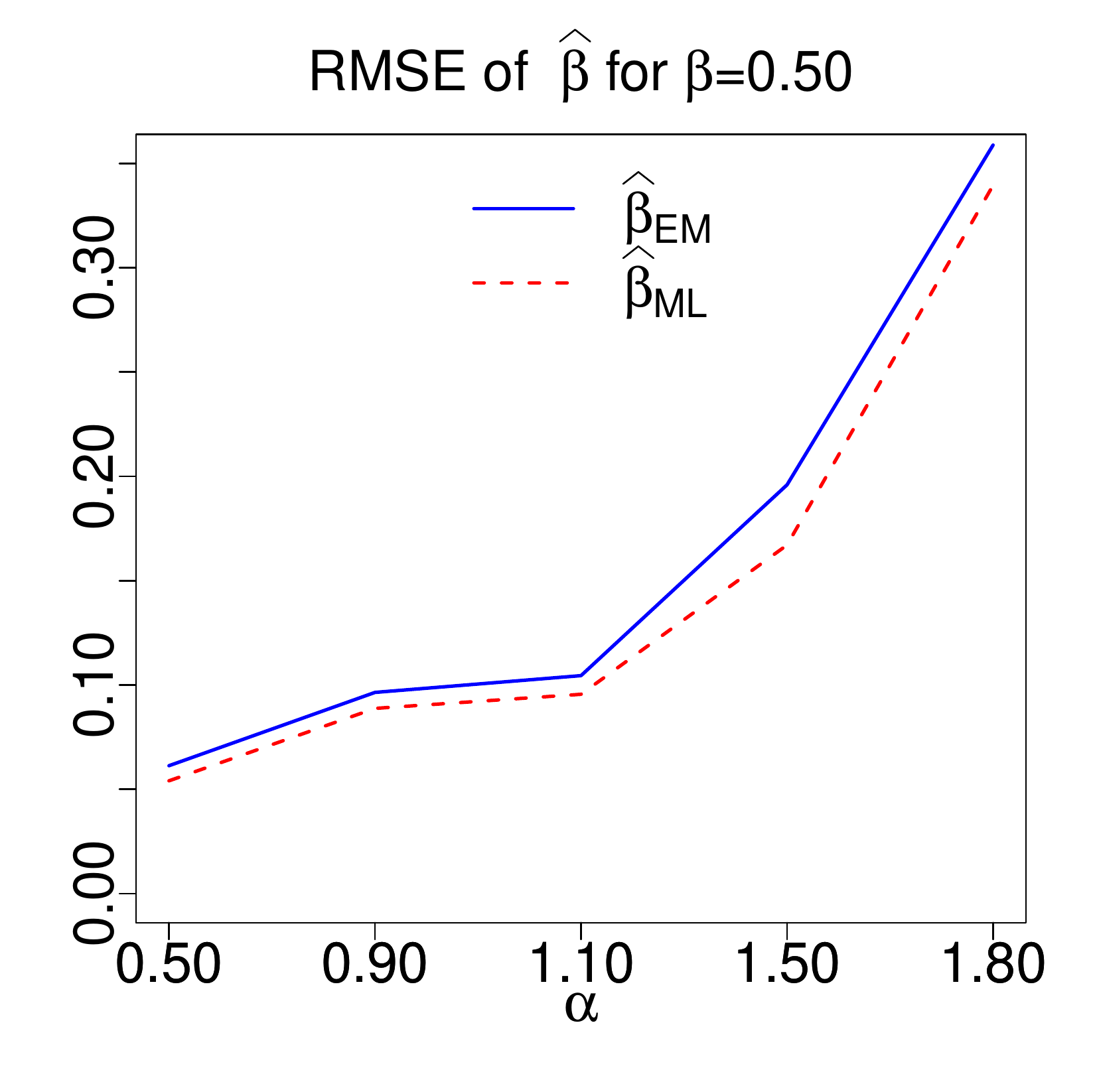}&
\includegraphics[width=40mm,height=40mm]{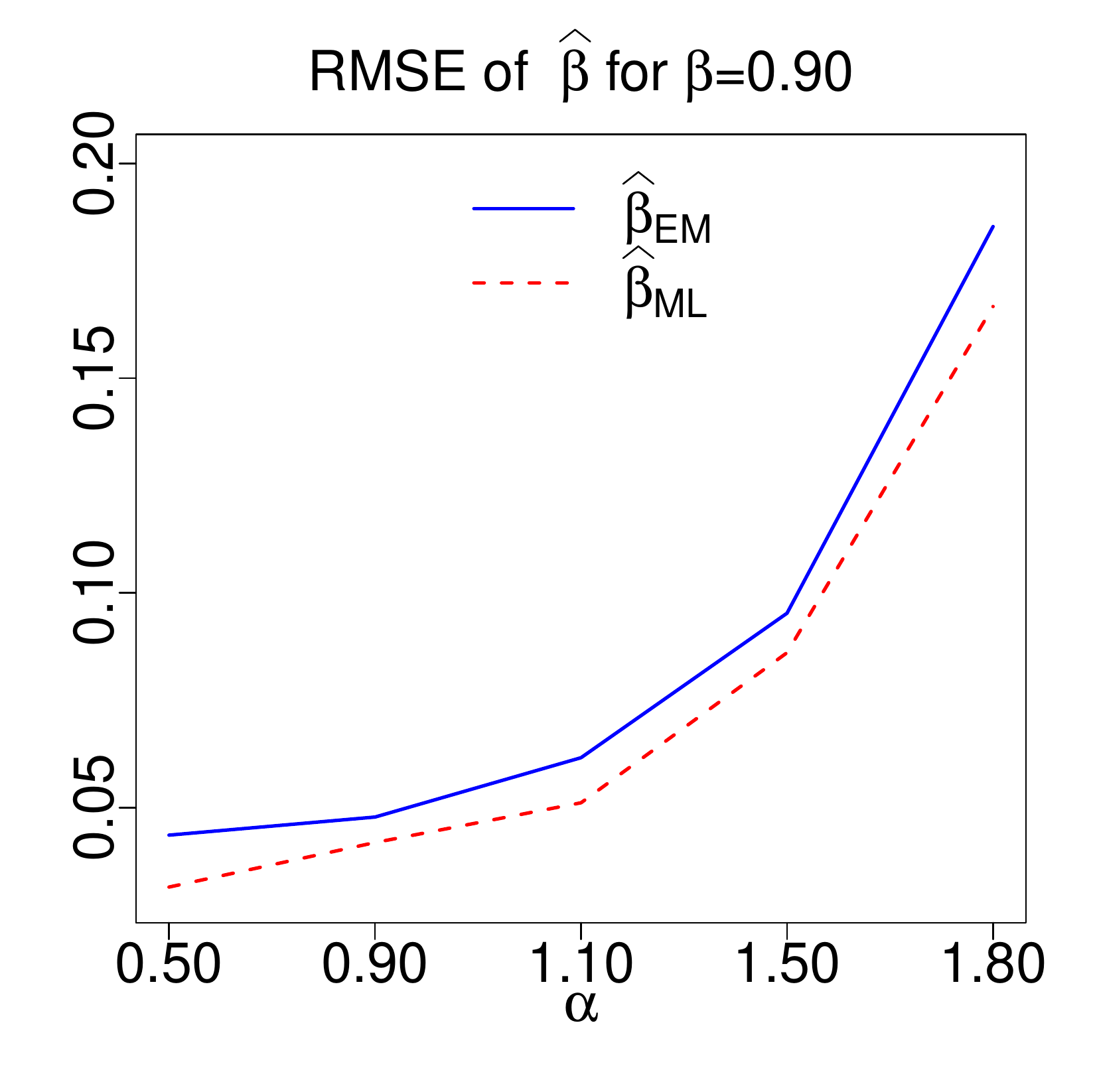}\\
\includegraphics[width=40mm,height=40mm]{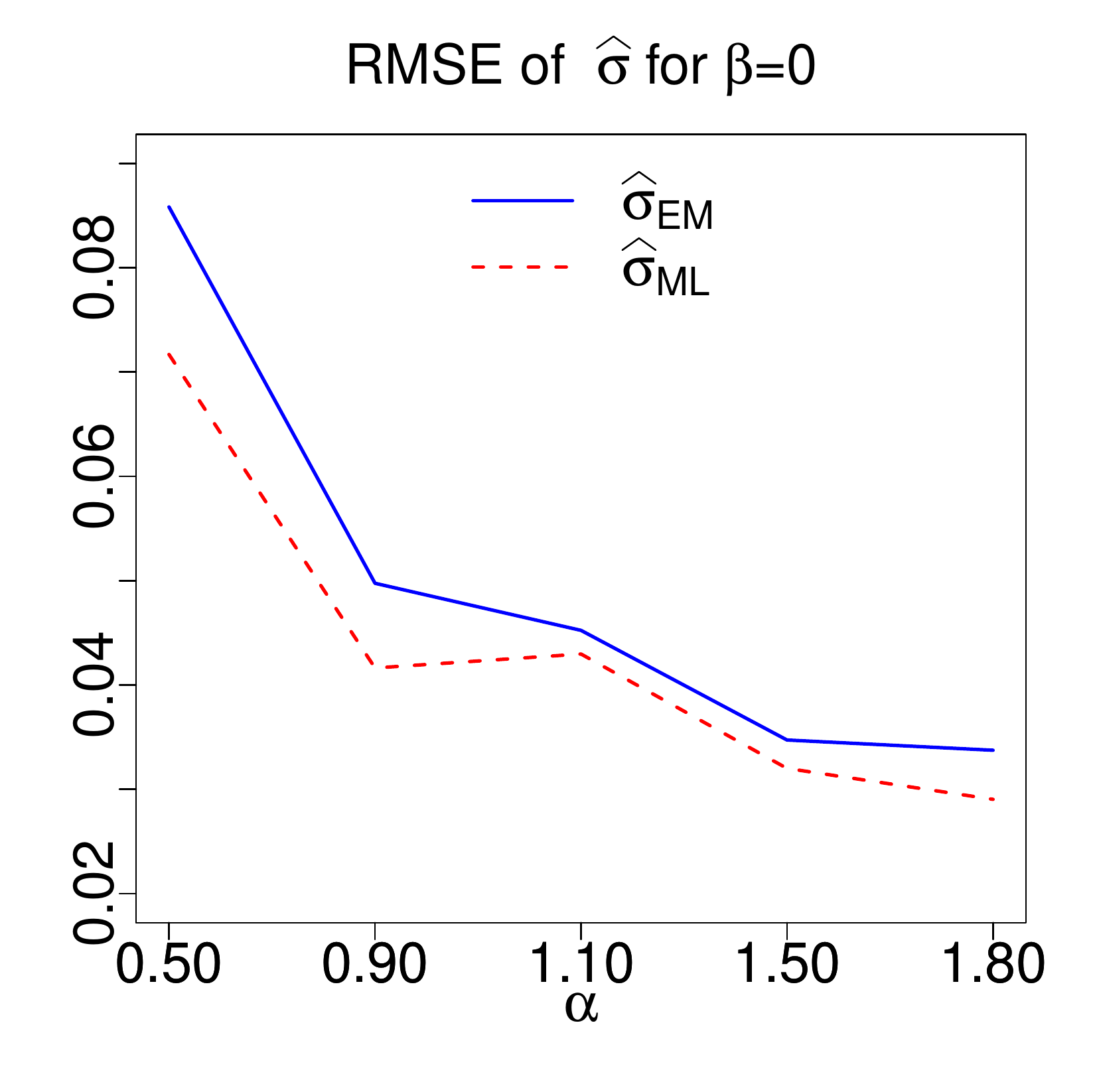}&
\includegraphics[width=40mm,height=40mm]{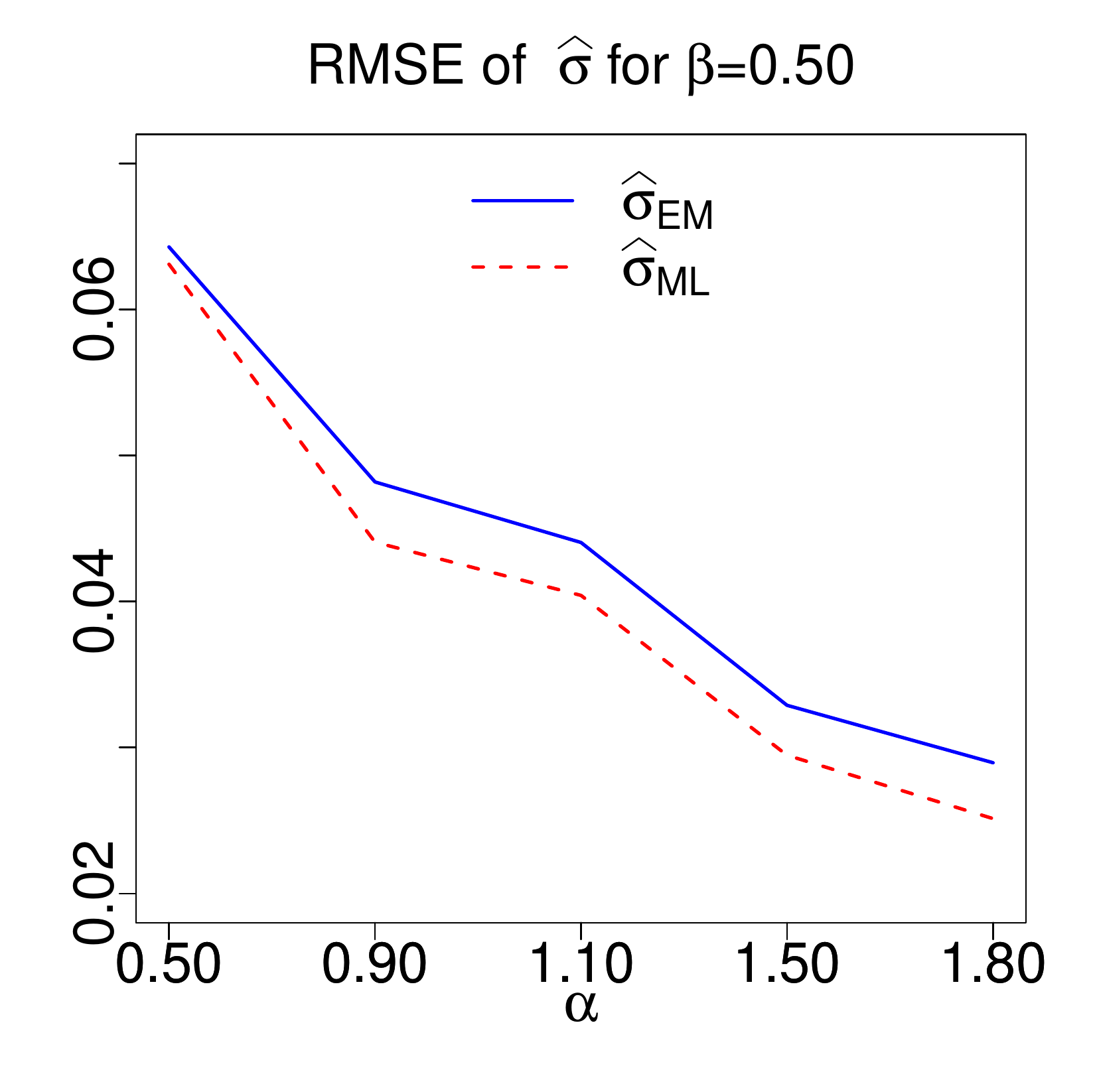}&
\includegraphics[width=40mm,height=40mm]{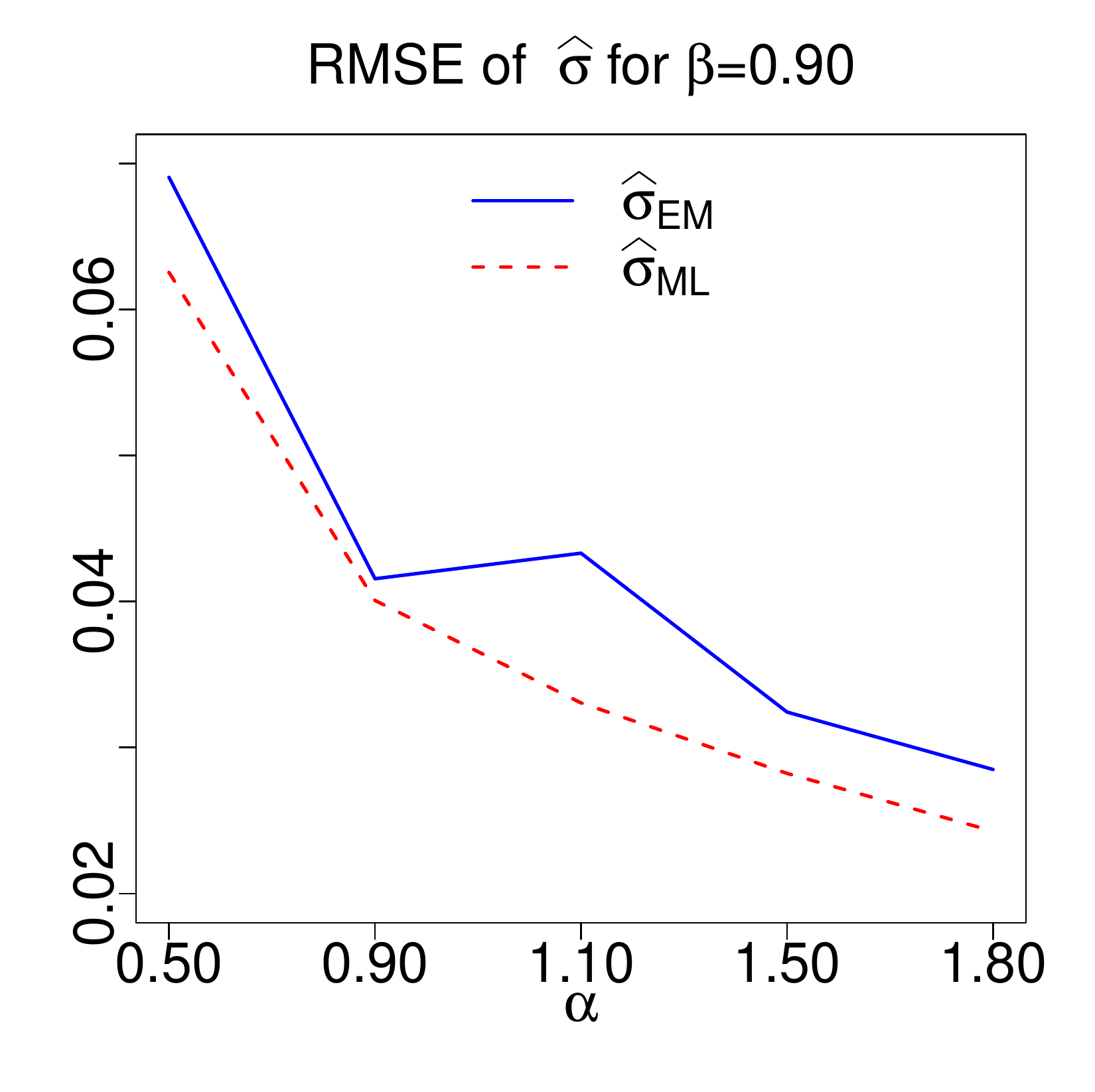}\\
\includegraphics[width=40mm,height=40mm]{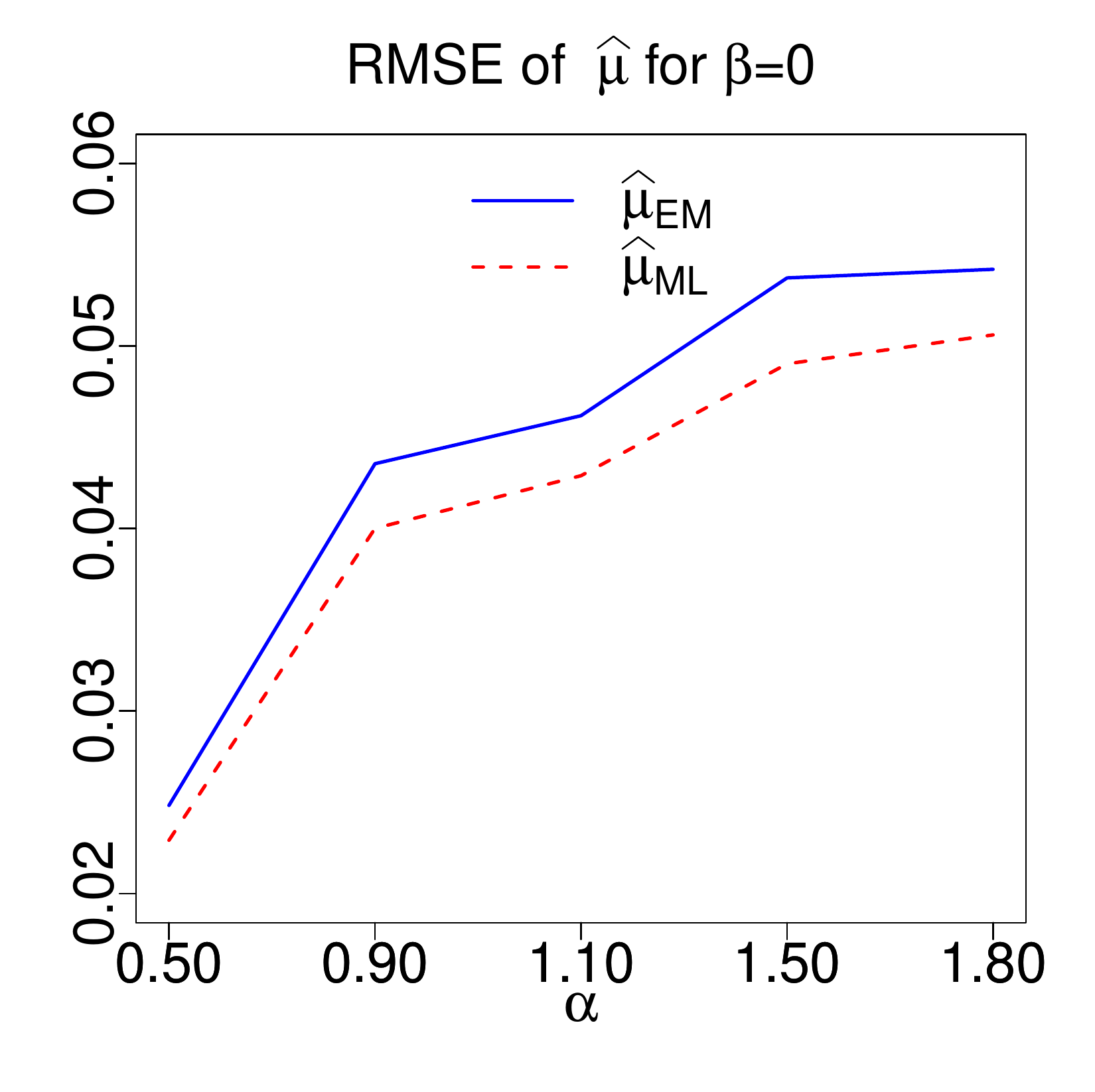}&
\includegraphics[width=40mm,height=40mm]{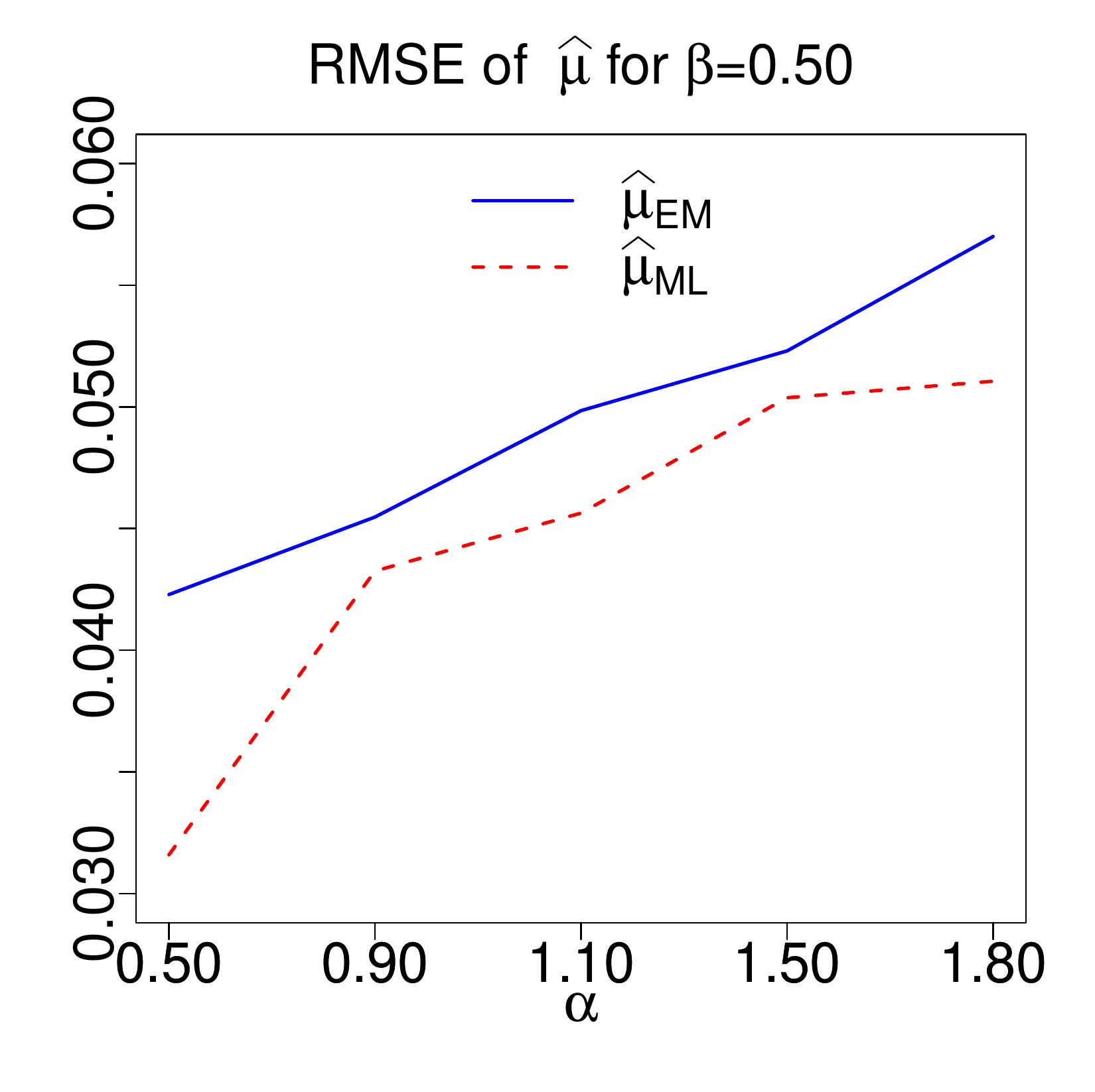}&
\includegraphics[width=40mm,height=40mm]{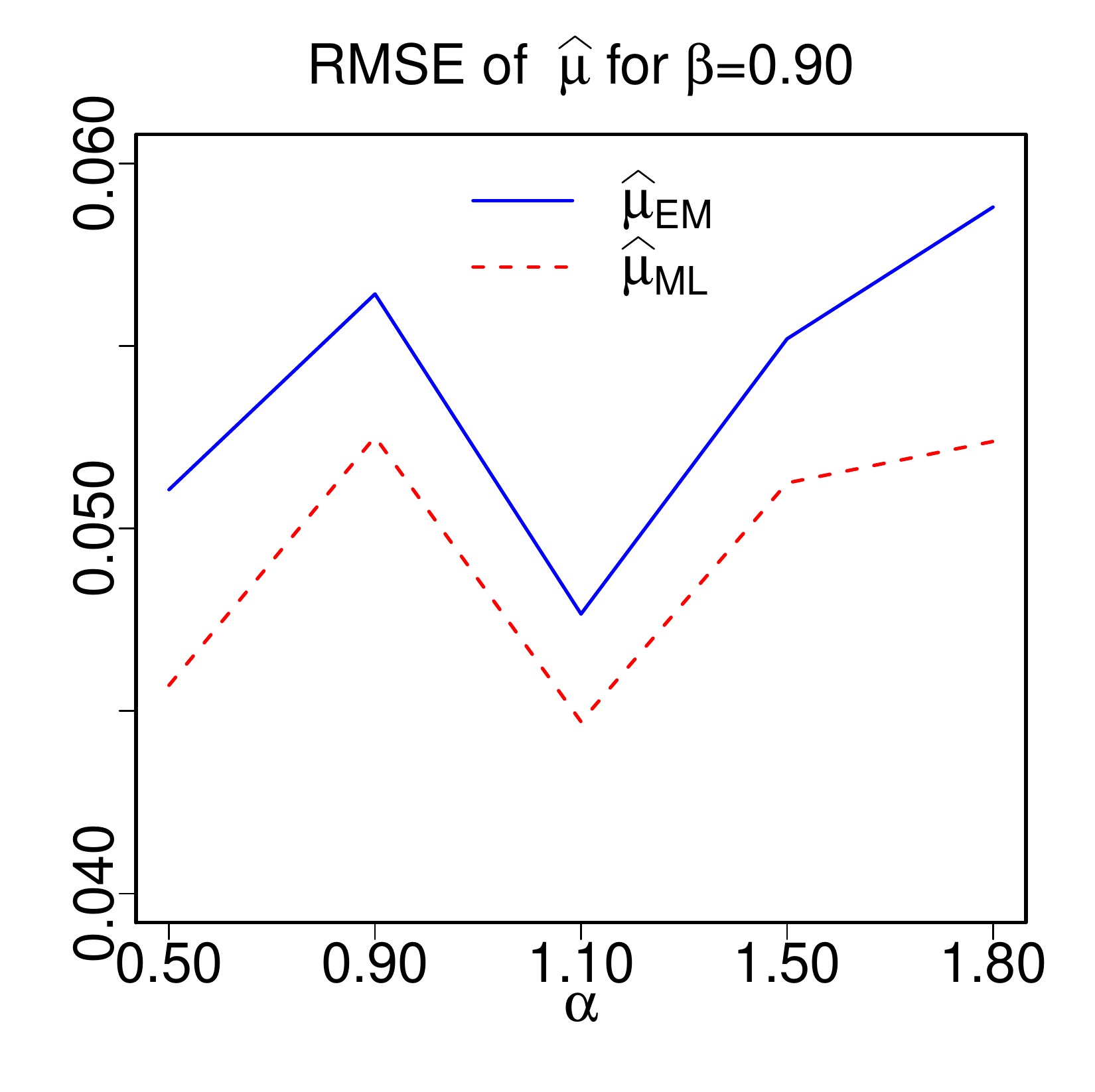}\\
\end{tabular}}
\caption{The RMSE of estimators obtained through the EM and ML approaches when $\sigma=0.5$ and $\mu=0$. In each sub-figure, the subscripts ML and EM indicate that the estimators $\hat{\alpha}$, $\hat{\beta}$, $\hat{\sigma}$, and $\widehat{\mu}$ are obtained using the EM algorithm (blue solid line) or the ML approach (red dashed line). The sub-figures in the first, second, and the third columns correspond to $\beta=0$, $\beta=0.50$, and $\beta=0.90$, respectively.}
\label{skewstable1}
\end{figure}

\begin{figure}[h!]
\resizebox{\textwidth}{!}
{\begin{tabular}{ccc}
\includegraphics[width=40mm,height=40mm]{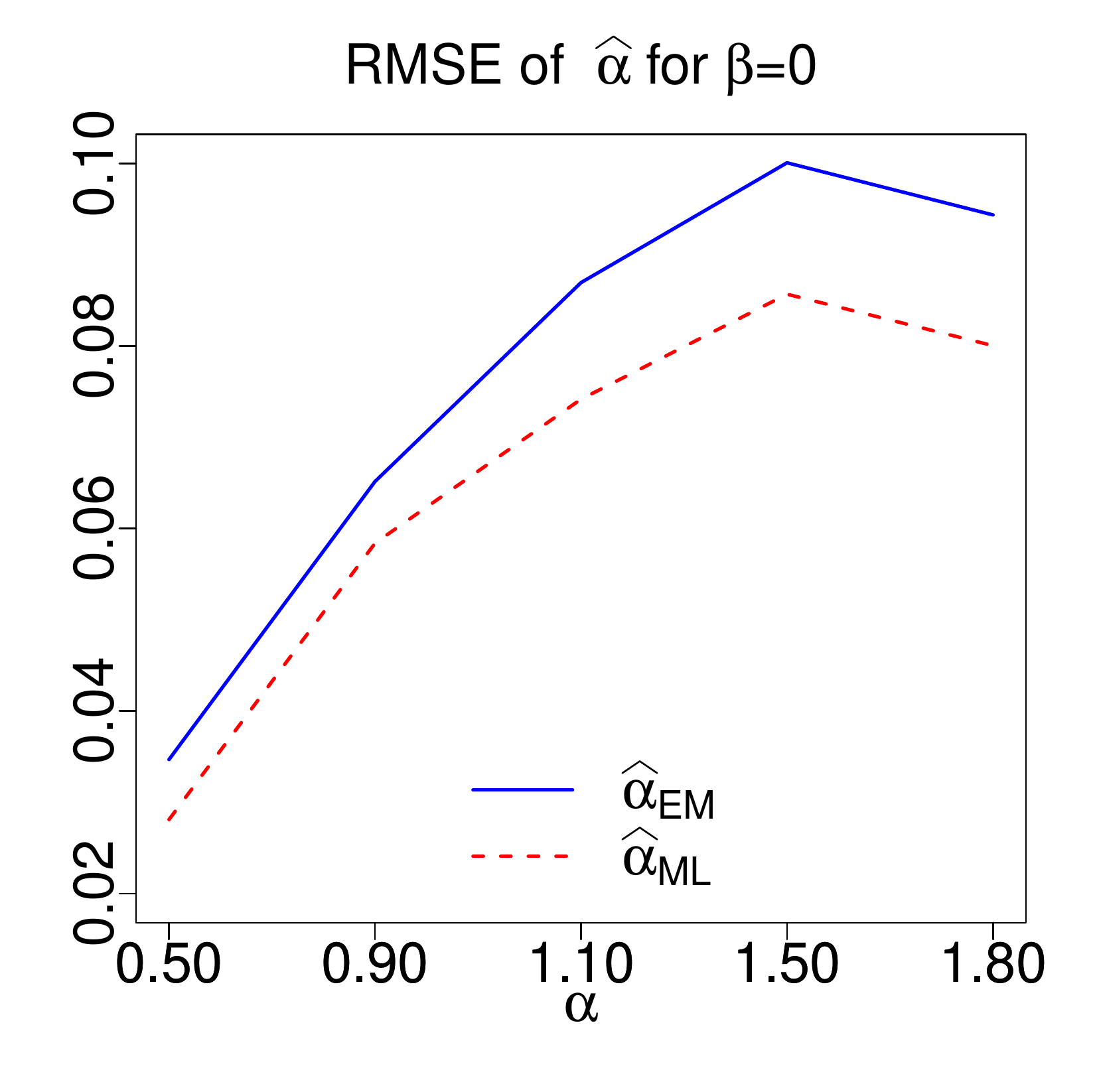}&
\includegraphics[width=40mm,height=40mm]{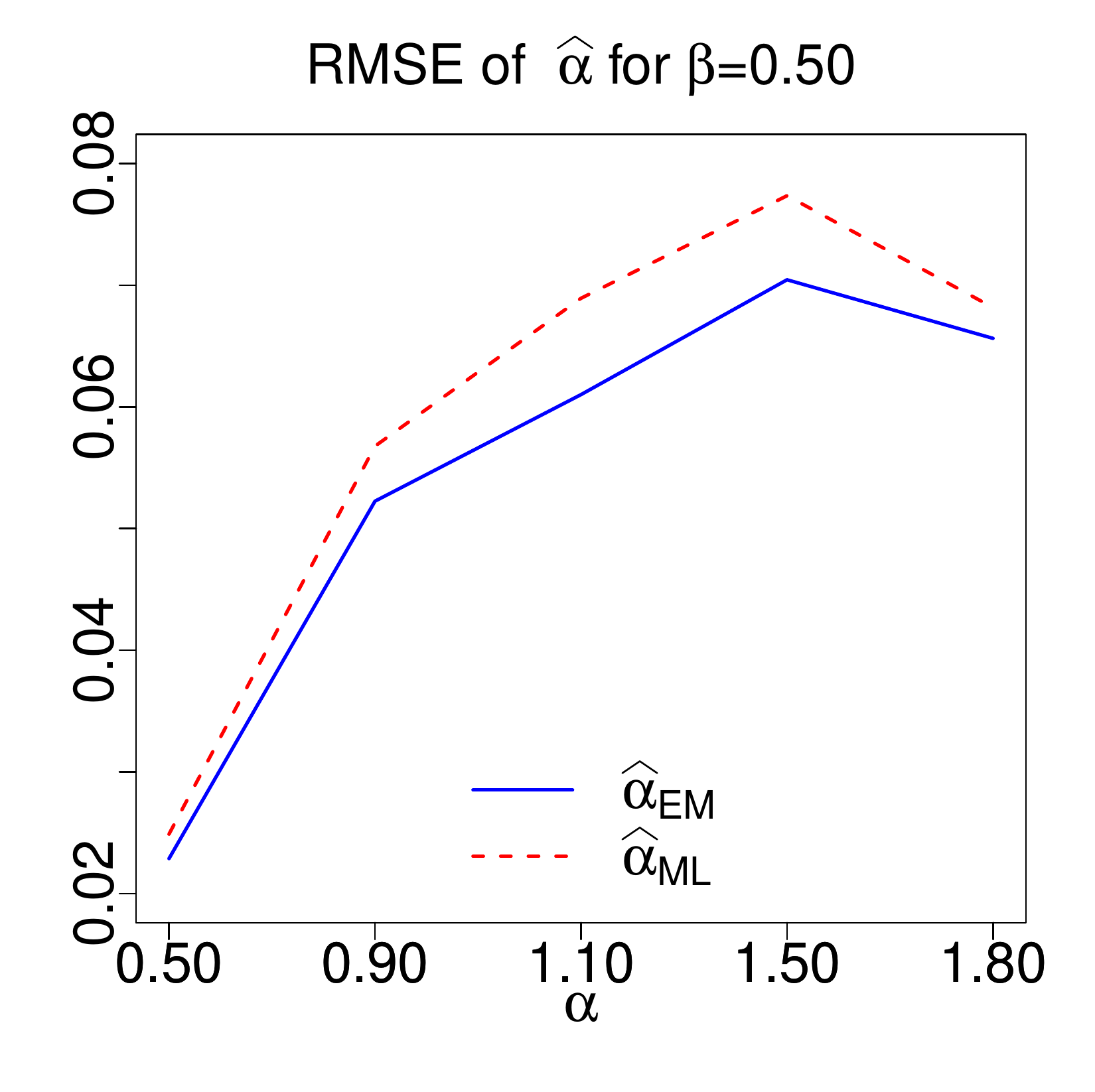}&
\includegraphics[width=40mm,height=40mm]{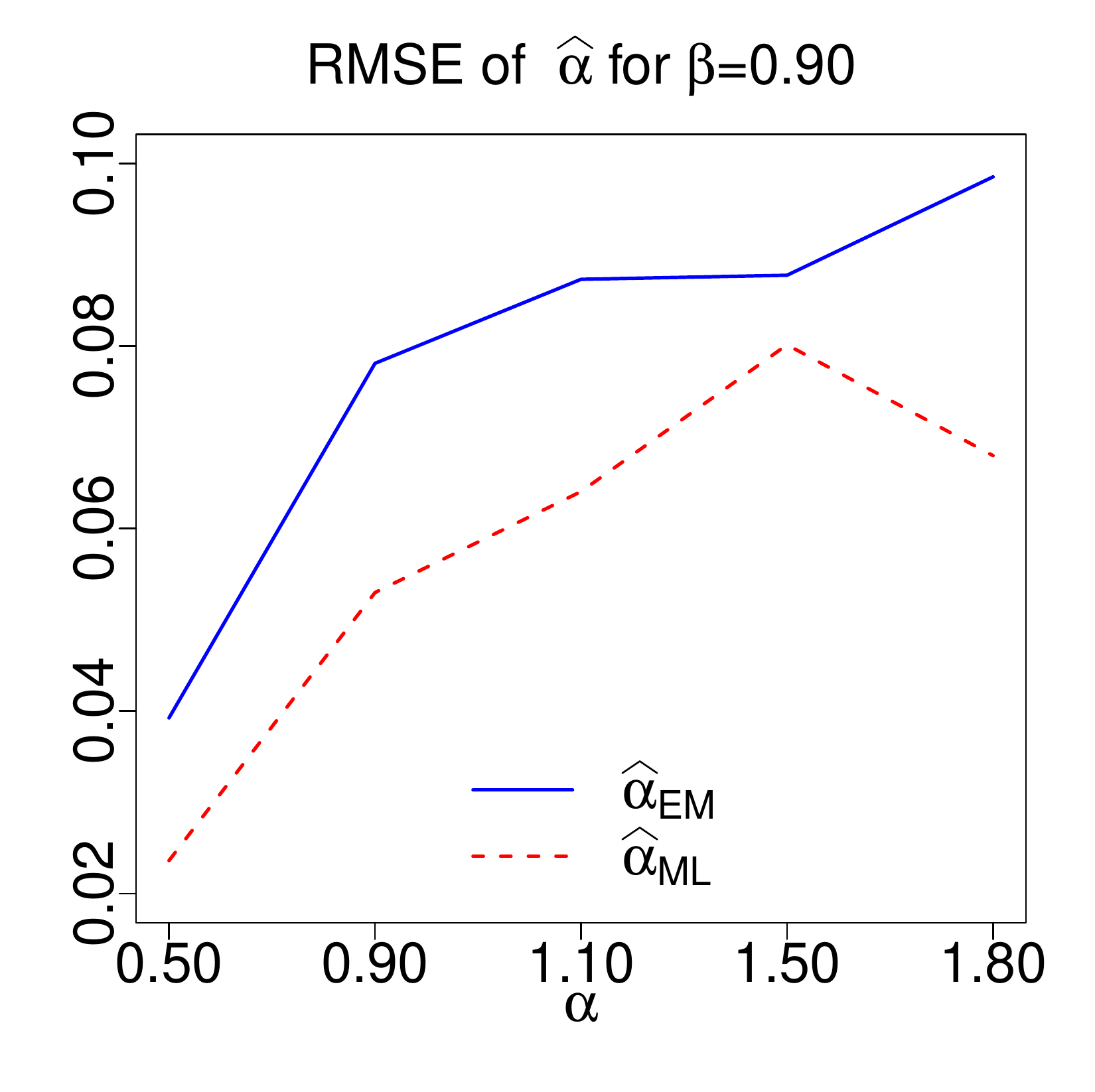}\\
\includegraphics[width=40mm,height=40mm]{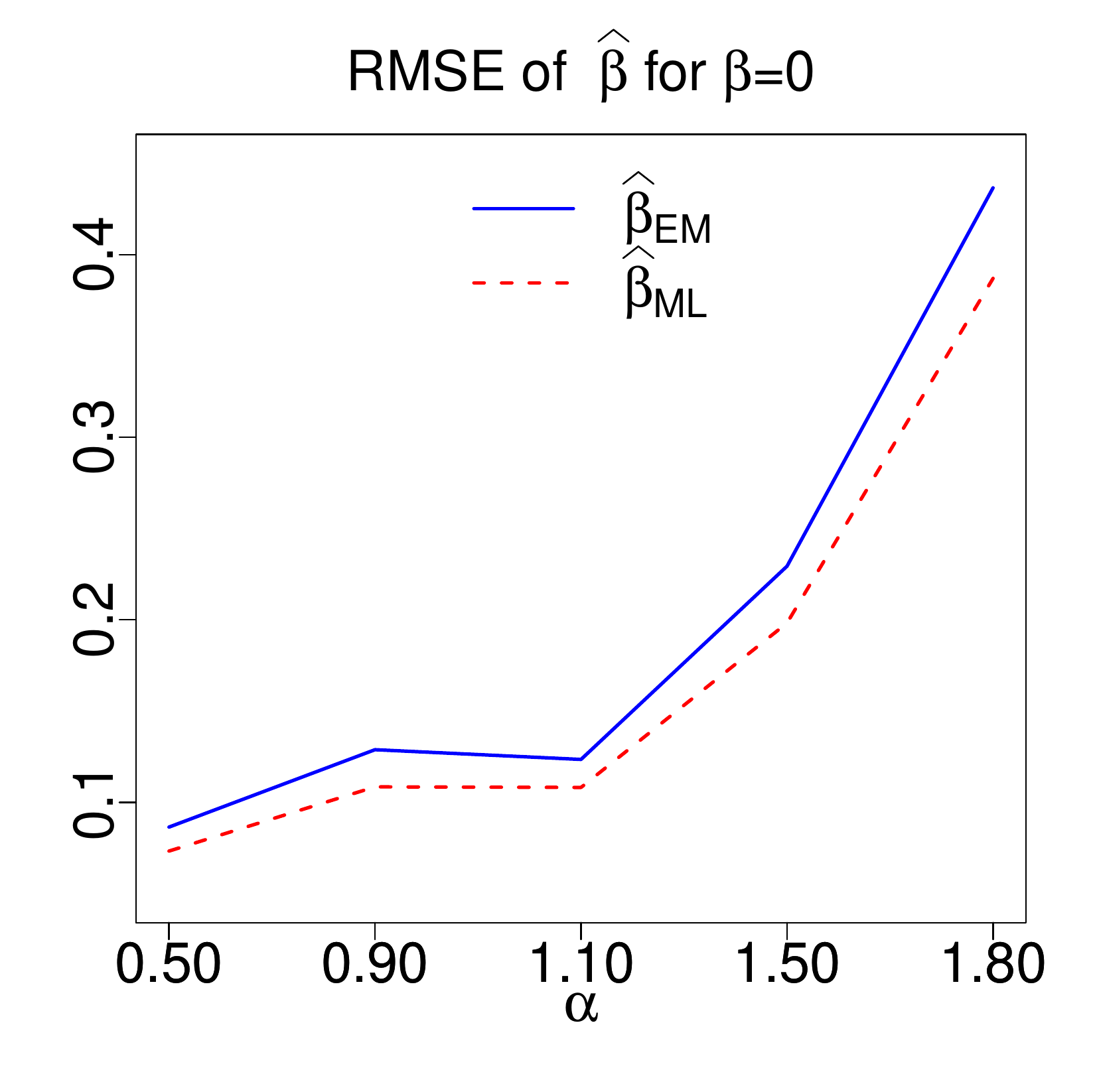}&
\includegraphics[width=40mm,height=40mm]{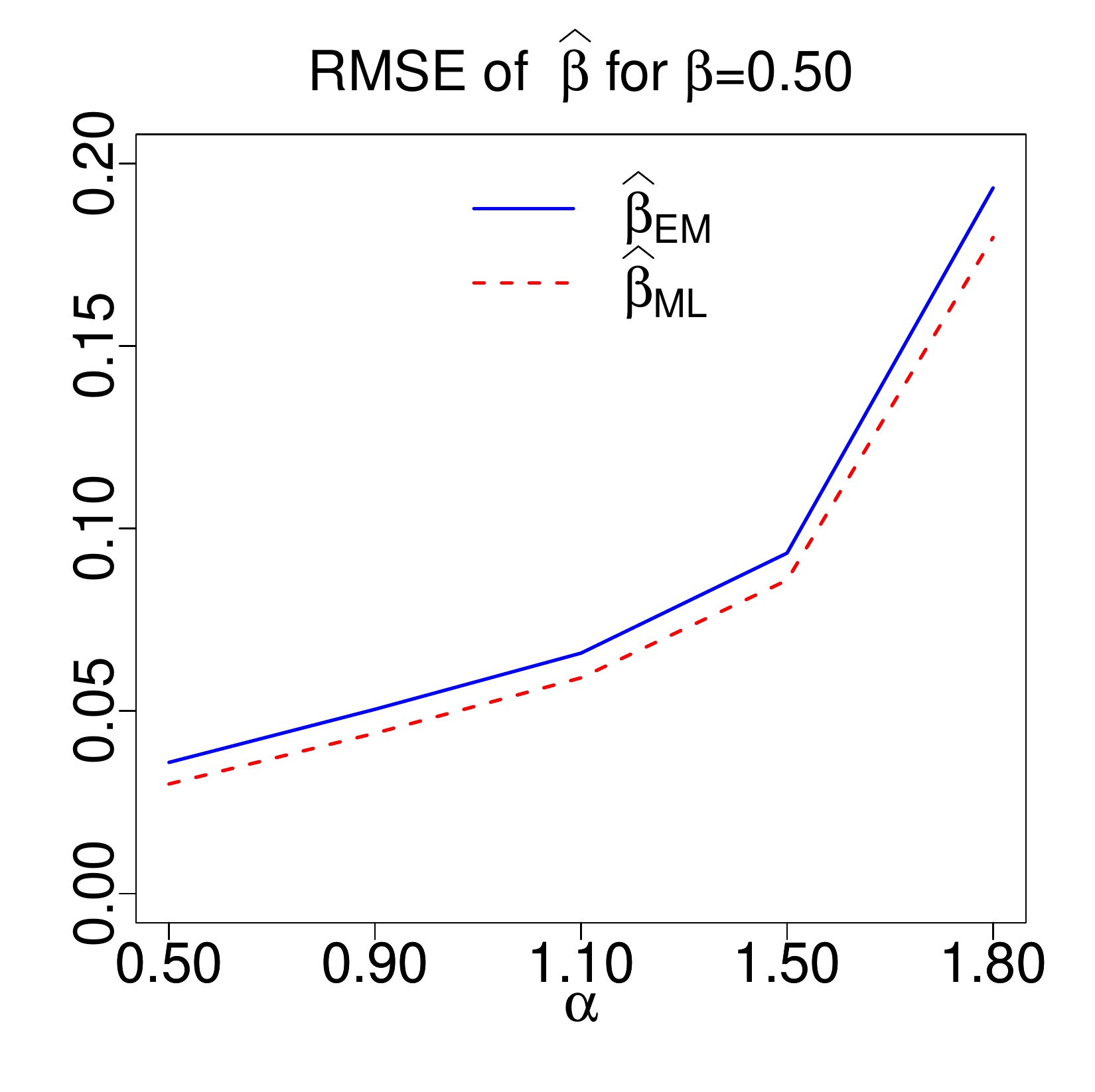}&
\includegraphics[width=40mm,height=40mm]{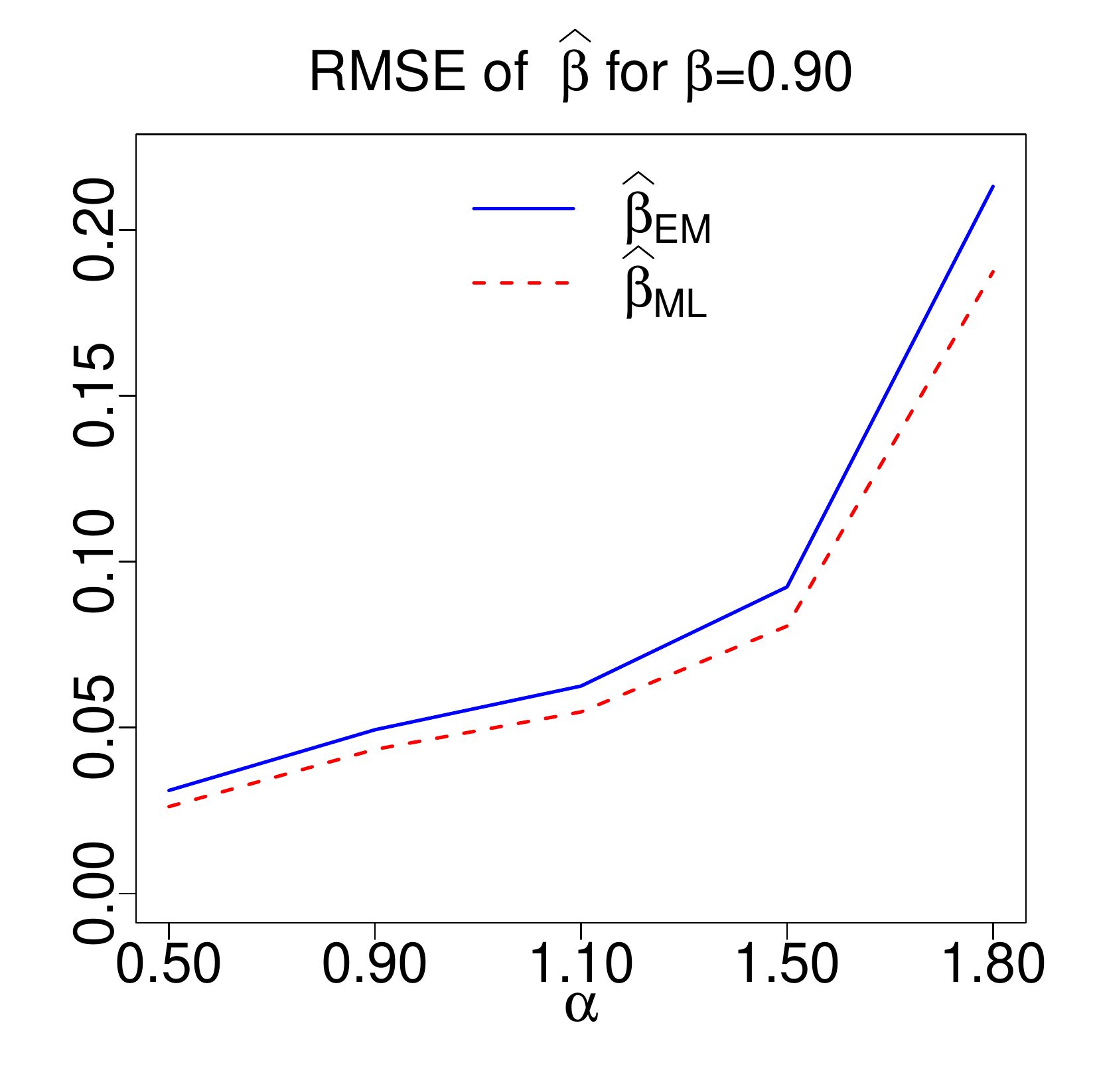}\\
\includegraphics[width=40mm,height=40mm]{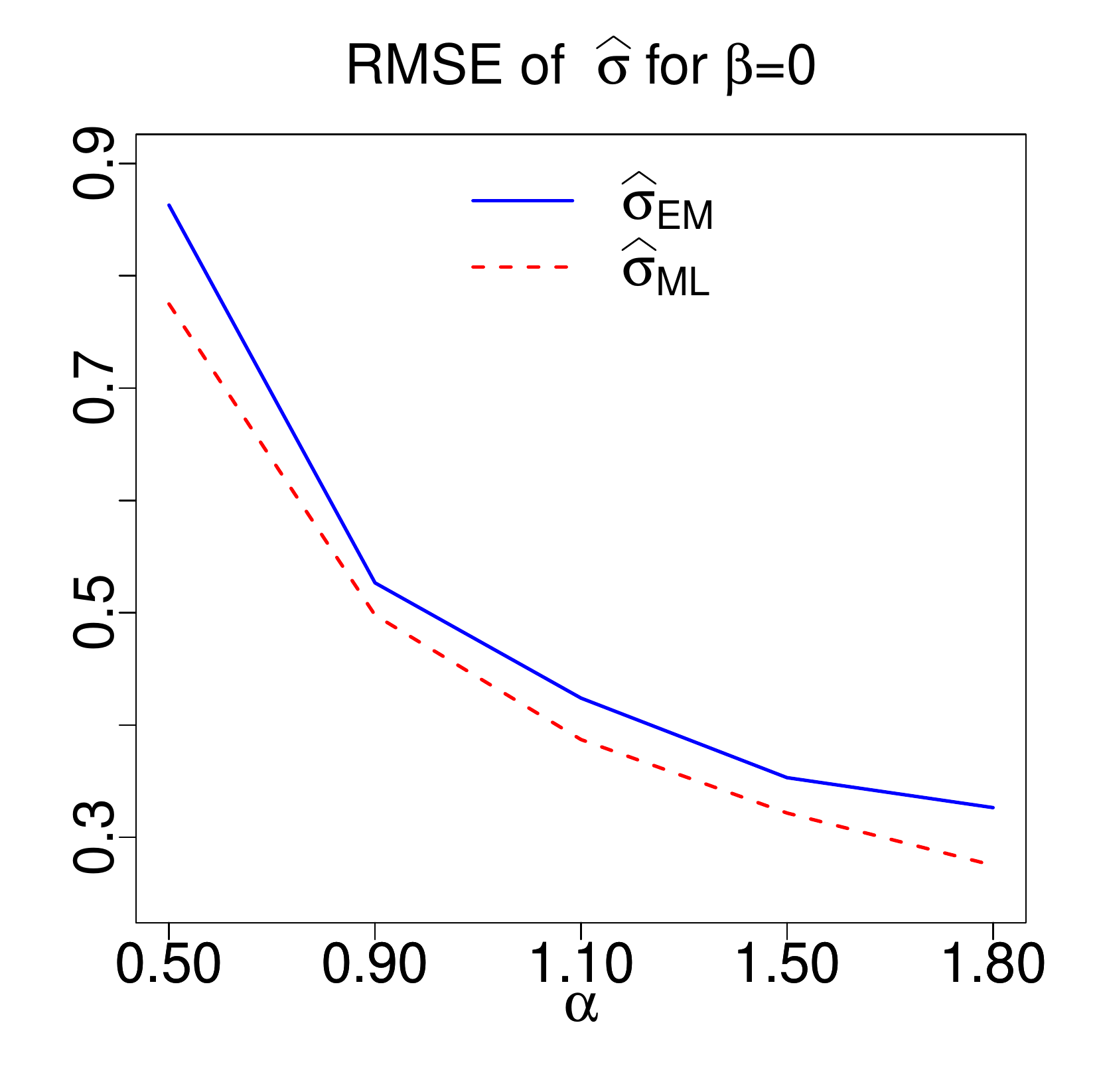}&
\includegraphics[width=40mm,height=40mm]{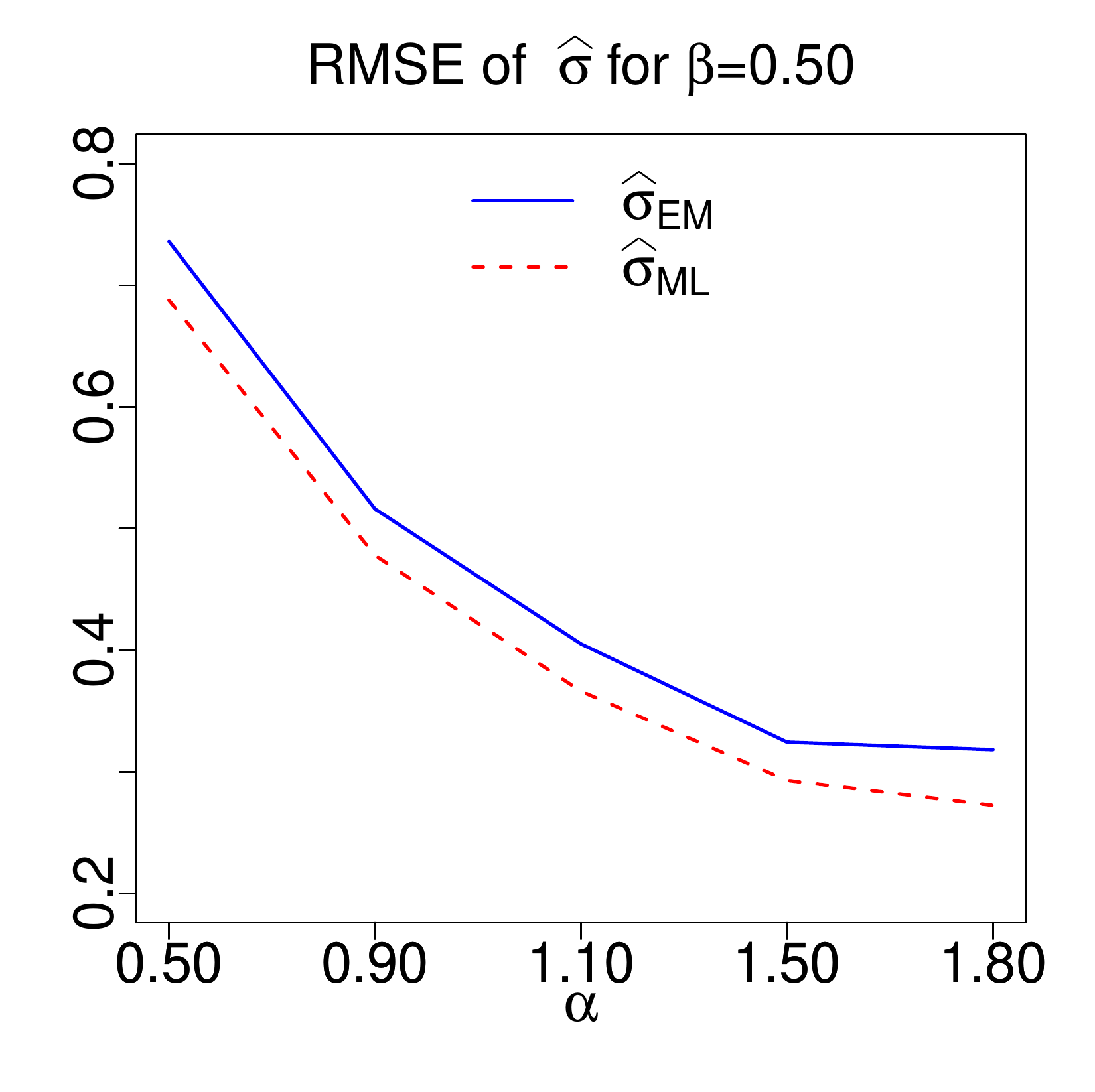}&
\includegraphics[width=40mm,height=40mm]{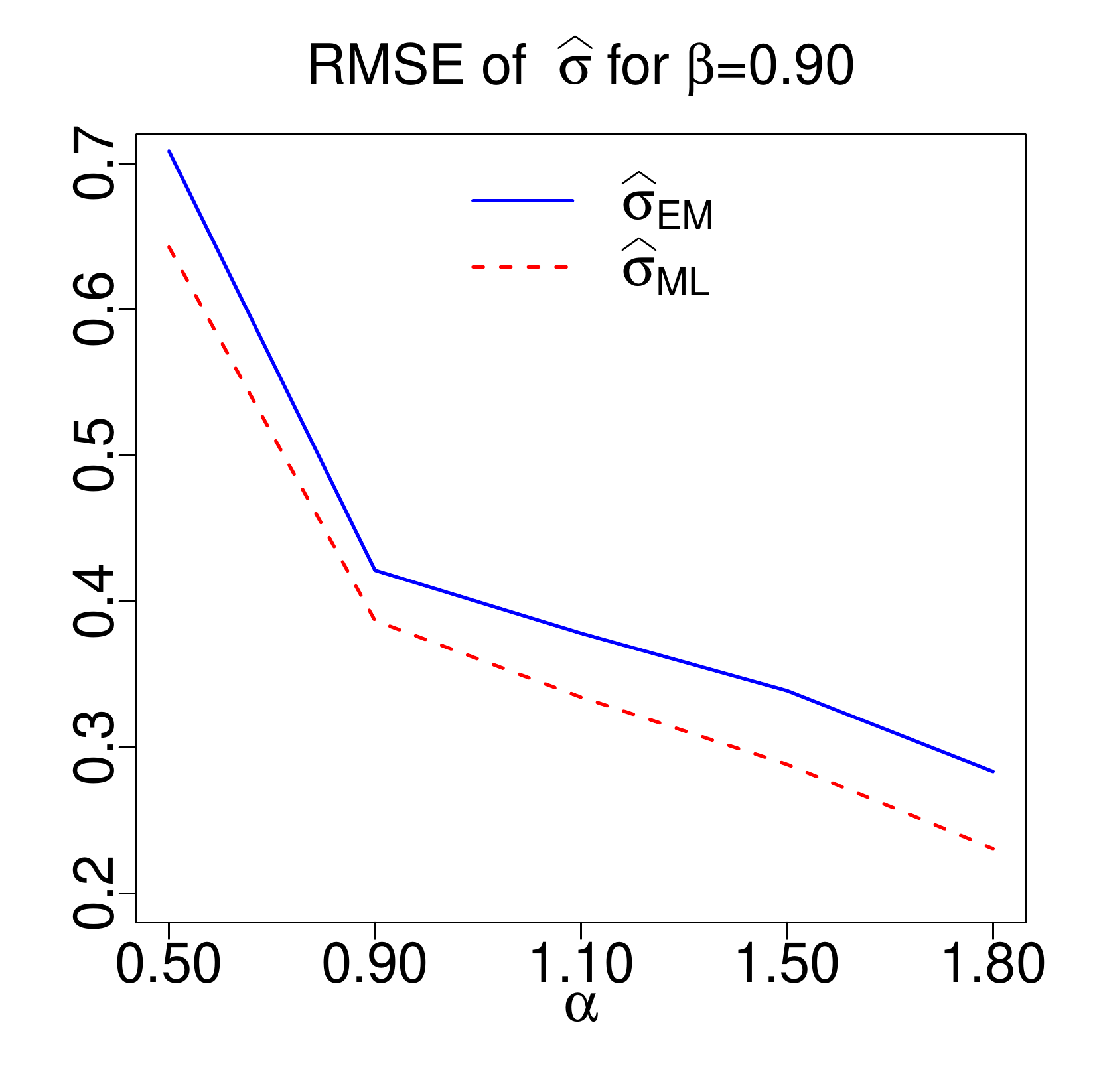}\\
\includegraphics[width=40mm,height=40mm]{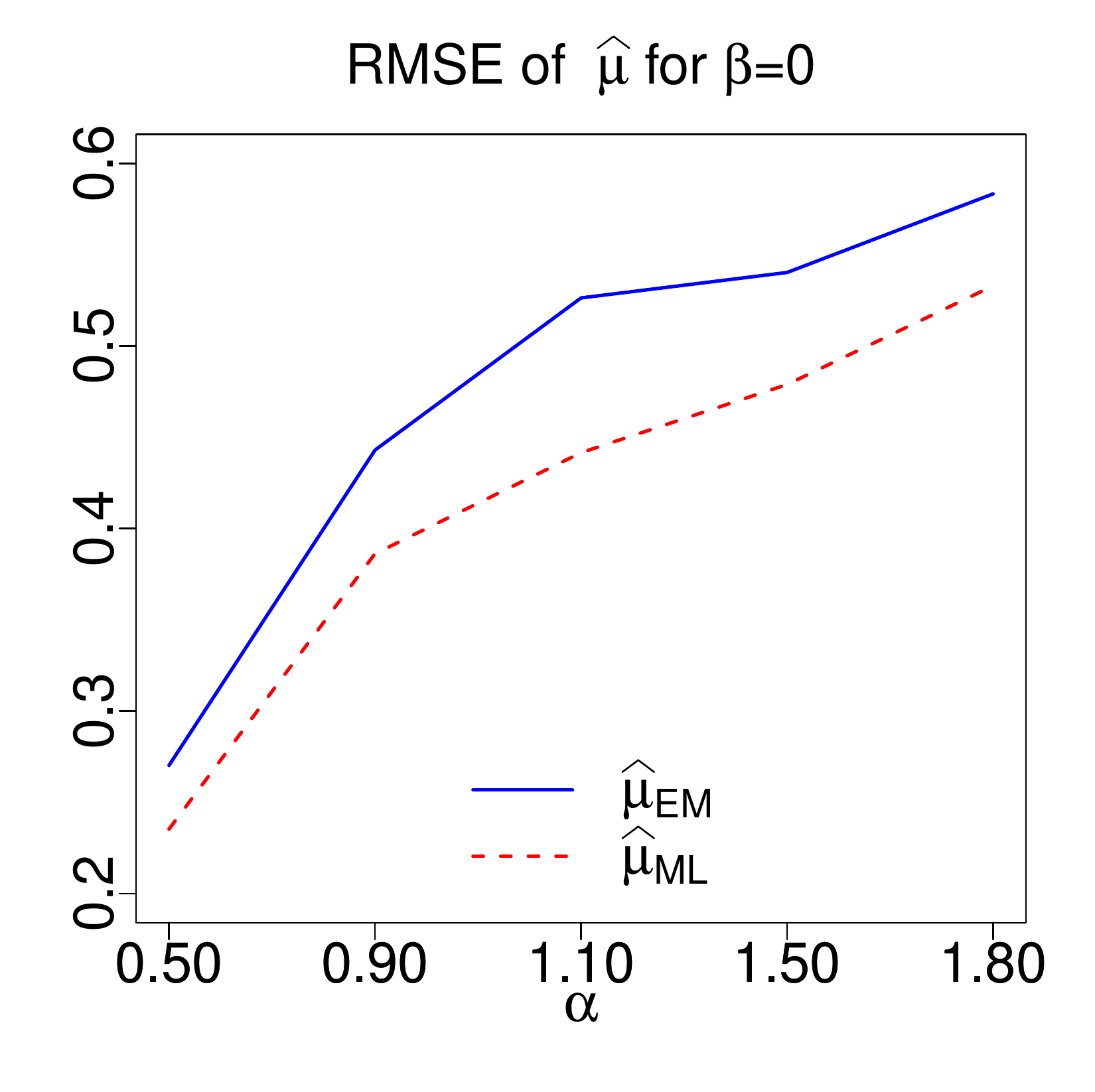}&
\includegraphics[width=40mm,height=40mm]{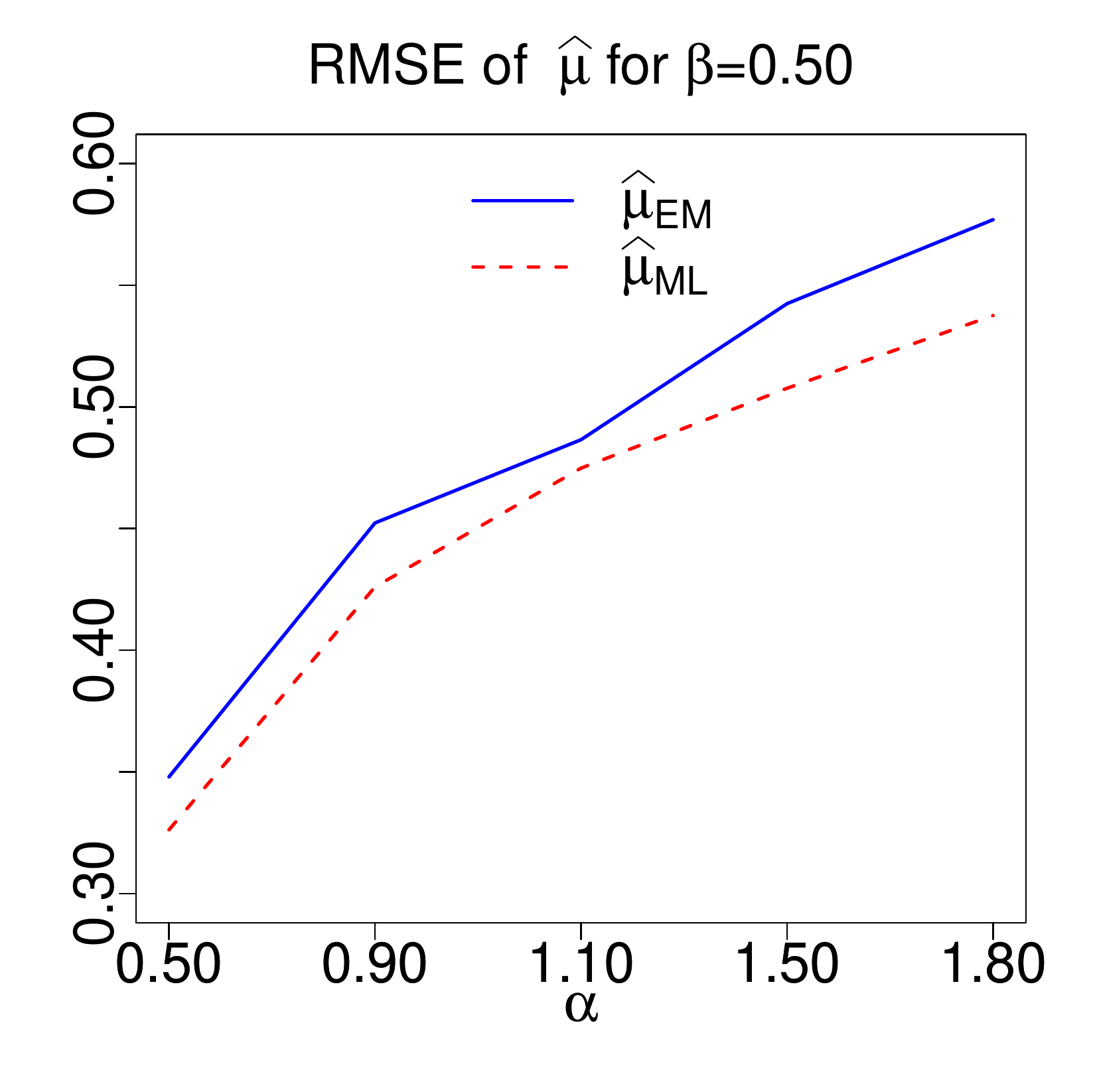}&
\includegraphics[width=40mm,height=40mm]{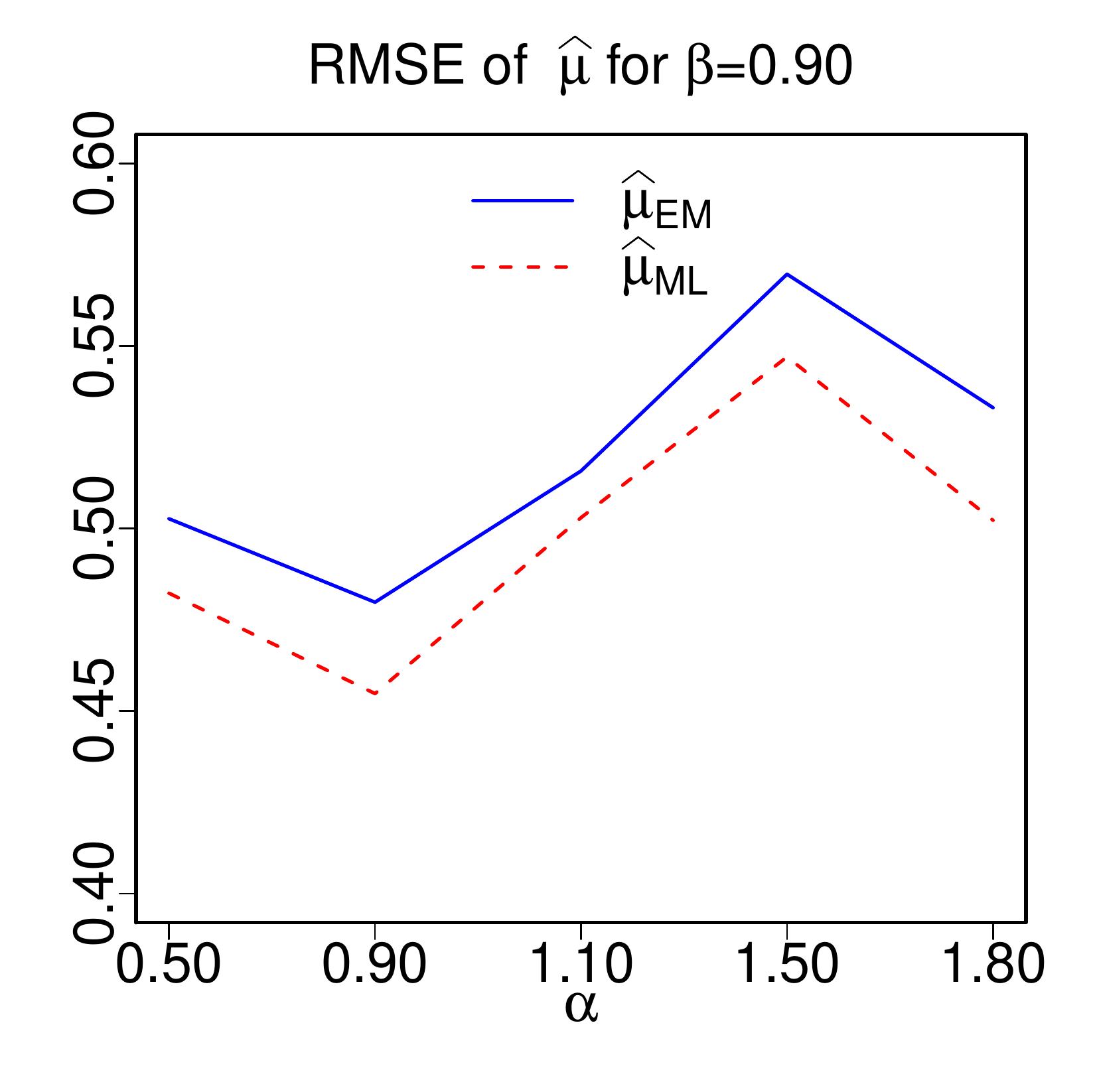}\\
\end{tabular}}
\caption{The RMSE of estimators obtained through the EM and ML approaches when $\sigma=5$ and $\mu=0$. In each sub-figure, the subscripts ML and EM indicate that the estimators $\hat{\alpha}$, $\hat{\beta}$, $\hat{\sigma}$, and $\widehat{\mu}$ are obtained using the EM algorithm (blue solid line) or the ML approach (red dashed line). The sub-figures in the first, second, and the third columns correspond to $\beta=0$, $\beta=0.50$, and $\beta=0.90$, respectively.}
\label{skewstable2}
\end{figure}

Using above representation, the methodology used for estimating the parameters of a symmetric stable distribution can be developed for estimating the parameters of a skewed stable distribution. To save the space, we refuse to describe the proposed EM algorithm. Performance of the proposed EM algorithm is proven via simulations. Some part of the simulations are shown in Figures \ref{skewstable1}-\ref{skewstable2}. These figures demonstrate that the proposed EM algorithm for skewed stable distribution works as well as the ML approach but just with a slight difference. During simulations, we found that the proposed EM algorithm generally outperforms the CF and SQ approaches and so removed by competitions.
\subsection{Parameter estimation for multivariate elliptically contoured stable distribution}
Estimating the parameters of a multivariate elliptically contoured stable distribution is computationally expensive. The parameters of this distribution, as described in Introduction, are estimated using the projection method. While the dispersion matrix estimated using projection method is not always positive definite \cite{nolan2013multivariate}, the estimated dispersion matrix using the EM algorithm \cite{teimouri2018parameter} is always positive definite. To show the performance of the EM algorithm, here we confine ourselves to the bivariate case. For higher dimensions the results are the same. Firstly, a simulation study is performed to compare the performance of the CF, EM, ML, and SQ approaches. Then, we apply all four methods to the real data. For simulation study, a sample of size $n=500$ vectors are simulated from a bivariate elliptically contoured stable distribution with $\alpha=1.5, 1.7, 1.9$, $\boldsymbol{\mu}=(\mu_1,\mu_2)^T=(0,0)^T$, and dispersion matrix 
\begin{equation*}
\Sigma=
\begin{pmatrix}
{\sigma_{11}}&{\sigma_{12}}\\ 
{\sigma_{21}}&{\sigma_{22}} 
\end{pmatrix}= 10 ^{-6} \times
\begin{pmatrix} 
{5.9552}&{4.0783}\\
{4.0783}&{13.9861}
\end{pmatrix}.
\end{equation*} 
Above setting of the dispersion matrix was used by \cite{rachev2003handbook}. We follow the CF, ML, and SQ approaches using the {STABLE} software. The boxplots based on 250 runs are depicted in Figure \ref{bivariate}. We note that the proposed EM algorithm is robust with respect to initial values. As it is seen, the proposed EM algorithm works as well as or better than the CF and SQ approaches in terms of bias (with respect to median) and length of the box.
For analyzing real data, we apply the CF, EM, ML, and SQ methods to S\&P500 and IPC indices. These indices are between 9 years of daily returns of 22 major worldwide market which, consists of 2535 observations from 1/4/2000 to 9/22/2009. This data set is available at the website of Yahoo finance. The empirical distribution of both S\&P500 and IPC indices show a symmetric pattern around the origin and corresponding scatterplot is roughly elliptical contoured. Furthermore, estimated tail indices after fitting stable distribution to the S\&P500 and IPC indices through the ML approach are 1.9003 and 1.9143, respectively, which are assumed to be equal. Thus, the assumption that $\boldsymbol{Z}=(S\&P500, IPC)^{T}$ follows an elliptically contoured stable distribution is acceptable, see \cite{nolan2013multivariate}. The results of estimating parameters through the CF, EM, ML, and SQ methods are shown in Table \ref{tab2}. The log-likelihood value indicates that the EM algorithm gives better performance than the CF and SQ approaches.
\begin{table}
\caption{\small{Estimation results for modelling $\boldsymbol{Z}=(S\&P500, IPC)^T$ by elliptically contoured stable distribution when the ML, CF, EM, and SQ approaches are applied to the data.}}
\label{tab2}
\centering
\small\addtolength{\tabcolsep}{.3pt}
\begin{tabular}{ccccc}
\hline &
\multicolumn{3}{c}{{{Parameters}}}\\
\cline{2-4}
\multicolumn{1}{c}{Method} &
\multicolumn{1}{c}{$\alpha$} &
\multicolumn{1}{c}{$\Sigma$} &
\multicolumn{1}{c}{$\boldsymbol{\mu}$}&
{{Log-likelihood}}\\
\hline
ML&
1.9073&$\begin{pmatrix}0.45396&0.41991\\0.41991&0.45626\end{pmatrix}$&$\begin{pmatrix}0.00528\\0.00697\end{pmatrix}$&-4967.706\\
\\ \vspace{.9pt}
CF&1.9530&$\begin{pmatrix}0.45668&0.41957\\0.41957&0.46008\end{pmatrix}$&$\begin{pmatrix}0.02505\\0.02437\end{pmatrix}$&-4978.067\\\\
EM&1.7116&$\begin{pmatrix}0.37830&0.34573\\0.34573&0.37589\end{pmatrix}$&$\begin{pmatrix}-0.02494\\-0.02215\end{pmatrix}$&-4975.581\\ \\
SQ&1.8179&$\begin{pmatrix}0.42751&0.35074\\0.35074&0.43147\end{pmatrix}$&$\begin{pmatrix}0.01559\\0.02313\end{pmatrix}$&-5176.572\\
\hline
\end{tabular}
\end{table}
Also, through a simulation study, we compare the execution times (in seconds) of the EM and ML methods. The results are given in Table \ref{tab3}.
\begin{table}[h!] 
\caption{\small{Average of the execution time in seconds for 100 runs when the EM and ML methods are applied to a $d$-dimensional elliptically contoured stable distribution. We run a program written in {R} on a 3.5 GHz Intel processor Core(TM) i7 using a 8 GB RAM. During simulation, $\Sigma$ is a positive definite matrix whose eigenvalues are randomly generated from a uniform distribution on (0, 2). The value inside of the parentheses corresponds to the ML method.}}
\label{tab3} 
\centering 
\small\addtolength{\tabcolsep}{1pt} 
\begin{tabular}{ccc} 
\hline
\multicolumn{1}{c}{\it{d}} & 
\multicolumn{1}{c}{$\alpha$=1.5} & 
\multicolumn{1}{c}{$\alpha$=1} \\ 
\hline 
5&181(1.2)&165(0.6)\\
10&202(3.3)&190(3.5)\\ 
20&220(7.6)&223(7.8)\\ 
50&245(42)&228(40)\\ 
100&416(193)&359(173)\\ 
\hline 
\end{tabular} 
\end{table} 
\subsection{Parameter estimation for multivariate strictly stable distribution}
\begin{figure}[h!]
\center
\includegraphics[width=65mm,height=75mm]{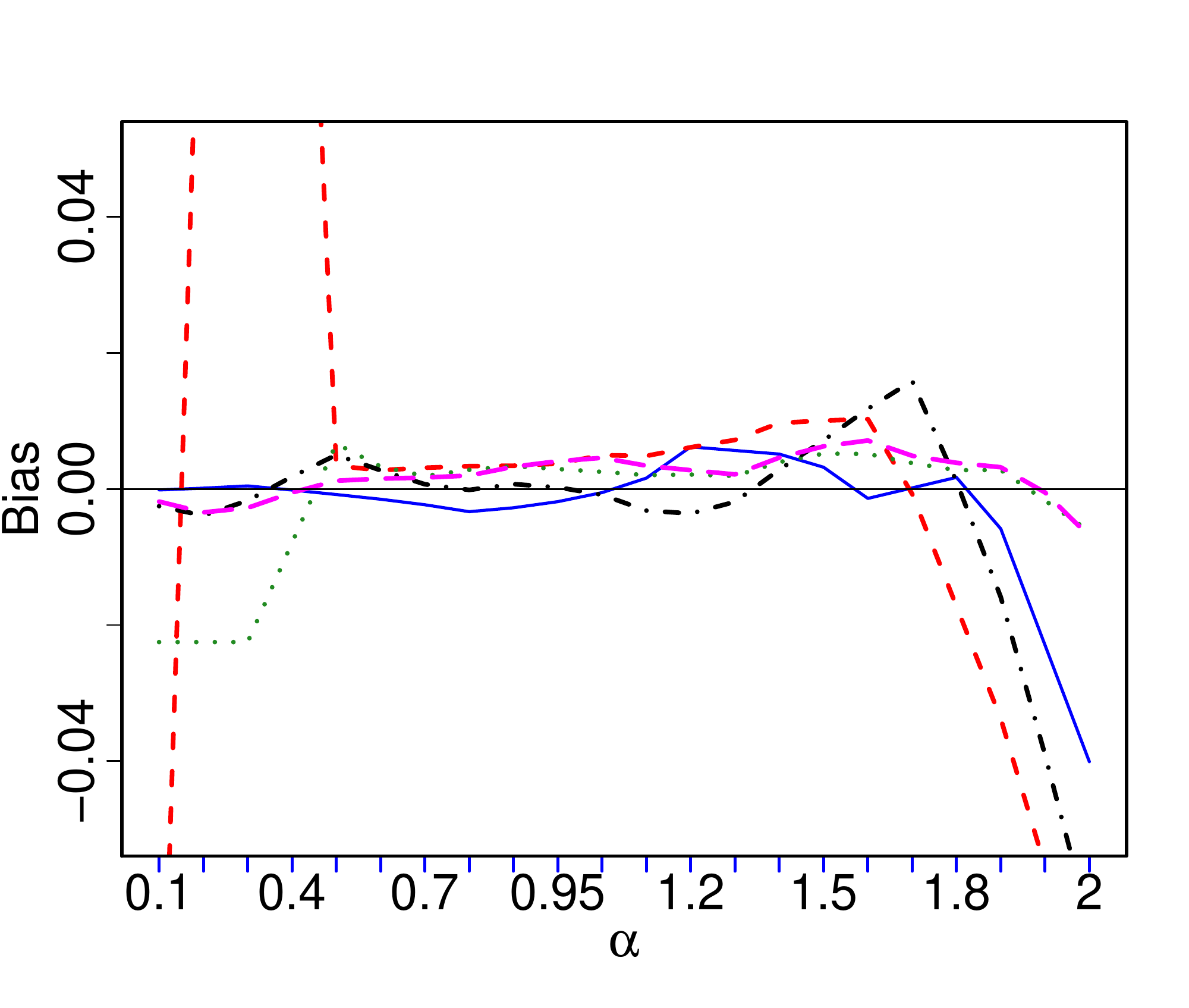} 
\includegraphics[width=65mm,height=75mm]{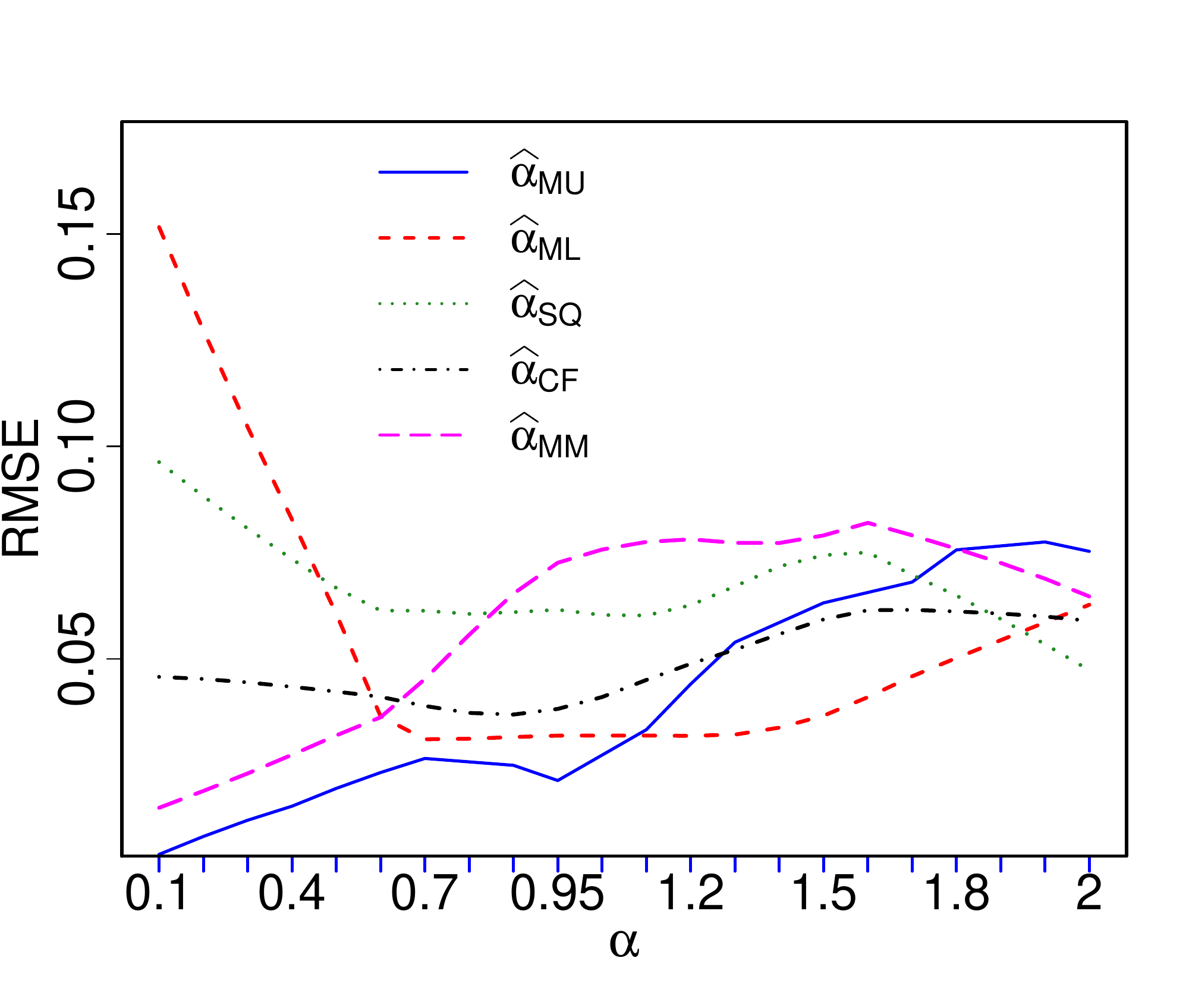}\\
\includegraphics[angle=90,width=65mm,height=75mm]{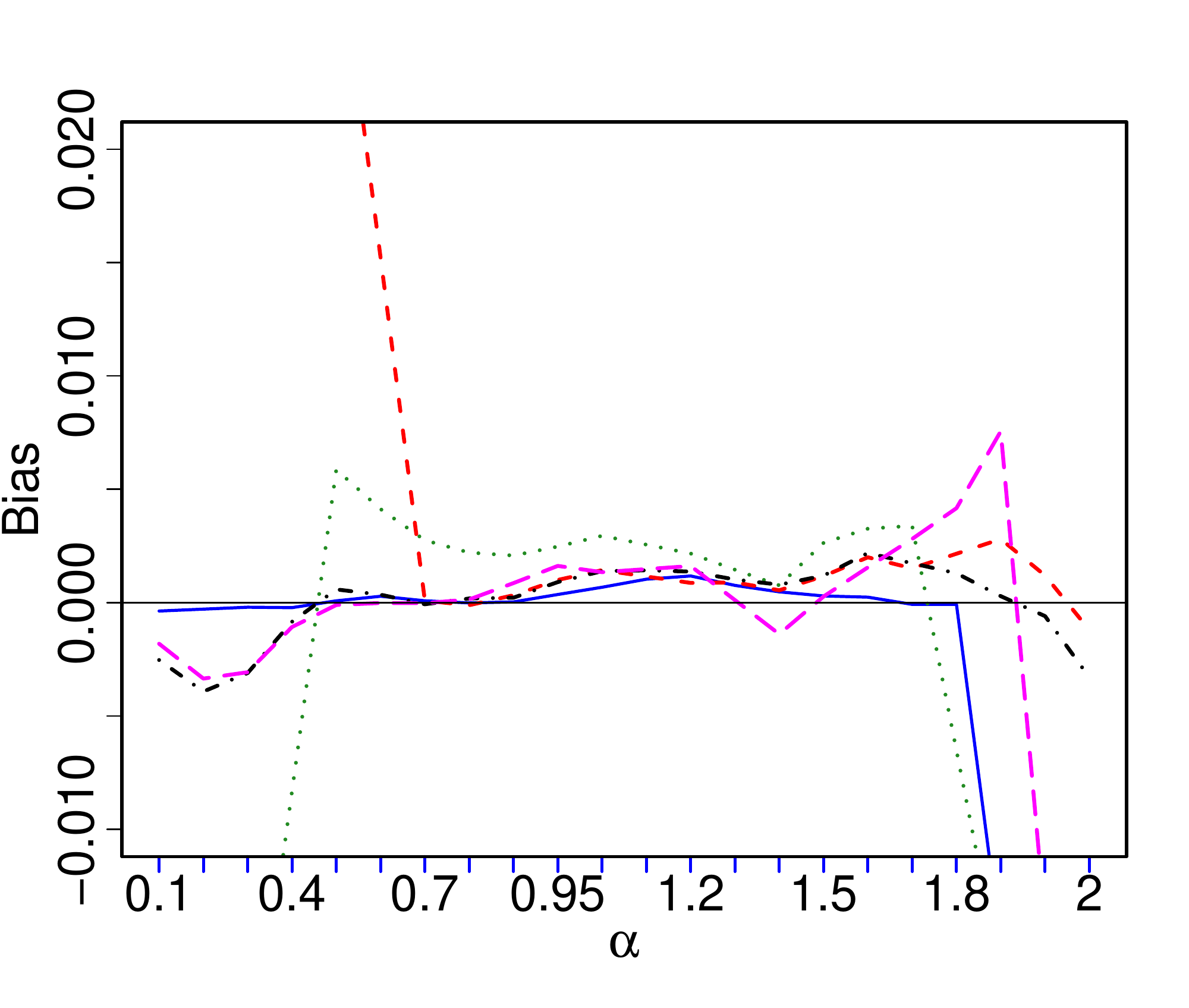} 
\includegraphics[width=65mm,height=75mm]{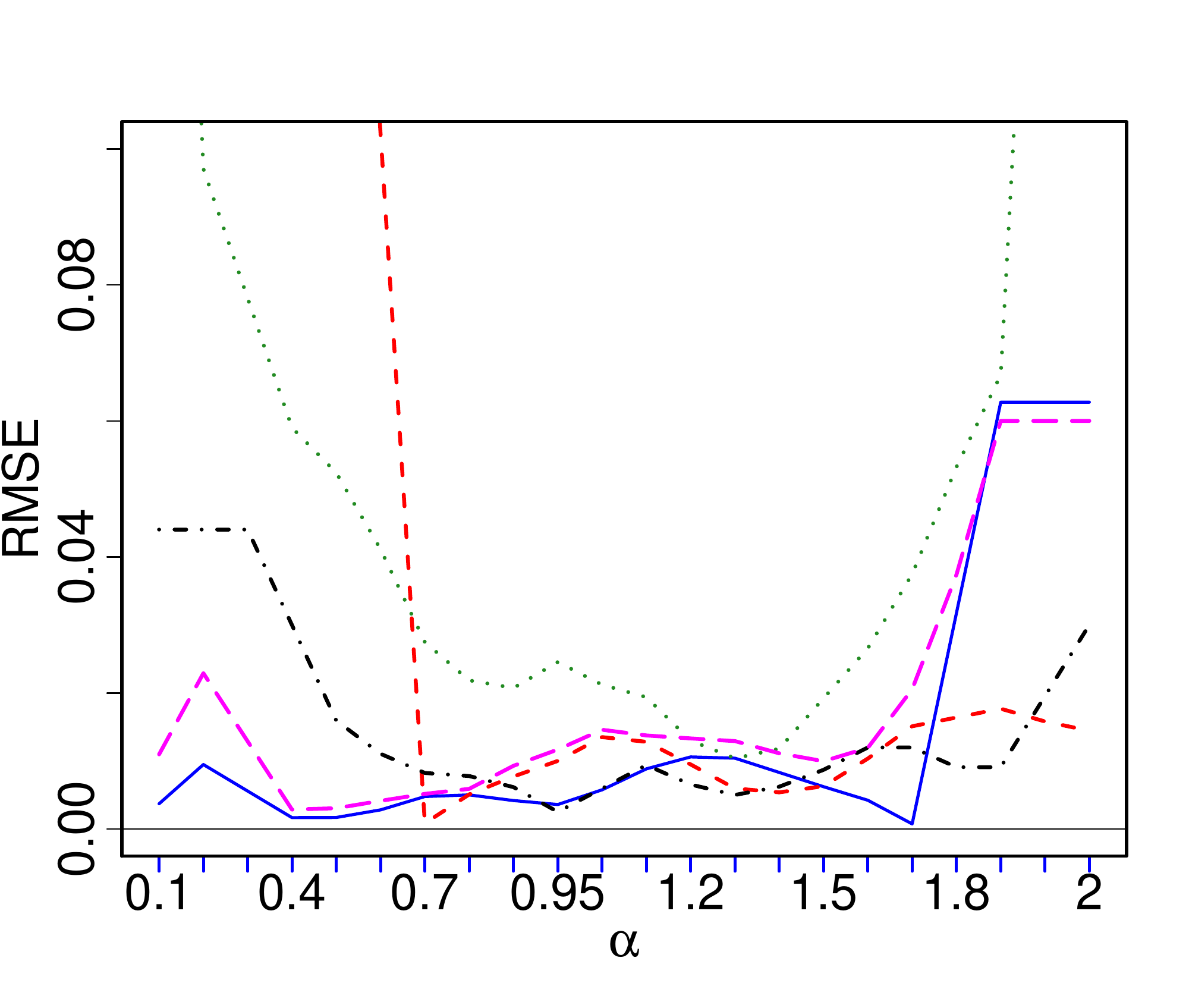} 
\caption{Biases and RMSEs of estimators ${\hat{\alpha}}_{MU}$, ${\hat{\alpha}}_{ML}$, ${\hat{\alpha}}_{SQ}$, ${\hat{\alpha}}_{CF}$, and ${\hat{\alpha}}_{MM}$. Top-left (bias when $n=500$), top-right (RMSE when $n=500$), bottom-left (bias when $n=5000$), and bottom-right (RMSE when $n=5000$).} 
\label{ustat} 
\end{figure}

Estimating the parameters of a multivariate stable distribution is a great challenge. This is because the multivariate stable distributions are a semi-parametric model, i.e. these models involve an infinite number of parameters. To overcome this problem, the spectral measure given in Definition \ref{def2} is discretized at $m$ points as
\begin{equation}\label{dis}
\Gamma(\cdot)\approx\sum_{j=1}^{m}\gamma_{j}I_{\boldsymbol{s}_{j}}(\cdot)
\end{equation}
where $\gamma_{j}$ is a mass at point $\boldsymbol{s}_{j}$ in ${\mathbb{S}}^{d}$ and $I_{\boldsymbol{s}_{j}}(.)$ is an indicator function at point $\boldsymbol{s}_{j}$; for $j=1,\dots,m$, see \cite{byczkowski1993approximation}. Assuming that the address of masses are known, discrete spectral measure given in (\ref{dis}) can be estimated through the projected CF, ML, or SQ approaches. Developing the idea proposed by \cite{fan2006parameter}, an estimator, shown here by ${\hat{\alpha}}_{MU}$, for the tail index of multivariate strictly stable distribution introduced by \cite{teimouri2017ustatistic}. Suppose the tail index estimator using CF, ML, and SQ methods are shown by ${\hat{\alpha}}_{CF}$, ${\hat{\alpha}}_{ML}$, ${\hat{\alpha}}_{SQ}$, respectively.
A comparison will be made between ${\hat{\alpha}}_{MU}$, ${\hat{\alpha}}_{CF}$, ${\hat{\alpha}}_{ML}$, ${\hat{\alpha}}_{SQ}$, and that introduced by \cite{mohammadi2015estimating}, here shown by ${\hat{\alpha}}_{MM}$, through a simulation study. The following observations can be made from Figure \ref{ustat}.
\begin{itemize} 
\item ${\hat{\alpha}}_{CF}$ and ${\hat{\alpha}}_{ML}$ show better performance than ${\hat{\alpha}}_{MU}$ for $\alpha>1.4$ in the sense of RMSE.
\item $\hat{\alpha}_{MU}$ is more efficient than all other estimators for $\alpha \leq 1.4$ in the sense of RMSE. 
\item $\hat{\alpha}_{MU}$ is more efficient than all other estimators for $\alpha \leq 1.75$ in the sense of bias for sample size $n=5000$. 
\item $\hat{\alpha}_{MU}$ shows better performance than $\hat{\alpha}_{SQ}$ in the sense of RMSE for sample size $n=5000$. 
\end{itemize} 
Also, an estimator of the discretized spectral measure can be constructed using ${\hat{\alpha}}_{MU}$, ${\hat{\alpha}}_{CF}$, ${\hat{\alpha}}_{ML}$, ${\hat{\alpha}}_{SQ}$, and ${\hat{\alpha}}_{MM}$. A comprehensive comparison was made between the performance of these estimators by \cite{teimouri2017ustatistic}. 
\section{The R package alphastable}
We develop the {\bf{alphastable}} package for generating stable random variables (in univariate and multivariate cases), generating truncated stable random variables (in univariate case), computing pdfs and cdfs (in univariate case), estimating the parameters of stable distribution (in univariate and multivariate cases), and modelling mixture of Cauchy and symmetric stable distributions. In what follows, we describe how to use the {\bf{alphastable}}. It depends on packages {\bf{mvtnorm}} (\verb+http://cran.r-project.org/web/packages/mvtnorm/index.html+), {\bf{nnls}} (\verb+https://cran.r-pr+ \verb+oject.org/web/packages/nnls/index.html+), and {\bf{stabledist}} (\verb+https:+ \verb+//cran.r-project.org/+ \verb+web/packages/stabledist/index.html+). The {\bf{alphastable}} package has been uploaded to comprehensive R archive network (CRAN), see \verb+https://CRAN.R-project.org/pac+ \verb+kage=alphastable+.
\subsection{Random generation}
\subsubsection{Random number generation from univariate stable distribution} 
Use the following command.\\
\verb|urstab(n,alpha,beta,sigma,mu,param)|\\
The available arguments are:\\
\verb|n|: the sample size\\
\verb|alpha|: the tail index parameter\\
\verb|beta|: the skewness parameter\\
\verb|sigma|: the scale parameter\\
\verb|mu|: the location parameter\\
\verb|param|: type of parameterization must be 0 or 1 for $S_0$ and $S_1$ parameterizations, respectively.\\\\
Example: we use the following command to simulate $n=200$ realizations from univariate stable distribution with parameters $\alpha=1.3$, $\beta=0.5$, $\sigma=2$, and $\mu=0$ in $S_0$ parameterization.
\begin{verbatim}
R>urstab(200,1.3,0.5,2,0,0)
\end{verbatim}
\subsubsection{Random number generation from truncated stable distribution} 
Use the following command.\\
\verb|urstab.trunc(n,alpha,sigma,mu,a,b,param)|\\
The available arguments are:\\
\\verb|n|: the sample size\\
\verb|alpha|: the tail index parameter\\
\verb|sigma|: the scale parameter\\
\verb|mu|: the location parameter\\
\verb|a|: the lower bound of truncation\\
\verb|b|: the upper bound of truncation\\
\verb|param|: type of parameterization must be 0 or 1 for $S_0$ and $S_1$ parameterizations, respectively.\\\\
Example: we use the following command to simulate $n=200$ realizations from truncated stable distribution with parameters $\alpha=1.3$, $\beta=0.5$, $\sigma=2$, $\mu=0$ which, is truncated over (-5,5) in $S_0$ parameterization.
\begin{verbatim}
R>urstab.trunc(200,1.3,0.5,2,0,-5,5,0)
\end{verbatim}
\subsubsection{Random vector generation from multivariate elliptically contoured stable distribution}
Use the following command.\\
\verb|mrstab.elliptical(n,alpha,Sigma,Mu)|\\
The available arguments are:\\
\verb|n|: the sample size\\
\verb|alpha|: the tail index parameter\\
\verb|Sigma|: the dispersion matrix\\
\verb|Mu|: the location vector\\\\
Example: we use the following command to simulate $n=200$ vectors from two-dimensional elliptically contoured stable distribution with parameters $\alpha=1.3$, $\Sigma=\begin{pmatrix}1&0.5\\0.5&1\end{pmatrix}$, and $\boldsymbol{\mu}=(0,0)^T$.
\begin{verbatim}
R>library("stabledist")
R>library("mvtnorm")
R>mrstab.elliptical(200,1.3,matrix(c(1,.5,.5,1),2,2),c(0,0))
\end{verbatim}
\subsubsection{Random vector generation from multivariate stable distribution}
Use the following command.\\
\verb|mrstab(n,m,alpha,Gamma,Mu)|\\
The available arguments are:\\
\verb|n|: the sample size\\
\verb|m|: the number of masses\\
\verb|alpha|: the tail index parameter\\
\verb|Gamma|: the mass vector\\
\verb|Mu|: the location vector\\\\
Example: we use the following command to simulate $n=200$ vectors from two-dimensional stable distribution with $\alpha=1.3$, the vector of masses $\boldsymbol{\gamma}=(0.1,0.5,0.5,0.1)^T$, and $\boldsymbol{\mu}=(0,0)^T$.

\begin{verbatim}
R>library("stabledist")
R>mrstab(200,4,1.3,c(0.1,0.5,0.5,0.1),c(0,0))
\end{verbatim}

\subsubsection{computing the pdf and cdf of univariate and multivariate stable distributions} 
Use the following command for computing the pdf of univariate stable distribution.\\\\
\verb|udstab(y,alpha,beta,sigma,mu,param)|\\
The available arguments are:\\
\verb|y|: a real value at which pdf is computed\\
\verb|alpha|: the tail index parameter\\
\verb|beta|: the skewness parameter\\
\verb|sigma|: the scale parameter\\
\verb|mu|: the location parameter\\
\verb|param|: type of parameterization must be 0 or 1 for $S_0$ and $S_1$ parameterizations, respectively.\\\\
Example: we use the following command to compute the pdf of a univariate stable distribution at point $y=2$ with parameters $\alpha=1.2$, $\beta=0.9$, $\sigma=1$, and $\mu=0$ in $S_1$ parameterization.
\begin{verbatim}
R>library("stabledist")
R>udstab(2,1.2,0.9,1,0,1)
[1] 0.0240643
\end{verbatim}
Use the following command for computing the cdf of univariate stable distribution.\\\\
\verb|upstab(y,alpha,beta,sigma,mu,param)|\\
The available arguments are:\\
\verb|y|: a real value at which cdf is computed\\
\verb|alpha|: the tail index parameter\\
\verb|beta|: the skewness parameter\\
\verb|sigma|: the scale parameter\\
\verb|mu|: the location parameter\\
\verb|param|: type of parameterization must be 0 or 1 for $S_0$ and $S_1$ parameterizations, respectively.\\\\
Example: we use the following commands to compute the cdf of a univariate stable distribution at point $y=2$ with parameters $\alpha=1.2$, $\beta=0.9$, $\sigma=1$, and $\mu=0$ in $S_1$ parameterization.
\begin{verbatim}
R>library("stabledist")
R>upstab(2,1.2,0.9,1,0,1)
[1] 0.9044019
\end{verbatim}
We use the following command for computing the pdf of multivariate elliptically contoured stable distribution.\\\\
\verb|mdstab.elliptical(z,alpha,Sigma,Mu)|\\
The available arguments are:\\
\verb|y|: a real vector at which pdf is computed\\
\verb|alpha|: the tail index parameter\\
\verb|Sigma|: the dispersion matrix\\
\verb|Mu|: the location vector\\\\
Example: we use the following command to compute the pdf of two-dimensional elliptically contoured stable distribution at point $\boldsymbol{z}=(5,5)^T$with parameters $\alpha=1.2$, $\Sigma=\begin{pmatrix}1&0.5\\0.5&1\end{pmatrix}$, and $\boldsymbol{\mu}=(0,0)^T$.
\begin{verbatim}
R>library("stabledist")
R>mdstab.elliptical(c(5,5),1.2,matrix(c(1,0.5,0.5,1),2,2),c(0,0))
[1] 0.0005279
\end{verbatim}
\subsection{Parameter estimation for Cauchy and mixture of Cauchy distributions}
Use the following command to estimate the parameters of Cauchy distribution.\\\\
\verb|ufitstab.cauchy(y,beta0,sigma0,mu0,param)|\\
The available arguments are:\\
\verb|y|: the vector of observations\\
\verb|beta0|: the initial value of skewness parameter to start the EM algorithm\\
\verb|sigma0|: the initial value of scale parameter to start the EM algorithm\\
\verb|mu0|: the initial value of location parameter to start the EM algorithm\\
\verb|param|: type of parameterization must be 0 or 1 for $S_0$ and $S_1$ parameterizations, respectively.\\\\
Example: we consider the large recorded intensities (in Richter scale) of the earthquake at seismometer locations in western North America between 1940 and 1980. The related features was reported by \cite{joyner1981peak}. Among the features, we focus on the 182 distances from the seismological measuring station to the epicenter of the earthquake (in km) as the variable of interest. This set of data can be found in package {\bf{nlme}}. Using the initial values as $\beta_{0}=0.5$, $\sigma_{0}=5$, and $\mu_{0}=10$, the EM estimations are $\hat{\beta}=0.91678$, $\hat{\sigma}=11.5981$, and $\hat{\mu}=16.9498$. The commands and related output are as the following.

\begin{figure}[h!]
\center
\includegraphics[width=80mm,height=80mm]{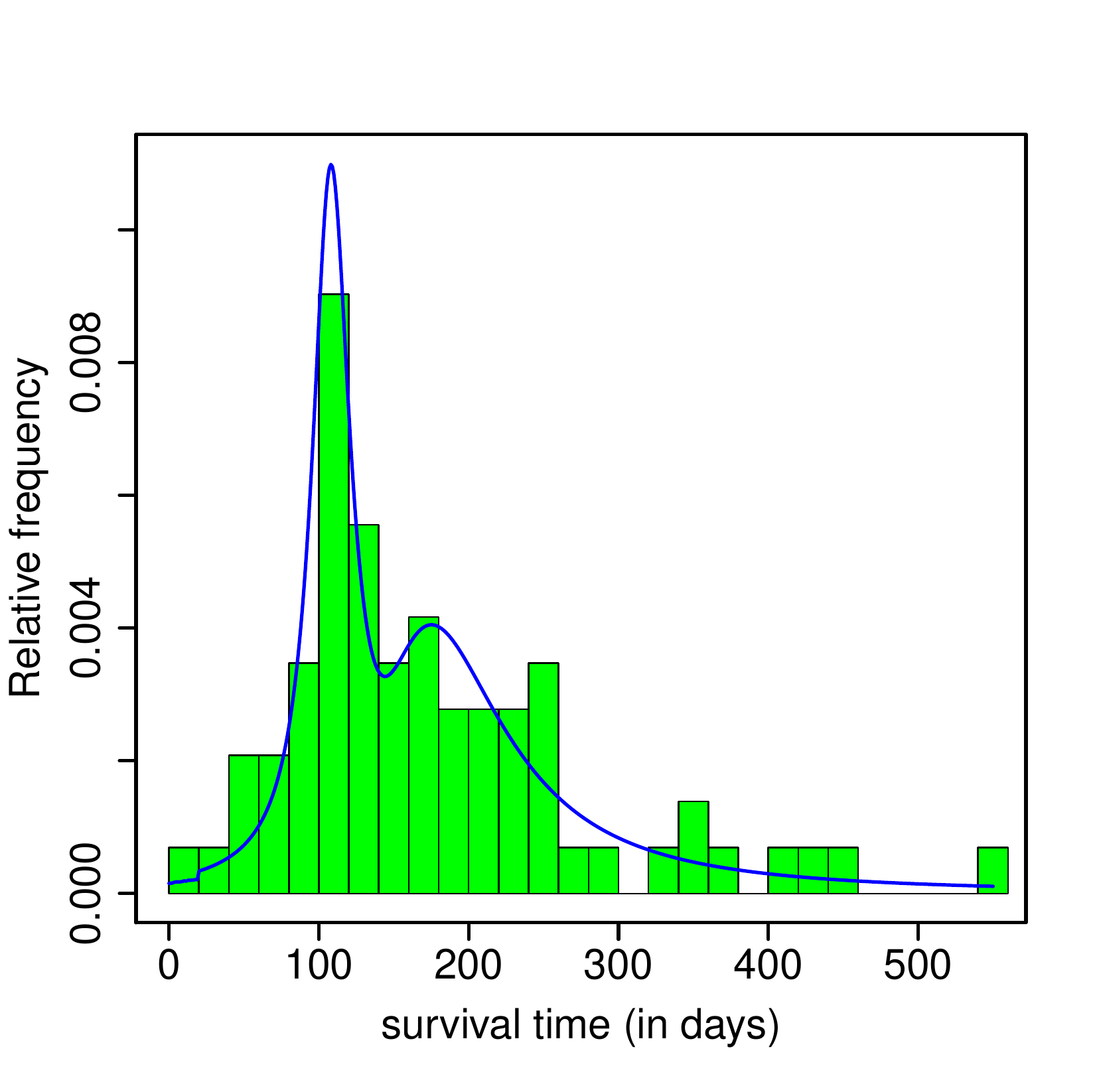}
\caption{Histogram of 72 guinea pigs survival times (in days). Fitted pdf of a two-component mixture of Cauchy distributions is shown by a blue solid curve.}
\label{pig}
\end{figure}

\begin{verbatim}
R>library("nlme")
R>y<-Earthquake[,3]
R>ufitstab.cauchy(y,0.5,5,10,0)
$beta
[1] 0.91678
$sigma
[1] 11.5981
$mu
[1] 16.9498
\end{verbatim}
The Kolmogorov-Smirnov (K-S) and Anderson-Darling (A-D) goodness-of-fit criteria are 0.0397 and 0.6980, respectively. \\\\
Use the following command for estimating the parameters of mixture of Cauchy distributions.\\\\
\verb|ufitstab.cauchy.mix(y,K,omega0,beta0,sigma0,mu0)|\\
The available arguments are:\\
\verb|y|: the vector of observations\\
\verb|K|: the number of components\\
\verb|omega0|: the initial value of weight vector to start the EM algorithm\\
\verb|beta0|: the initial value of skewness vector to start the EM algorithm\\
\verb|sigma0|: the initial value of scale vector to start the EM algorithm\\
\verb|mu0|: the initial value of location vector to start the EM algorithm\\\\
Example: We use the survival times in days of 72 guinea pigs infected with different doses of tubercle bacilli, see \cite{bjerkedal1960acquisition}. To implement the EM algorithm the initial values are: $\underline{\omega}_{0}=(0.65,0.35)$, $\underline{\beta}_{0}=(0.20,0.05)$, $\underline{\sigma}_{0}=(20,50)$, and $\underline{\mu}_{0}=(95,210)$. The EM estimations are $\underline{\hat{\omega}}=(0.5327,0.4673)$, $\underline{\hat{\beta}}=(0.05160,0.8233)$, $\underline{\hat{\sigma}}=(15.6626,39.1202)$, and $\underline{\hat{\mu}}=(108.5602,194.5814)$. The commands and related output are given by the following.
\begin{verbatim}
R>pig<-c(10,33,44,56,59,72,74,77,92,93,96,100,100,102,105,107,107,108,108,108,
+         109,112,121,122,122,124,130,134,136,139,144,146,153,159,160,163,163,
+         168,171,172,176,113,115,116,120,183,195,196,197,202,213,215,216,222,
+         230,231,240,245,251,253,254,255,278,293,327,342,347,361,402,432,458,
+         555)
R>library("stabledist")
R>ufitstab.cauchy.mix(pig,2,c(0.65,0.35),c(0.20,0.05),c(20,50),c(95,210))
$omega
[1] 0.5327 0.4673
$beta
[1] 0.05160 0.8233
$sigma
[1] 15.6626 39.1202
$mu
[1] 108.5602 194.5814
\end{verbatim}
The corresponding K-S and A-D goodness-of-fit statistics are 0.04882 and 0.25062, respectively. Figure \ref{pig} displays fitted two-component Cauchy distribution to the guinea pigs survival times data.
\subsection{Estimating the parameters of zero-location symmetric stable distribution using U-statistic}
Use the following command.\\\\
\verb|ufitstab.ustat(y)|\\
The available argument is:\\
\verb|y|: the vector of observations\\\\
Example: Suppose the CAC40 (Bourse de Paris) data discussed in the previous section follow a zero-location symmetric stable distribution. The commands and related output for applying the U-statistic to these data are given as follows.
\begin{verbatim}
R>CAC40<-c(-0.020807,0.003035,-0.036817,...,-0.054140,-14.87798)
R>library("stabledist")
R>ufitstab.sym(CAC40)
$alpha
[1] 1.4643
$sigma
[1] 0.4725
\end{verbatim}
The K-S and A-D statistics are 0.0290 and 4.0935, respectively.
\subsection{Estimating the parameters of symmetric and mixture of symmetric stable distributions}
\begin{figure}[h!]
\center
\includegraphics[width=80mm,height=80mm]{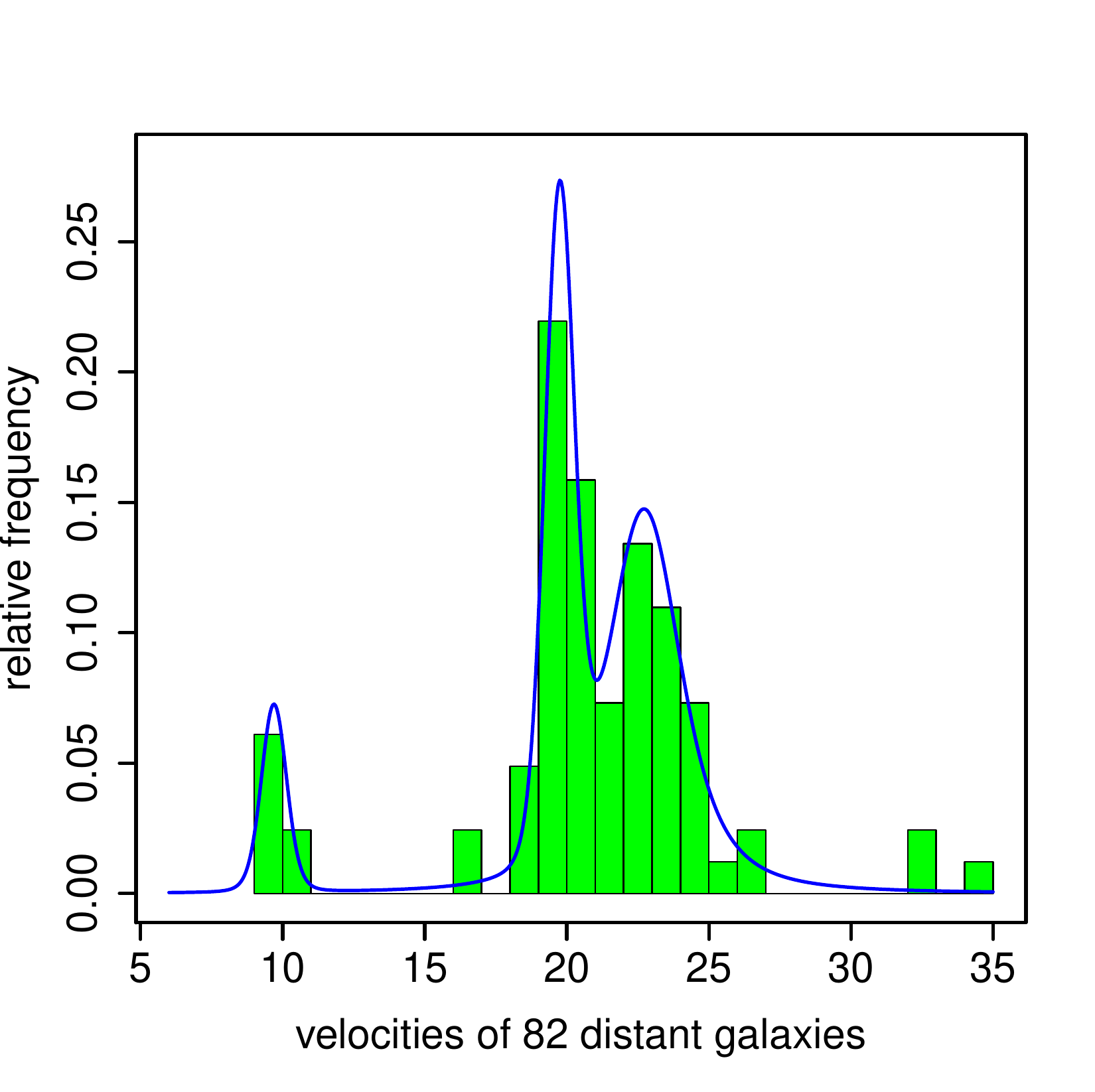}
\caption{Histogram of velocities of 82 distant galaxies. Fitted density function of three-component mixture of symmetric stable distributions is shown by a blue curve.}
\label{galaxy}
\end{figure}

We use the following command for estimating the parameters of symmetric stable distribution.\\\\
\verb|ufitstab.sym(y,alpha0,sigma0,mu0)|\\
The available arguments are:\\\\
\verb|y|: the vector of observations\\
\verb|alpha0|: the initial value of tail index parameter to start the EM algorithm\\
\verb|sigma0|: the initial value of scale parameter to start the EM algorithm\\
\verb|mu0|: the initial value of location parameter to start the EM algorithm\\\\
Example: we apply the EM algorithm to CAC40 (Bourse de Paris) data. Using the initial values as $\alpha_{0}=1.2$, $\sigma_{0}=1$, and $\mu_{0}=1$, the EM estimations are $\hat{\alpha}=1.7898$, $\hat{\sigma}=0.5561$, and $\hat{\mu}=0.0111$. The commands and related output are given as follows.
\begin{verbatim}
R>CAC40<-c(-0.020807,0.003035,-0.036817,...,-0.054140,-14.87798)
R>library("stabledist")
R>ufitstab.sym(CAC40,1.2,1,1)
$alpha
[1] 1.7898
$sigma
[1] 0.5561
$mu
[1] 0.0111
\end{verbatim}
The K-S and A-D statistics are 0.0192 and 1.6430, respectively. We use the following command to estimate the parameters of SSMM.\\\\
\verb|ufitstab.sym.mix(y,k,omega0,alpha0,sigma0,mu0)|\\
The available arguments are:\\\\
\verb|y|: the vector of observations\\
\verb|k|: the number of components\\
\verb|omega0|: the initial value of weight vector to start the EM algorithm\\
\verb|alpha0|: the initial value of tail index vector to start the EM algorithm\\
\verb|sigma0|: the initial value of scale vector to start the EM algorithm\\
\verb|mu0|: the initial value of location vector to start the EM algorithm\\\\
Example: For validating the performance of the proposed EM algorithm we use galaxy (velocities of 82 distant galaxies, diverging from our own galaxy) data. These data are available at \verb+https://people.maths.bris.ac.uk/mapjg/mixdata+. The commands and related output are given by the following.
\begin{verbatim}
R>galaxy<-c(9.172,9.350,9.483,9.558,9.775,10.227,10.406,16.084,16.170,18.419,
+             18.552,18.600,18.927,19.052,19.070,19.330,19.343,19.349,19.440,
+             19.473,19.529,19.541,19.547,19.663,19.846,19.856,19.863,19.914,
+             19.918,19.973,19.989,20.166,20.175,20.179,20.196,20.215,20.221,
+             20.415,20.629,20.795,20.821,20.846,20.875,20.986,21.137,21.492,
+             21.701,21.814,21.921,21.960,22.185,22.209,22.242,22.249,22.314,
+             22.374,22.495,22.746,22.747,22.888,22.914,23.206,23.241,23.263,
+             23.484,23.538,23.542,23.666,23.706,23.711,24.129,24.285,24.289,
+             24.366,24.717,24.990,25.633,26.960,26.995,32.065,32.789,34.279)
R>library("stabledist")
R>y<-galaxy
R>ufitstab.sym.mix(y,3,c(0.1,0.35,0.55),c(1.2,1.2,1.2),c(1,1,1),c(8,20,22))
$omega
[1] 0.0842 0.3405 0.5753
$alpha
[1] 1.7781 1.8494 1.3695
$sigma
[1] 0.3307 0.3836 1.1388
$mu
[1] 9.6941 19.7442 22.7213
\end{verbatim}
The K-S and A-D statistics are 0.0372 and 0.2040, respectively. The fitted pdf to the galaxy data is shown in Figure \ref{galaxy}. We note that the SSMM outperforms the other mixture models including: skew normal (\cite{lin2007finite, lee2013mixtures, basso2010robust}), slash normal \cite{basso2010robust}, $t$ \cite{peel2000robust}, skew $t$ (\cite{lee2013mixtures, vrbik2014parsimonious}), and GH \cite{browne2015mixture}, see \cite{teimouri2018algorithm}. For applying aforementioned models to the galaxy data, we use packages {\bf{mixsmsn}} \cite{prates2013mixsmsn} and {\bf{MixGHD}} \cite{MixGHD}. Contrary to the SSMM, these models may give different estimation results when they are applied several times to the galaxy data. This privilege of SSMM put it into the class of models which, are appropriate for robust mixture modelling. 
\subsection{Estimating the parameters of skewed stable distribution}
\begin{figure}[h!]
\center
\includegraphics[width=80mm,height=80mm]{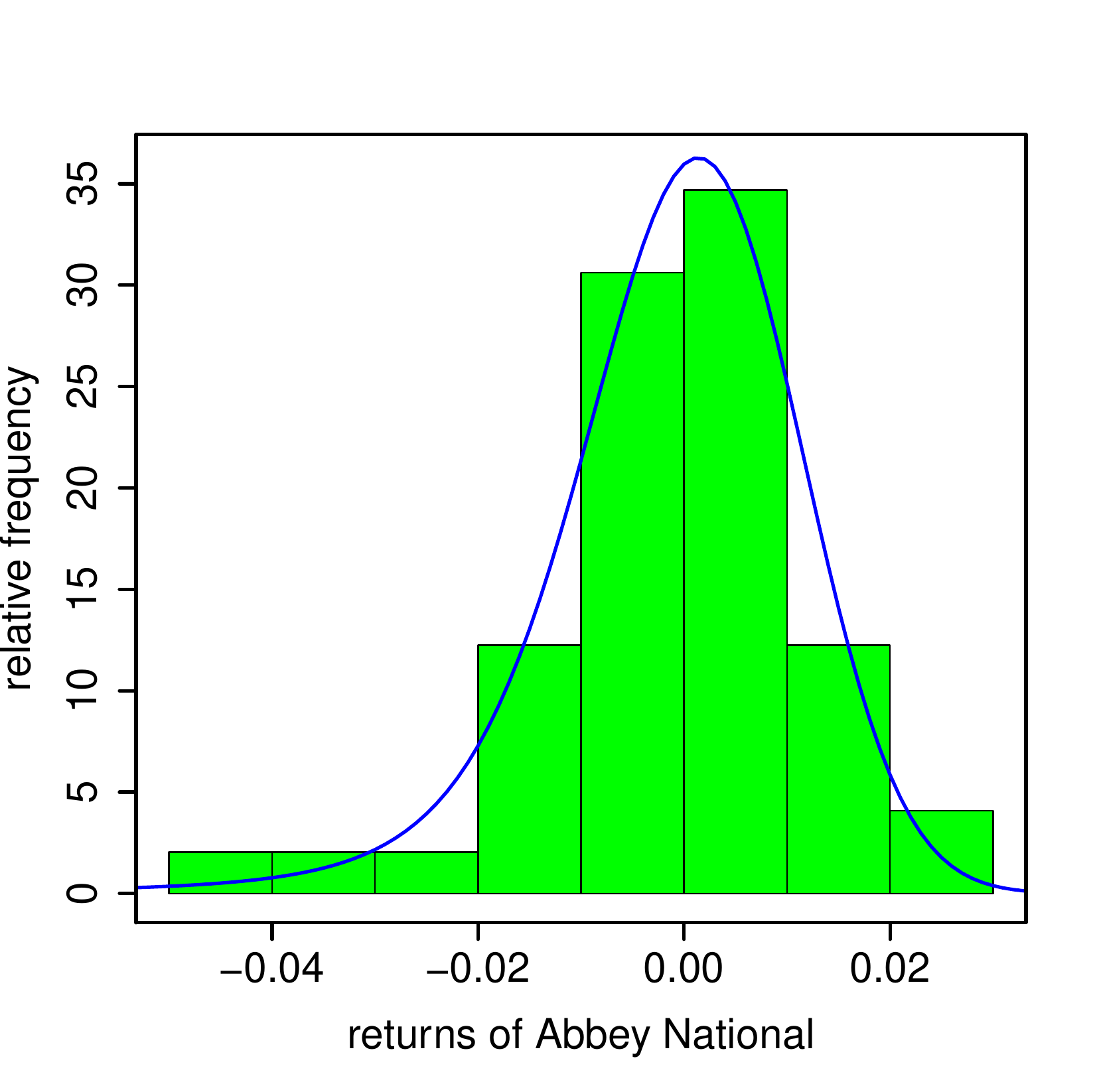}
\caption{Histogram of 49 Abbey National returns. Fitted density function of stable distribution, shown by blue curve, captures clearly the histogram of data.}
\label{abbey}
\end{figure}

Use the following command.\\\\
\verb|ufitstab.skew(y,alpha0,beta0,sigma0,mu0,param)|\\
The available arguments are:\\\\
\verb|y|: the vector of observations\\
\verb|alpha0|: the initial value of tail index parameter to start the EM algorithm\\
\verb|beta0|: the initial value of skewness parameter to start the EM algorithm\\
\verb|sigma0|: the initial value of scale parameter to start the EM algorithm\\
\verb|mu0|: the initial value of location parameter to start the EM algorithm\\
\verb|param|: type of parameterization must be 0 or 1 for $S_0$ and $S_1$ parameterizations, respectively.\\\\
Example: We use the daily price returns of Abbey National shares between 31/7/91 and 8/10/91 (including $n=50$ business days). By assuming that $p_{t}$ denotes the price at $t$th day, the price return at $t$th day is defined as $(p_{t-1}-p_{t})/p_{t-1}$; for $t=2,\dots,n$, see \cite{buckle1995bayesian}. Using the initial values as $\alpha_{0}=0.8$, $\beta_{0}=0$, $\sigma_{0}=0.25$, and $\mu_{0}=0.25$ and \verb|param|=1, the EM-based estimations are $\hat{\alpha}=1.8046$, $\hat{\beta}=-0.9999$, $\hat{\sigma}=0.0078$, and $\hat{\mu}=0.0009$. The commands and related output are as the following.
\begin{verbatim}
R> price<-c(296,296,300,302,300,304,303,299,293,294,294,293,295,287,288,297,
+           305,307,304,303,304,304,309,309,309,307,306,304,300,296,301,298,
+           295,295,293,292,307,297,294,293,306,303,301,303,308,305,302,301,
+           297,299)
R>y<-c()
R>n<-length(price)
R>for(i in 2:n){y[i]<-(price[i-1]-price[i])/price[i-1]}
R>library("stabledist")
R>ufitstab.skew(y[2:n],0.8,0,0.25,0.25,1)
$alpha
[1] 1.8046
$beta
[1] -0.9999
$sigma
[1] 0.0078
$mu
[1] 0.0009
\end{verbatim}
The corresponding log-likelihood statistics for the ML and EM estimations are 146.3868 and 147.0238, respectively. Figure \ref{abbey} displays the fitted stable pdf to the Abbey National returns data.
\subsection{Estimating the parameter of multivariate elliptically contoured stable distribution}
Use the following command.\\\\
\verb|mfitstab.elliptical(z,alpha0,Sigma0,Mu0)|\\
The available arguments are:\\\\
\verb|z|: an $n\times d$ vector of observations\\
\verb|alpha0|: the initial value of tail index parameter to start the EM algorithm\\
\verb|Sigma0|: the initial value of dispersion matrix to start the EM algorithm\\
\verb|Mu0|: the initial value of location vector to start the EM algorithm\\\\
Example: we follow for applying the EM algorithm to $\boldsymbol{Z}=(S$\&$P500, IPC)^{T}$ introduced in the previous section. The initial values are $\alpha=1.2$, $\Sigma=\begin{pmatrix}0.75&0.25\\0.25&0.75\end{pmatrix}$, and $\boldsymbol{\mu}=(0.5,0.5)^T$. The commands and related output are given by the following.
\begin{verbatim}
R>SandP<-c(-0.01602344,-0.01331844,1.51226709,...,-0.78849818,0.39571979)
R>IPC<-c(-0.03497474,-0.00589434,0.98289177,...,-0.42876127,0.21059706)
R>z<-cbind(SandP,IPC)
R>library("stabledist")
R>mfitstab.elliptical(z,1.2,matrix(c(0.75,0.25,0.25,0.75),2,2),c(0.5,0.5))
$alpha
[1] 1.7315
$Sigma
[,1] [,2]
[1,] 0.3995 0.3648
[2,] 0.3648 0.3968
$mu
[1] -0.0275 -0.0245
\end{verbatim}
The estimated tail index, dispersion matrix, and location vector are $\hat{\alpha}=1.7315$ and $\Sigma=\begin{pmatrix}0.3995&0.3648\\0.3648&0.3968\end{pmatrix}$, and $\hat{\boldsymbol{\mu}}=(-0.0275,-0.0245)^T$, respectively.
The log-likelihood statistics correspond to the ML and EM estimations are -4967.706 and -4975.571, respectively. It should be noted that the EM algorithm outperforms the CF and SQ approaches. 
\subsection{Estimating the parameters of strictly multivariate stable distribution}
\begin{figure}[h!]
\centering
\includegraphics[width=80mm,height=80mm]{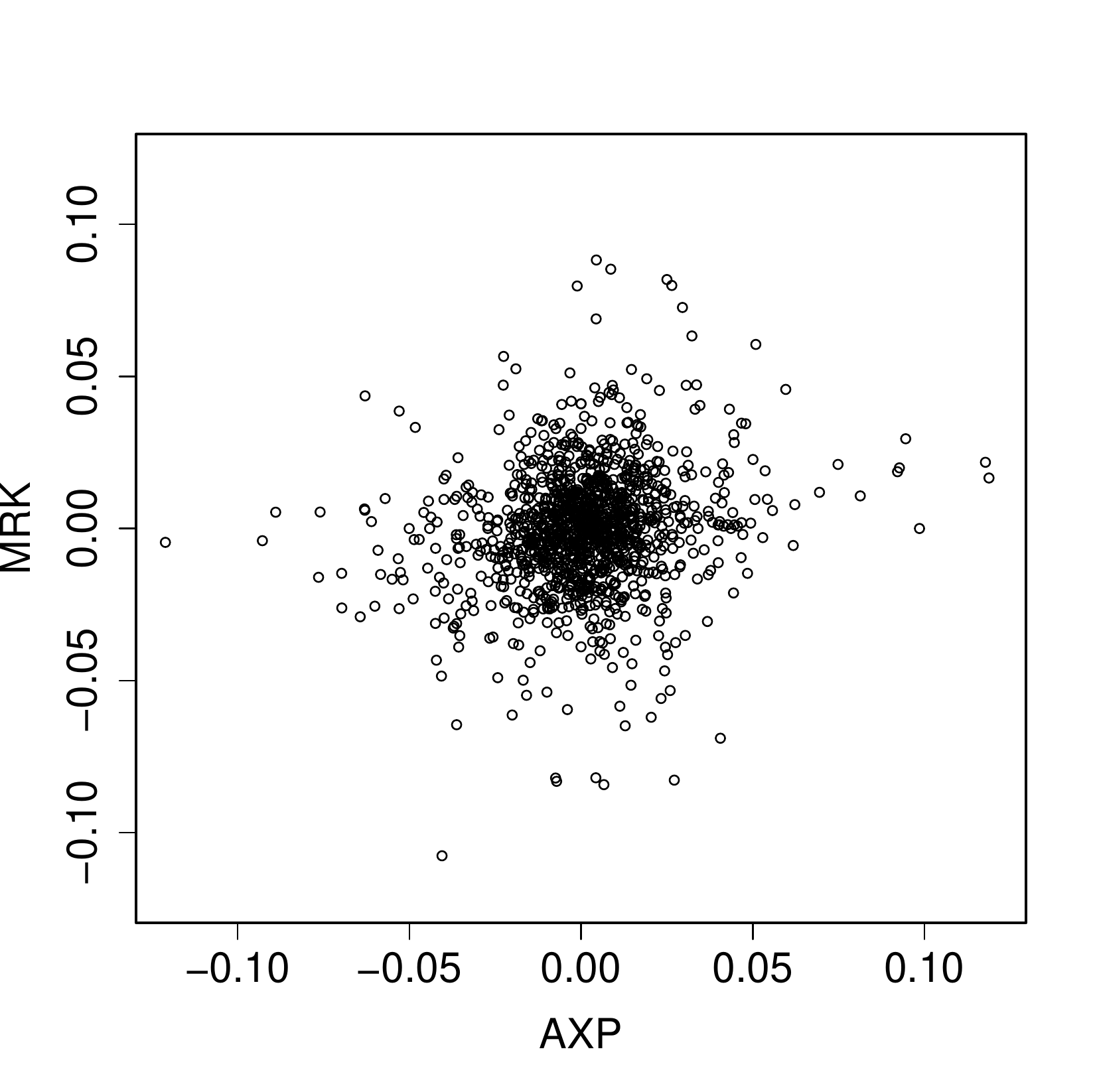}
\caption{Scatterplot for daily log-return percent of AXP versus MRK.}
\label{axp}
\end{figure}

Use the following command.\\\\
\verb|mfitstab.ustat(x,m,method=method)|\\
The available arguments are:\\\\
\verb|x|: an $n\times 2$ vector of observations\\
\verb|m|: the number of masses\\
\verb|method|: must be 1 or 2 which, corresponds to the method given by \cite{teimouri2017ustatistic} and
\cite{mohammadi2015estimating}, respectively\\\\
Example: Here, we focus on daily log-return (in percent) of 1247 closing prices of AXP (American Express Company) and MRK (Merck \& Co. Inc.), see \cite{nolan2013multivariate}. This set of data are between January 3, 2000 and December 31, 2004. The scatterplot of $\boldsymbol{X}=(AXP, MRK)^T$ is shown in Figure \ref{axp}. For $m=4$, the masses of the spectral measure are addressed by $\boldsymbol{s}_j=\left(\cos(2\pi(j-1)/m), \sin (2\pi(j-1)/m)\right)^{T}$; for $j=1,\dots, m$.
The commands and related report are given by the following.
\begin{verbatim}
R>AXP<-c(0.007512315,-0.005194817,-0.018692133,...,-0.007746755,-0.000507314)
R>MRK<-c(-0.020927520,0.016074450,-0.014427908,...,-0.004621792,-0.009931801)
R>x<-cbind(AXP,MRK)
R>library("nnls")
R>mfitstab.ustat(x,4,1)
$alpha
[1] 1.5881
$mass
[1] 0.000000 0.000176 0.000627 0.000019
R>mfitstab.ustat(x,4,2)
$alpha
[1] 1.6983
$mass
[1] 0.000000 0.000203 0.000590 0.000000
\end{verbatim}
As it is seen, the estimated discrete spectral measure using the method proposed by \cite{teimouri2017ustatistic} and
\cite{mohammadi2015estimating} are $\hat{\boldsymbol{\gamma||}}=(0.000000, 0.000176, 0.000627, 0.000019)^T$ and $\hat{\boldsymbol{\gamma}}=(0.000000, 0.000203, 0.000590, 0.000000)^T$, respectively.

\section{Conclusions}
In this paper we introduce an {R} package called {\bf{alphastable}} which, provides an environment for: 1- generating random numbers from univariate, truncated, and multivariate stable distributions. 2- computing the probability density function of univariate and multivariate elliptically contoured stable distributions, 3- computing the distribution function of univariate stable distributions, 4- estimating the parameters of univariate symmetric stable, univariate Cauchy, mixture of Cauchy, mixture of univariate symmetric stable, multivariate elliptically contoured stable, and multivariate strictly stable distributions. For computing the pdf and cdf of univariate stable distribution, we use asymptotic series which yield accurate results in their convergence regions. Using an estimator of the tail index based on U-statistic, the {\bf{alphastable}} estimates the parameters of multivariate strictly stable distribution. Our proposed package uses the EM algorithm to estimate the parameters of univariate symmetric stable distribution, univariate Cauchy distribution, $d$-dimensional elliptically contoured stable distribution, mixture of univariate Cauchy distributions, and mixture of univariate symmetric stable distributions. In just mentioned cases, the EM algorithm not only is robust with respect to the initial values but also performs accurately for entire parameter space. Other packages that developed freely for {R} environment, cannot be applied for estimating the parameters of mixture of stable and multivariate stable distributions. The present work is limited to the mixture of univariate symmetric stable distributions. This methodology can be developed for mixture of multivariate elliptically contoured stable distributions. We declare that all computations have been performed on a machine with 3.5 GHz Core(TM) i7-2700K Intel(R) processor and 8 GB of RAM. Our package depends on packages {\bf{mvtnorm}}, {\bf{nnls}}, and {\bf{stabledist}} which, are available on CRAN at \verb+https://cran.r-project.org/web/packages/stabledist/index.html+, \verb+https://cran.r+ \verb+-project.org/web/packages/mvtnorm/index.html+, and \verb+https://cran.+ \verb+r-pro+ \verb+ject.org/web/packages/nnls/index.html+, respectively. The {\bf{alphastable}} package has been uploaded to the comprehensive R archive network (CRAN), see \verb+https://CRAN.R-pro+ \verb+ject.org/pac+ \verb+kage=alphastable+.
\bibliography{refs}
\bibliographystyle{abbrv}
\newpage{}
\begin{figure} [h!]
\setlength\tabcolsep{.02pt} 
\resizebox{\textwidth}{!}{\begin{tabular}{cccccc} 
\begin{rotate}{90}~~~~\tiny{~~~~~~~~$\hat{\alpha}$}\end{rotate}& 
\includegraphics[width=25mm,height=25mm]{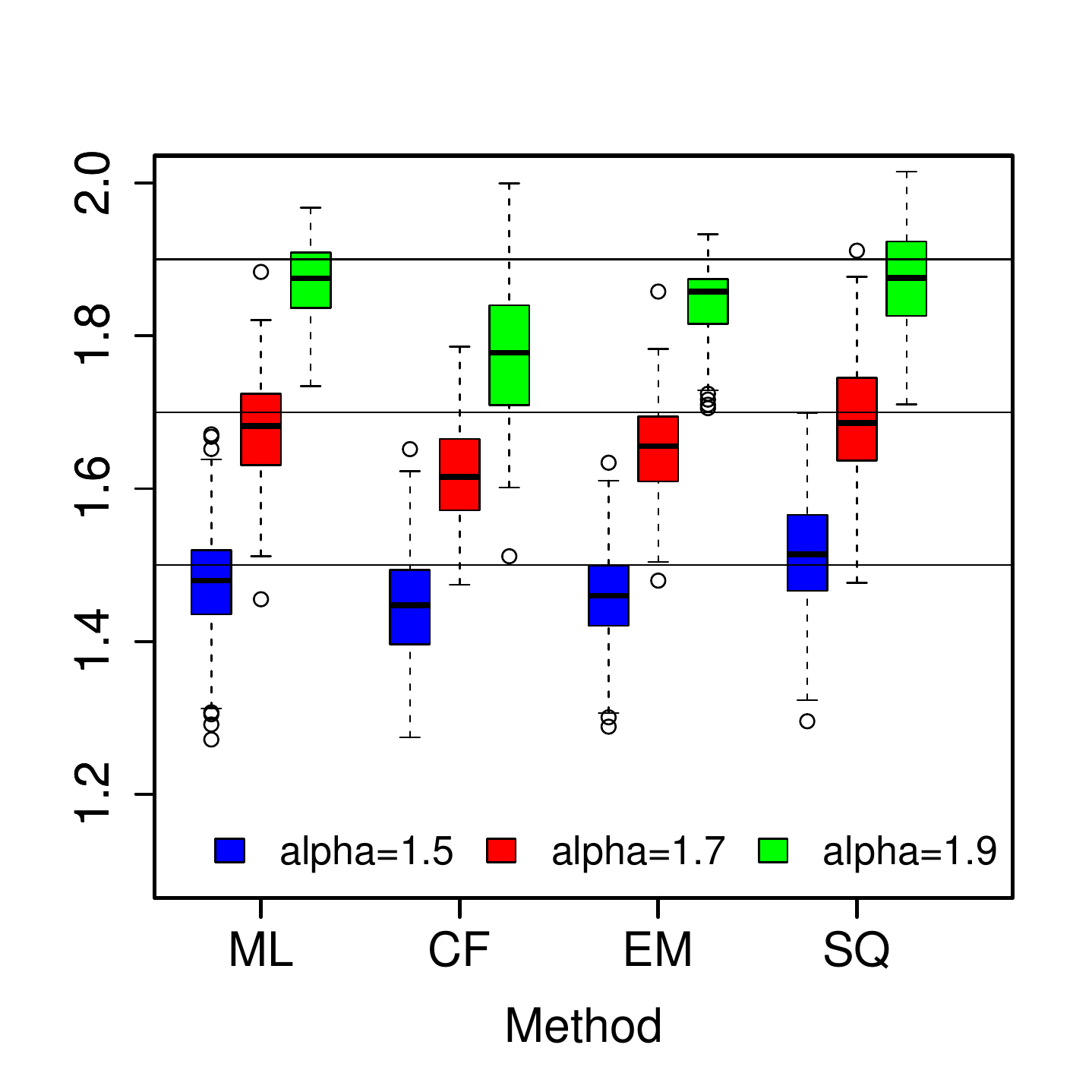}& 
\begin{rotate}{90}~~~\tiny{~~~~~~~~$\hat{\mu}_1$}\end{rotate}& 
\includegraphics[angle=90,width=25mm,height=25mm]{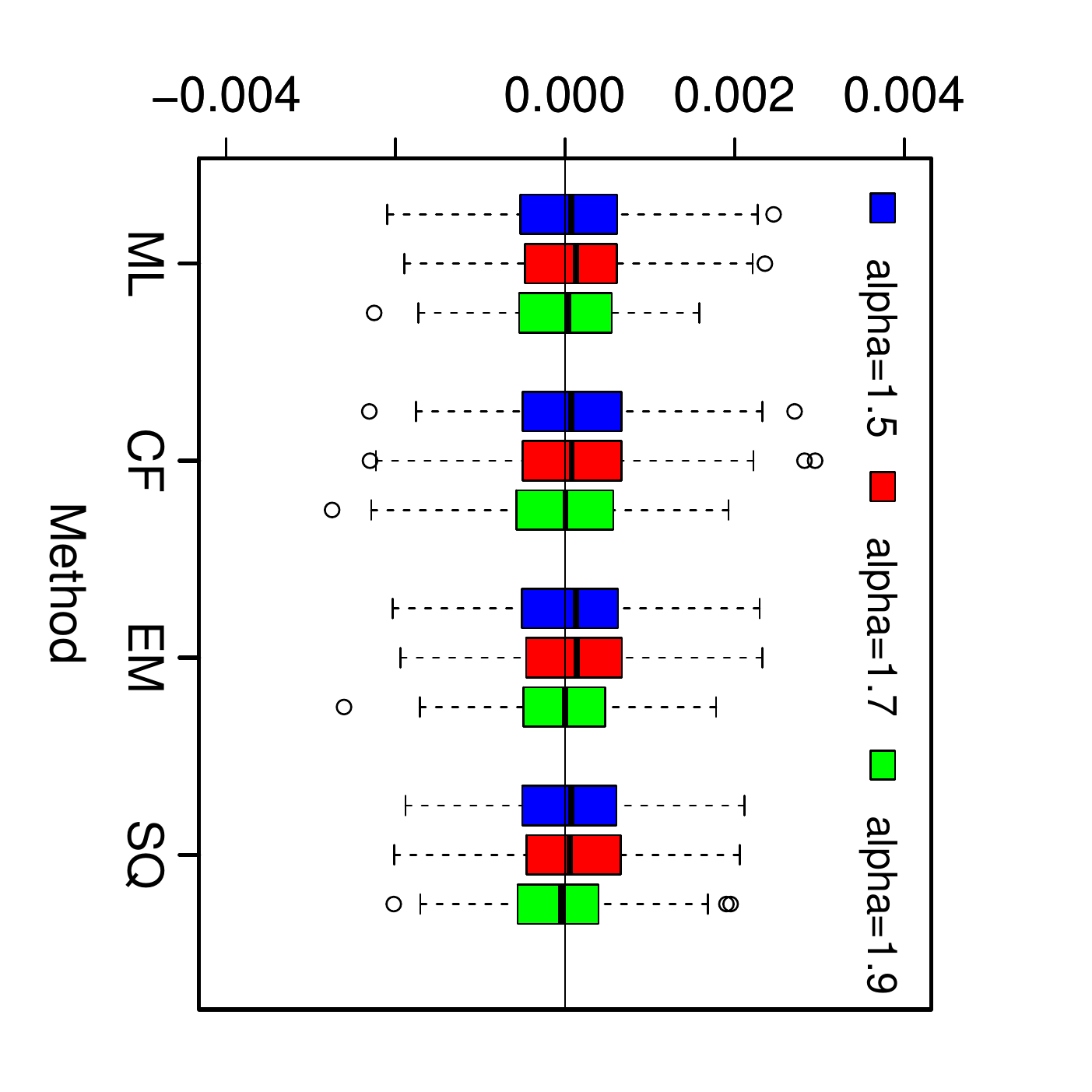}& 
\begin{rotate}{90}~~~\tiny{~~~~~~~~$\hat{\mu}_2$}\end{rotate}& 
\includegraphics[angle=90,width=25mm,height=25mm]{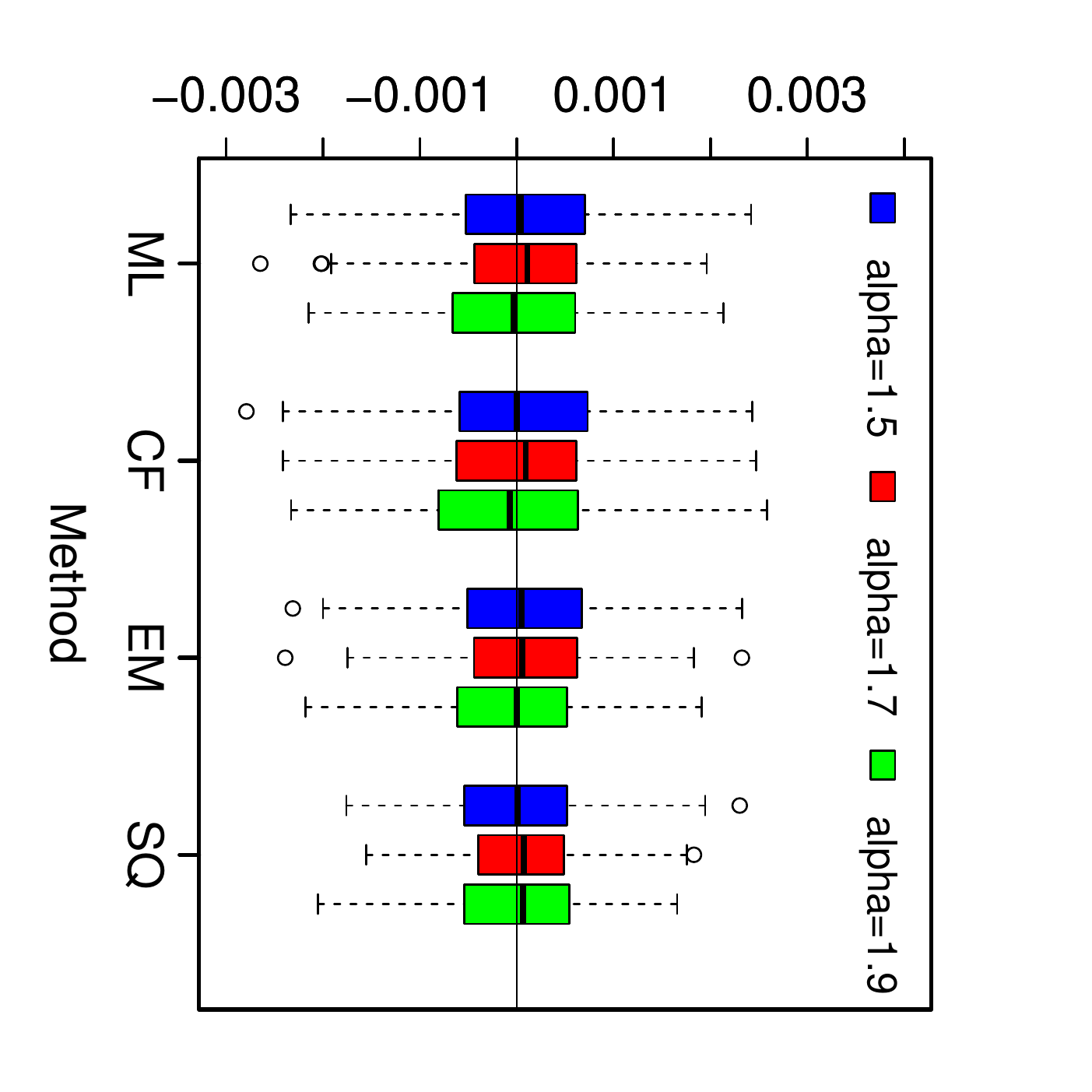}\\ 
\begin{rotate}{90}~~~\tiny{~~~~~~~~$\hat{\sigma}_{11}$}\end{rotate}& 
\includegraphics[angle=90,width=25mm,height=25mm]{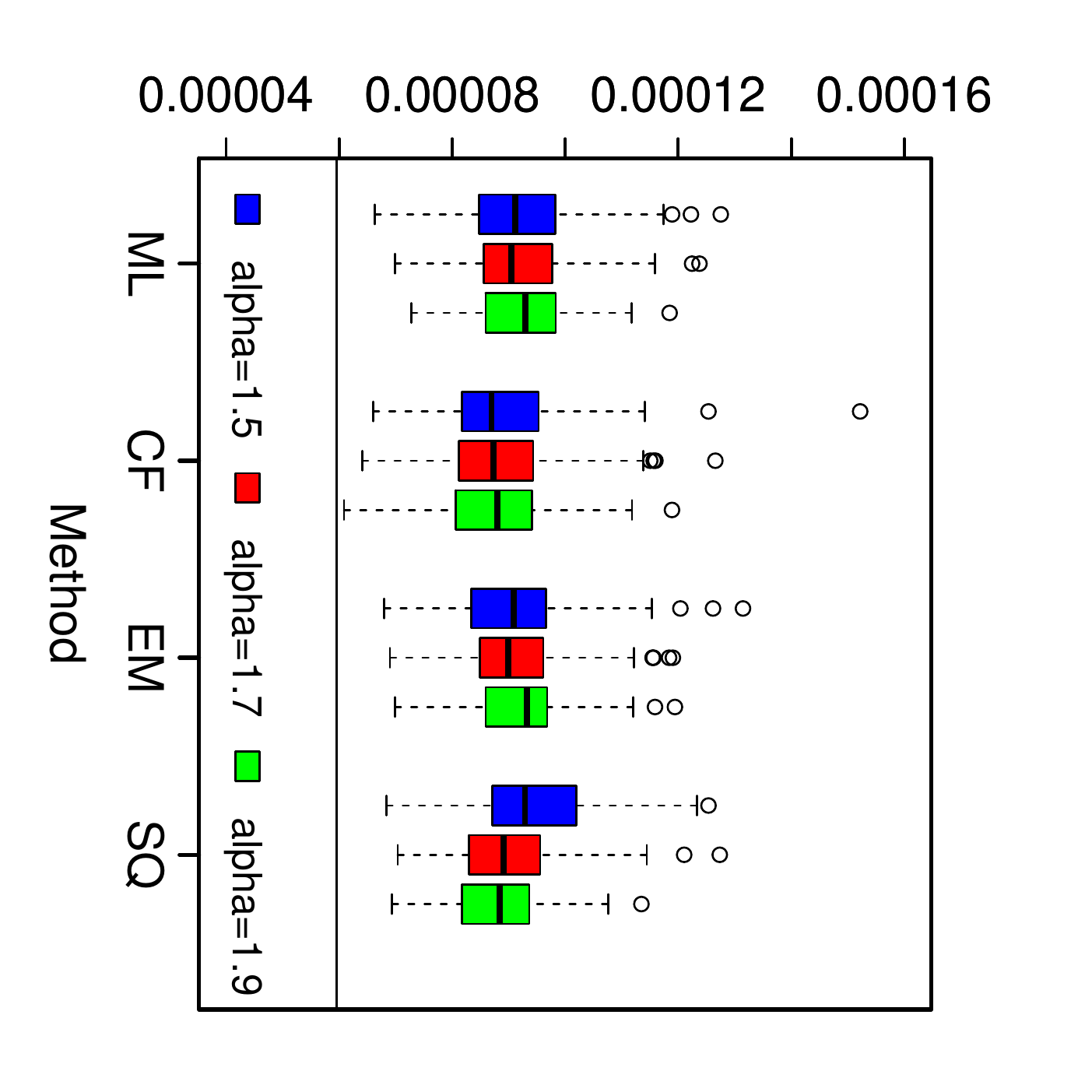}& 
\begin{rotate}{90}~~~\tiny{~~~~~~~~$\hat{\sigma}_{12}$}\end{rotate}& 
\includegraphics[width=25mm,height=25mm]{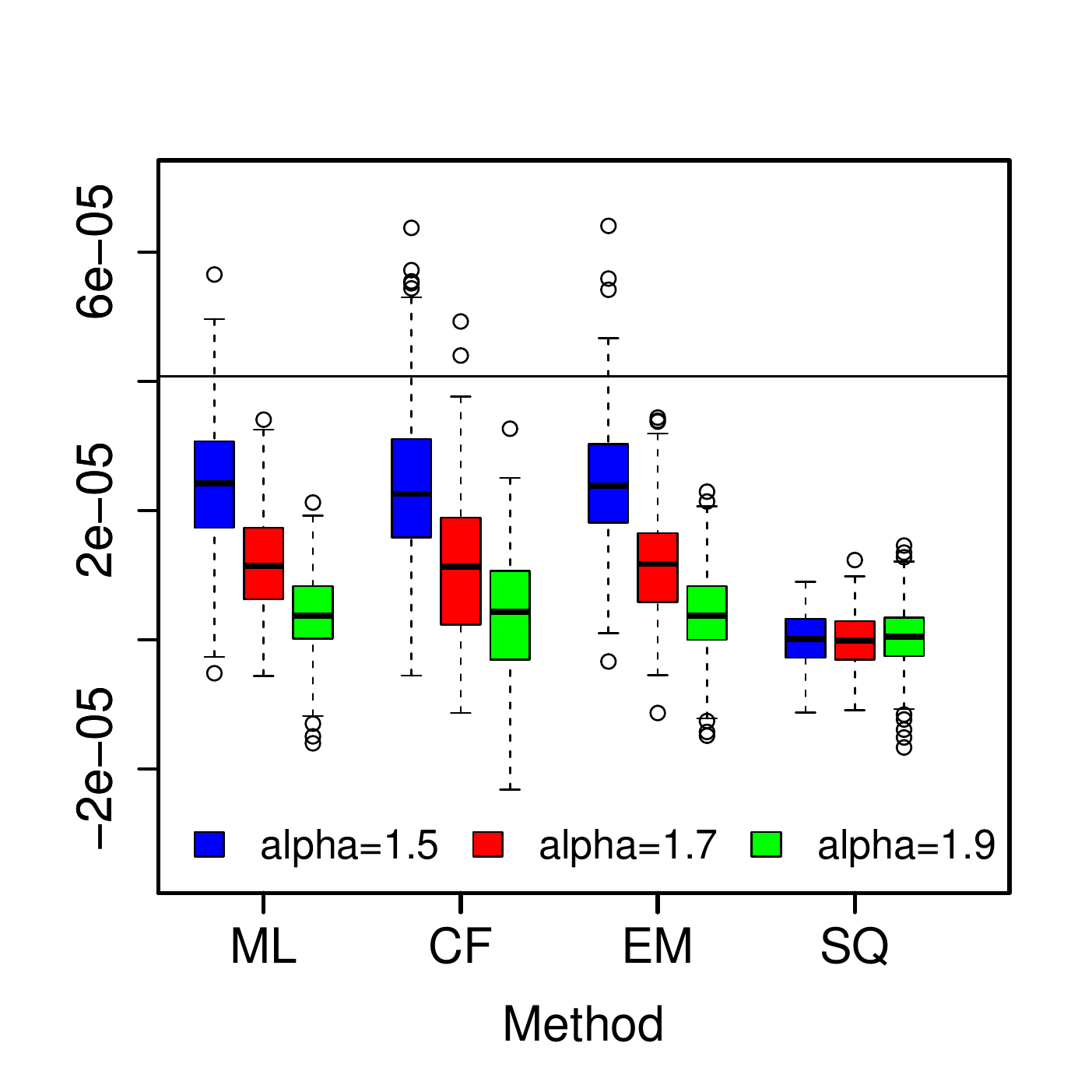}& 
\begin{rotate}{90}~~~\tiny{~~~~~~~~$\hat{\sigma}_{22}$}\end{rotate}& 
\includegraphics[angle=90,width=25mm,height=25mm]{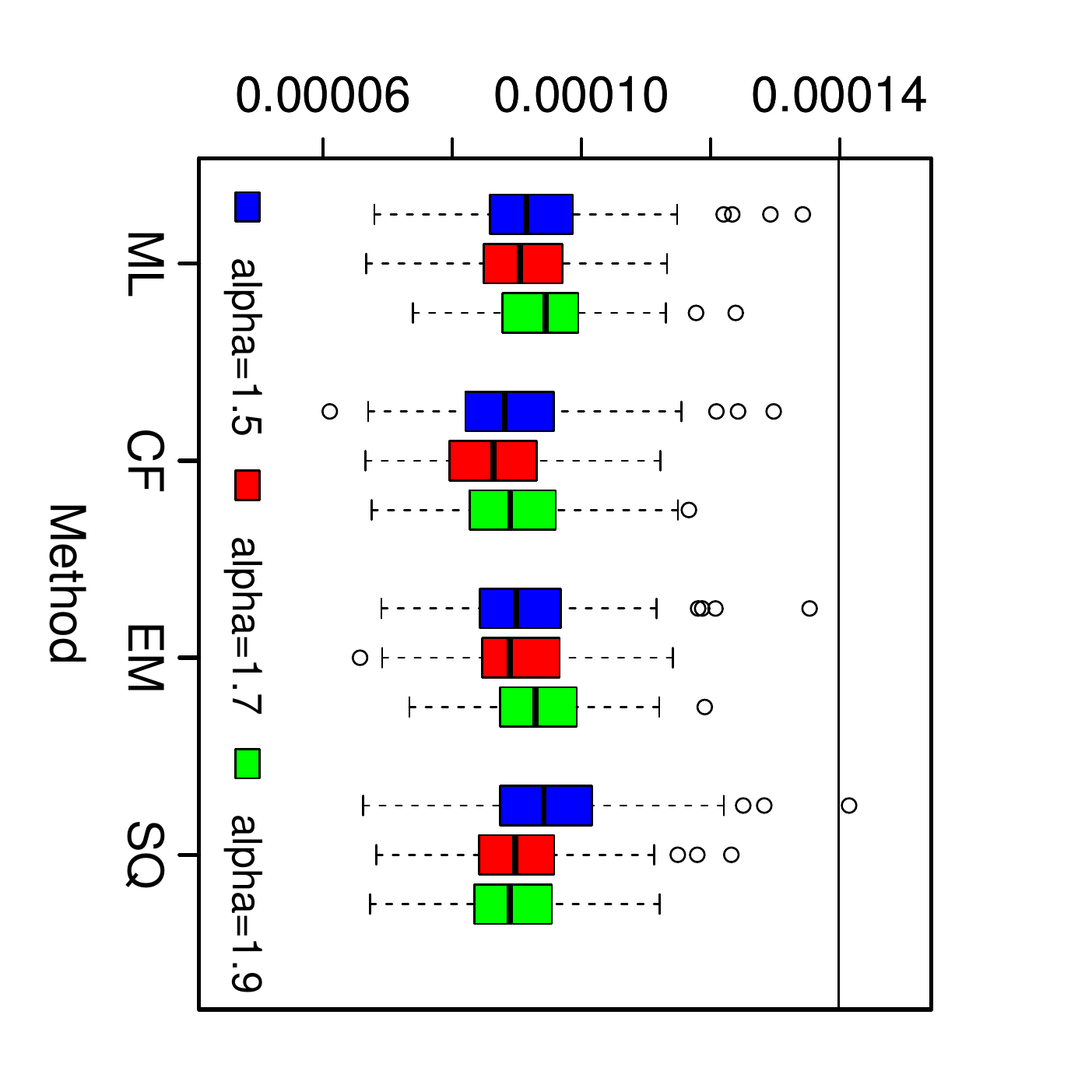}\\ 
\end{tabular}} 
\caption{Boxplot of the ML-, CF-, EM-, and SQ-based estimators of $\alpha$, $\boldsymbol{\mu}=(\mu_1,\mu_2)^{T}$, and entries of $\Sigma$ when $n=500$ vectors generated from a bivariate elliptically contoured stable with $\alpha=1.5,1.7, 1.9$, $\boldsymbol{\mu}=(\mu_1,\mu_2)^T=(0,0)^T$, $\sigma_{11}=0.000059552$, $\sigma_{12}=0.000040783$, and $\sigma_{22}=0.000139861$. In each sub-figure, the horizontal line denotes the true value of the parameter. Each boxplot is constructed based on $N=250$ runs. The used color scheme under each method for boxplots is: blue, red, and green for $\alpha$=1.5, 1.7, and 1.9, respectively.} 
\label{bivariate} 
\end{figure} 
\end{document}